\documentclass{revtex4} 

\def\nn{\nonumber} 
\usepackage{graphicx,amsmath,amssymb}
\begin{document} 

\title{Relativistic models of magnetars: structure and deformations}

\author{A. Colaiuda $^{1,2}$, V. Ferrari $^1$, L. Gualtieri$^1$,
J.A. Pons$^3$} 

\affiliation{$^1$ Dipartimento di Fisica ``G.Marconi'', 
 Universit\` a di Roma ``La Sapienza''
and Sezione INFN  ROMA1, 00185 Roma, Italy\\ 
$^2$ Institut f\"ur Theoretische Physik, 72076 T\"ubingen, 
Germany\\
$^3$ Departament de F\'{\i}sica Aplicada, Universitat d'Alacant,
03080 Alacant, Spain}

\begin{abstract} 
We find numerical solutions of the coupled system of
Einstein-Maxwell's equations with a linear approach, in which the
magnetic field acts as a perturbation of a spherical neutron star.  In
our study, magnetic fields having both poloidal and toroidal
components are considered, and higher order multipoles are also
included.  We evaluate the deformations induced by different field
configurations, paying special attention to those for which the star
has a prolate shape.  We also explore the dependence of the stellar
deformation on the particular choice of the equation of state and on
the mass of the star.  Our results show that, for neutron stars with
mass $M=1.4~M_\odot$ and surface magnetic fields of the order of
$10^{15}$ G, a quadrupole ellipticity of the order of
$10^{-6}-10^{-5}$ should be expected.  Low mass neutron stars are in
principle subject to larger deformations (quadrupole ellipticities up
to $10^{-3}$ in the most extreme case).  The effect of quadrupolar
magnetic fields is comparable to that of dipolar components. A
magnetic field permeating the whole star is normally needed to obtain
negative quadrupole ellipticities, while fields confined to the crust
typically produce positive quadrupole ellipticities.
\end{abstract} 

\maketitle 

%%%%%%%%%%%%%%%%%%%%%%%%%%%%%%%%%%%%%%%%%%%%%%%%%%%%%%%%%%%%%%%%%%%%%%%%%%% 
\section{Introduction} 
%%%%%%%%%%%%%%%%%%%%%%%%%%%%%%%%%%%%%%%%%%%%%%%%%%%%%%%%%%%%%%%%%%%%%%%%%%% 
The measured periods and spin down rates of soft-gamma repeaters (SGR)
and of anomalous X-ray pulsars (AXP), and the observed X-ray
luminosities of AXP, indicate that these neutron stars have extremely
high magnetic fields, as large as $10^{14}-10^{15}$ G
\cite{SGRAXP,WT}.  Furthermore, if these sources are the central
engine of gamma-ray bursts, as suggested in \cite{GRB}, their magnetic
field might even be larger.  Up to now, about ten highly magnetized
neutron stars, the ``magnetars'', have been identified in our Galaxy,
but their actual number may be larger, and it has been suggested that
a fraction of pulsars ( $\gtrsim 10\%$ \cite{KOUV}) would possibly
become magnetars at some stage of evolution.  The discovery of
magnetars has triggered a growing interest in the study of the
structure, dynamics and evolution of neutron stars with large magnetic
fields, and has raised a number of interesting issues.  For example,
quasi-periodic oscillations have been detected in the aftermath of the
giant flares of SGR 1806-20 and SGR 1900+14, and it is not clear
whether they are associated to crustal modes, or to modes of the
magnetic field (or both); if the spacing between the observed
frequencies would be explained, one may gain information on the
internal structure of the star \cite{FLARES}.

In addition, magnetars may be interesting sources of gravitational
waves, especially if they possess a toroidal magnetic field; indeed,
as suggested by Jones and Cutler \cite{Jones,Cutler}, a large toroidal
component tends to distort the star into a prolate shape, leading
to a secularly unstable object: the wobble angle between the angular
momentum and the star's magnetic axis would grow on a dissipation
timescale, until they become orthogonal. This may produce a copious
flux of gravitational waves, potentially detectable by the advanced
version of gravitational wave detectors LIGO and VIRGO \cite{Cutler}.

In order to understand magnetars' structure and dynamics, it is
necessary to model their equilibrium configuration in the framework of
general relativity, including both poloidal and toroidal magnetic
field components.  Toroidal fields should form during the first
seconds after core collapse, when the star is likely to be rapidly and
differentially rotating: the fluid motion would drag the poloidal
field lines creating large toroidal fields \cite{GRB}; in addition,
convective motions prevailing in the early life of a neutron star
could also create toroidal fields by dynamo
processes\cite{SGRAXP,BR,Oron02}. These toroidal components are
expected to survive when the proto-neutron star cools down and the
crust forms.  We also stress that large toroidal components contribute
to explain the giant flares in current models of SGR's \cite{SGRAXP}.
In \cite{FR} it has been shown that a purely poloidal magnetic field
is unstable, and decays on a timescale much shorter than the star's
life (see also \cite{BS} and references therein, and \cite{PG07});
however, as discussed in \cite{BS}, a magnetic field configuration
with prevailing toroidal component is also expected to be unstable on
a short timescale.  Thus, both toroidal and poloidal magnetic fields
have to be included to construct accurate, and stable, models of
magnetars.

In recent literature, magnetars equilibrium configurations have been
studied by solving Einstein-Maxwell's equations, coupled with the
Hydrodynamics equations, in full general relativity
\cite{BBGN,BG,CPL}. However, the numerical schemes used in most cases
require circularity of the space-time, i.e. the existence of two
hypersurface-orthogonal Killing vectors, and this assumption
automatically excludes toroidal magnetic fields, since they break
circularity.  Therefore, in \cite{BBGN,BG,CPL} only poloidal magnetic
fields have been considered.

A different approach has been used in \cite{IS,KOK,KOKrot}, where
equilibrium configurations have been studied using a perturbative
techniques, i.e.  solving Einstein-Maxwell-Hydrodynamics equations,
linearized about a spherically symmetric background, and expanding the
perturbed equations in tensor harmonics.  Toroidal fields have been
included in the analysis only in \cite{IS}, but this work is based on
very restrictive assumptions: the magnetic field is assumed to vanish
outside the star.  Poloidal and toroidal fields have also been
considered in the framework of Newtonian gravity in a recent work
\cite{Haskell}.

In this paper we construct equilibrium configurations of neutron stars
with strong magnetic fields, in general relativity.  Since magnetars
rotate very slowly, we restrict to non-rotating stars.  However,
rotation can play an important role in the early phases of the stellar
evolution, therefore it will be included in future developments of
this work.  We follow a perturbative approach, generalizing the work
of \cite{KOK} to include toroidal magnetic fields, with a magnitude
comparable with that of the poloidal fields.  We start solving the
relativistic Grad-Shafranov equation, to which Maxwell's equations can
be reduced, in the background of a non rotating star; we impose a set
of boundary conditions which correspond to different magnetic field
configurations, and construct the corresponding stress-energy tensor.
The magnetic field perturbs the star, which is consequently deformed;
to compute the stellar structure and its deformation, we then solve
the Einstein-Maxwell-Hydrodynamics equations linearized about the
spherically symmetric background of the non rotating star, having the
electromagnetic and the fluid stress-energy tensors as a source.  We
compare the deformation induced by a magnetic field with that which
would be produced by rotation, and find that effect of magnetic fields
is dominant for magnetars as SGR and AXP. We discuss how the magnetic
field profile and the corresponding stellar deformation depend on the
stellar mass and on the equation of state of the fluid composing the
star, comparing different stellar models. In current literature, only
the $l=1$ multipole of the electromagnetic potential is usually
considered. In this paper we also solve the relevant equations for the
$l=2$ multipole.

The main features of the perturbative approach are described in
Section \ref{model}. The results of the numerical integrations of the
relativistic Grad-Shafranov equation, and of the equations of stellar
perturbations, are reported and commented in Section \ref{results} for
different field configurations and different stellar models.  In
Section \ref{results} we also discuss the effects of the $l=2$
multipole.  Conclusions are drawn in Section \ref{conclusions}.

%%%%%%%%%%%%%%%%%%%%%%%%%%%%%%%%%%%%%%%%%%%%%%%%%%%%%%%%%%%%%%%%%%%%%%%%%%% 
\section{Structure of a stationary, axisymmetric neutron star 
with poloidal and toroidal magnetic fields}\label{model}
%%%%%%%%%%%%%%%%%%%%%%%%%%%%%%%%%%%%%%%%%%%%%%%%%%%%%%%%%%%%%%%%%%%%%%%%%%% 
In what follows we shall assume that the magnetized fluid composing
the non rotating neutron star can be described within the framework of
ideal magneto-hydrodynamics (MHD), i.e.  that there is no separation
of charge currents flowing through the star. It should be mentioned
that, although this assumption is appropriate inside the fluid core,
it may not apply to the stellar solid crust.  The magnetic field, and
the deformation it induces on the star, are treated as stationary and
axisymmetric perturbations of a spherically symmetric background. We
consider perturbations up to order $O(B^2)$.  We shall follow the
notation and the formalism introduced by Konno, Obata and Kojima in
\cite{KOK}, generalized to include toroidal magnetic fields.

Before proceeding with the perturbative approach, in the next
subsection we shall summarize some general properties of stationary,
axisymmetric magnetized stars, which will be useful in subsequent
sections. These properties and their proofs can be found in the
literature, but are scattered in different papers \cite{Carter},
\cite{Oron}; here we report them in a unified and consistent way.

%%%%%%%%%%%%%%%%%%%%%%%%%%%%%%%%%%%%%%%%%%%%%%%%%%%%%%%%%%%%
\subsection{Some properties of stationary, 
axially symmetric magnetized stars}\label{nonper}
%%%%%%%%%%%%%%%%%%%%%%%%%%%%%%%%%%%%%%%%%%%%%%%%%%%%%%%%%%%%
We consider a stationary, axisymmetric space-time describing a
magnetized star, with coordinates
\begin{equation}
x^\mu=(t,x^a,\phi)~~~~~(a=1,2)
\end{equation}
where $\eta=\partial/\partial t$ and $\xi=\partial/\partial\phi$ are
Killing vectors. The coordinates $x^a$ can be, for instance, spherical
coordinates ($r,\theta$), or cylindrical coordinates ($r,z$).  Any
stationary, axisymmetric quantity, such as the vector potential or the
fluid 4-velocity, are independent of $t$ and $\phi$, i.e.:
$A_\mu=A_\mu(x^a)$, $u^\mu=u^\mu(x^a)$.

The electric and magnetic field are defined as
\begin{equation}
E_\mu\equiv F_{\mu\nu}u^\nu~,~~~
B_\mu\equiv-\frac{1}{2}\epsilon_{\alpha\beta\gamma\delta}u^\beta 
F^{\gamma\delta}\,,
\end{equation}
where $F_{\mu\nu}\equiv\partial_\mu A_\nu-\partial_\nu A_\mu$ It may
be noted that since $\partial_t A_\mu=\partial_\phi A_\mu=0$,
$F_{t\phi}=0$.  We define the local angular velocity of the fluid as
\begin{equation}
\Omega(x^a)\equiv\frac{d\phi}{dt}=\frac{u^\phi}{u^t} \,.
\end{equation}
The components $u^a$ of the fluid velocity are called {\it meridional
currents}. Furthermore, we define the quantities
\begin{equation}
\Phi(x^a)\equiv\eta^\mu A_\mu=A_t~,~~~
\Psi(x^a)\equiv\xi^\mu A_\mu=A_\phi\,.
\end{equation}
The ideal MHD hypothesis implies that the electric field, measured by
a comoving observer, vanishes:
\begin{equation}
E_\mu=F_{\mu\nu}u^\nu=0\label{infcond}\,.
\end{equation}
An axially symmetric magnetic field is {\it poloidal} if its non
vanishing space-components are
\begin{equation}
\left(B^a,0\right)\,;
\end{equation}
it is  {\it toroidal} if
\begin{equation}
\left(0,0, B^\phi\right)\,.
\end{equation}
%%%%%%%%%%%%%%%%%%%%%%%%%%%%%%%%%%%%%%%%%%%%%%%%%%%%%%%%%%%%
\subsubsection{Vanishing meridional currents}
%%%%%%%%%%%%%%%%%%%%%%%%%%%%%%%%%%%%%%%%%%%%%%%%%%%%%%%%%%%%
If  meridional currents vanish $u^a=0$, and
eq. (\ref{infcond}) gives
\begin{equation}
E_a=F_{at}u^t+F_{a\phi}u^\phi=(F_{at}+\Omega F_{a\phi})u^t=0
\qquad\Rightarrow\qquad
\frac{F_{1t}}{F_{2t}}=\frac{F_{1\phi}}{F_{2\phi}}=-\Omega\,.\label{int1}
\end{equation}
As $F_{at}=\partial_a\Phi$ and $F_{a\phi}=\partial_a\Psi$,
eq. (\ref{int1}) becomes
\begin{equation}
\partial_a\Phi=-\Omega\partial_a\Psi\,.
\label{partder}
\end{equation}
Furthermore, from (\ref{int1}) we have
\begin{equation}
\partial_1\Phi\partial_2\Psi-\partial_2\Phi\partial_1\Psi=0\,,
\end{equation}
which implies (assuming the domain where $\Phi$ and $\Psi$ are defined
is simply connected) that $\Phi=\Phi(\Psi)$. From eq. (\ref{partder})
then it follows
\begin{equation}
\frac{d\Phi}{d\Psi}=-\Omega\,.
\end{equation}
As shown by Carter (see theorem 7 of \cite{Carter}, and its corollary;
see also \cite{BGSM}), if the space-time is stationary and
axisymmetric, and if meridional currents are zero, then
$A_a=0$. Therefore, the vector potential is
\begin{equation}
A_\mu=(\Phi,0,0,\Psi)\,,
\end{equation}
the electromagnetic tensor becomes
\begin{equation}
F_{\mu\nu}=\left(\begin{array}{c|c|c}
0 & \Omega\Psi_{,a} & 0 \\\hline
-\Omega\Psi_{,a} & 0 & \Psi_{,a} \\\hline
0 & -\Psi_{,a} & 0 \\
\end{array}\right)\,,
\end{equation}
and the magnetic field is 
\begin{equation}
B^\alpha=\epsilon^{\alpha\beta\mu\nu}u_{\beta}F_{\mu\nu}=
\left(0,B^a,0\right)\,,
\end{equation}
since $g^{at}=g^{a\phi}=0$ when $u^a=0$ \cite{Carter}.  Thus, if
meridional currents vanish the magnetic field is poloidal.
%%%%%%%%%%%%%%%%%%%%%%%%%%%%%%%%%%%%%%%%%%%%%%%%%%%%%%%%%%%%
\subsubsection{Non-vanishing meridional currents}
%%%%%%%%%%%%%%%%%%%%%%%%%%%%%%%%%%%%%%%%%%%%%%%%%%%%%%%%%%%%
Let us now consider the general case $u^a\neq 0$. eq. (\ref{infcond})
gives
\begin{eqnarray}
E_t&=&F_{ta}u^a=-u^a\partial_a\Phi=-u^\mu\partial_\mu\Phi=-\frac{d\Phi}{d\tau}
=0\nn\\
E_\phi&=&F_{\phi a}u^a=-u^a\partial_a\Psi=-u^\mu\partial_{\mu}\Psi
=-\frac{d\Psi}{d\tau}
=0\,,
\end{eqnarray}
i.e., $\Phi,\Psi$ are constant along the fluid flow. Then
\begin{eqnarray}
&&u^1\Phi_{,1}+u^2\Phi_{,2}=0\nn\\
&&u^1\Psi_{,1}+u^2\Psi_{,2}=0\,,
\label{int2}
\end{eqnarray}
which implies $\Phi=\Phi(\Psi)$. We introduce the quantity
\begin{equation}
\bar\Omega(\Psi)\equiv-\frac{d\Phi}{d\Psi}\,,
\end{equation}
so that
\begin{equation}
\partial_{a}\Phi=-\bar\Omega\partial_{a}\Psi\,.
\end{equation}
Notice that in general $\bar\Omega\neq\Omega$. Indeed
\begin{equation}
E_a=-\left(\partial_a\Phi+\Omega\partial_a\Psi\right)u^t+F_{ab}u^b
=(\bar\Omega-\Omega)\partial_a\Psi u^t+F_{ab}u^b=0\,\qquad a,b=1,2\,;
\label{OOb}
\end{equation}
thus, if $F_{ab}\neq0$, then $\bar\Omega\neq\Omega$. 

From eq. (\ref{int2}) it also follows 
\begin{equation}
\Psi_{,2}=-\frac{u^1}{u^2}\Psi_{,1}\,,\label{psia0}
\end{equation}
which, differentiated with respect to $x^1$, gives
\begin{equation}
-\left(\frac{u^1}{u^2}\right)_{,1}\Psi_{,1}=
\frac{u^a}{u^2}\partial_a\Psi_{,1}\,.
\label{conta0}
\end{equation}
Using the {\it continuity equation}
\begin{equation}
u^\alpha_{~,\alpha}=-\frac{d}{d\tau}\ln(\sqrt{-g}n)\,,
\end{equation}
where $n$ is the baryon number density,
eq. (\ref{conta0}) can be transformed as follows
\begin{equation}
\frac{d}{d\tau}\ln(\Psi_{,1})=u^a\partial_{a}\ln(\Psi_{,1})
=-u^2\left(\frac{u^1}{u^2}\right)_{,1}
=-u^\alpha_{~,\alpha}+\frac{u^\alpha u^2_{,\alpha}}{u^2}
=\frac{d}{d\tau}\ln\left(nu^2\sqrt{-g}\right)\,.
\end{equation}
If we now define
\begin{equation}
C\equiv\frac{\Psi_{,1}}{nu^2\sqrt{-g}}\,,\label{definC}
\end{equation}
we find
\begin{equation}
\frac{d}{d\tau}C=u^aC_{,a}=0\quad\Rightarrow\quad
u^1C_{,1}+u^2C_{,2}=0\,,
\end{equation}
which, together with eq. (\ref{psia0})
implies that $C$ is a function of $\Psi$ only, i.e.
$C=C(\Psi)$.

By replacing
$\Psi_{,1}=C(\Psi)n u^2\sqrt{-g}$ in eq. (\ref{OOb}) we find
\begin{equation}
F_{12}=-(\bar\Omega-\Omega)C n u^t \sqrt{-g}\,.\label{F12C}
\end{equation}
Then, if $\Omega$ and $\bar\Omega$ do not coincide,  $F_{12}\neq0$;
consequently,  the magnetic field has both poloidal and toroidal components.

A possible interpretation of  $\bar\Omega$
is the following (see for instance \cite{IS}).
From eq. (\ref{F12C}) we find
\[
\Omega = \bar\Omega + \frac{F_{12}}{C n u^t \sqrt{-g}}\,,
\]
from which we see that the fluid angular velocity $\Omega$
has two contributions: the first, $\bar\Omega$, can interpreted as due to the
stellar rotation, the second is clearly due to the electromagnetic field. 
Although this interpretation is purely conventional,
since we are considering a non rotating star, we shall assume
$\bar\Omega=0$, and consequently
$A_0=\Phi=0$. Thus the form of the vector potential is
\begin{equation}
A_\mu(r,\theta)=\left(0,A_r,A_\theta,\Psi\right)\,.
\label{vectorpot}
\end{equation}
%%%%%%%%%%%%%%%%%%%%%%%%%%%%%%%%%%%%%%%%%%%%%%%%%%%%%%%%%%%%
\subsubsection{Electromagnetic current and Lorentz Force}
%%%%%%%%%%%%%%%%%%%%%%%%%%%%%%%%%%%%%%%%%%%%%%%%%%%%%%%%%%%%
The Lorentz force is defined as
\begin{equation}
f_{\mu}\equiv F_{\mu\nu}J^\nu\,,
\label{deflorentz}
\end{equation}
where the electromagnetic current $J^\mu$ is given by
\begin{equation}
J^\mu=\frac{1}{4\pi\sqrt{-g}}\left(\sqrt{-g}F^{\mu\nu}\right)_{,\nu}\,.
\end{equation}
The stress-energy tensor of a perfect fluid with an electromagnetic
field is
\begin{equation}
T^{\mu\nu}=T^{\mu\nu}_{fluid}+T^{\mu\nu}_{em}
\end{equation}
where
\begin{equation}
T^{\mu\nu}_{fluid}=(\rho+p)u^\mu u^\nu+pg^{\mu\nu}\,,\qquad
T^{\mu\nu}_{em}=\frac{1}{4\pi}\left(F^{\mu\alpha}F^{\nu}_{~\alpha}
-\frac{1}{4}g^{\mu\nu}F^{rho\sigma}F_{\rho\sigma}\right)\,.
\end{equation}
By projecting the equation $T^{\mu\nu}_{~~;\nu}=0$ orthogonally to
$u^\mu$, we find the relativistic Euler equation in presence of
a magnetic field:
\begin{equation}
(\rho+p)a_\mu+p_{,\mu}+u_\mu u^\nu p_{,\nu}-f_\mu=0\,,\label{Euler}
\end{equation}
where $a_\mu=u^\nu u_{\mu;\nu}$.
Let us now consider the $\phi$ component of this equation. 
Under the  stationarity and axisymmetry assumption it becomes
\begin{equation}
(\rho+p)a_\phi+u_\phi u^ap_{,a}-f_\phi=0\,; \label{blippo}
\end{equation}
being
\begin{equation}
a_\phi=u^\mu u_{\phi;\mu}=u^a u_{\phi,a}-u^\mu
u^\nu\Gamma_{\phi\mu\nu}=u^a u_{\phi,a}+\frac{1}{2}u^\mu
u^\nu g_{\mu\nu,\phi}=u^a u_{\phi,a}\,,
\end{equation}
using the first law of thermodynamics,
$u^ap_{,a}=\frac{\rho+p}{n}u^an_{,a}$, eq. (\ref{blippo}) gives
\begin{equation}
f_\phi=\frac{\rho+p}{n}u^a\left(nu_\phi\right)_{,a}\,.
\label{fphi}
\end{equation}
If meridional currents are zero,  $f_\phi=0$. 
%%%%%%%%%%%%%%%%%%%%%%%%%%%%%%%%%%%%%%%%%%%%%%%%%%%%%%%%%%%%
\subsection{The equations for the vector potential $A^\mu$}
%%%%%%%%%%%%%%%%%%%%%%%%%%%%%%%%%%%%%%%%%%%%%%%%%%%%%%%%%%%%
The background geometry of the star in coordinates $(t,r,\theta,\phi)$
is 
\begin{eqnarray}
ds^2&=&-e^{\nu(r)}dt^2+e^{\lambda(r)}dr^2+r^2(d\theta^2+\sin^2\theta d\phi^2)
=g^{(0)}_{\mu\nu}dx^\mu dx^\nu\label{g0}\\
u^{(0)\mu}&=&(e^{-\nu/2},0,0,0)\,,
\label{u0}
\end{eqnarray}
where $\nu(r),\lambda(r)$ are the solution of Einstein's equations for
an assigned equation of state.  If a magnetic field is present, from
the expression of $F_{\mu\nu}$ in terms of the electric and magnetic
field $F_{\mu\nu}=u_\mu E_\nu -u_\nu E_\mu
+\epsilon_{\mu\nu\alpha\beta}u^\alpha B^\beta$ we see that, if
$E_\mu=0$, then $F_{\mu\nu} =O(B)$; consequently, also the vector
potential $A_\mu$ is of order $O(B)$.  By using a function
$\Lambda(r,\theta)$ such that $\Lambda_{,\theta}=A_\theta$, we can
gauge away the $\theta$ component of the vector potential
(\ref{vectorpot}). By introducing the function $\Sigma(r,\theta)\equiv
e^{\frac{\nu-\lambda}{2}}(A_r-\Lambda_{,r})$, it then becomes
\begin{equation}
A_\mu=\left(0,e^{\frac{\lambda-\nu}{2}}\Sigma,0,\Psi\right)\,,
\label{vectorpot1}
\end{equation}
with $\Sigma$ and $\Psi$ of order $O(B)$.

The magnetic field induces motion in the fluid, and consequently
induces a perturbation on the components of the four-velocity $\delta
u^\alpha$, on the pressure and energy density ($\delta p$ and $\delta
\rho$, respectively), and on the metric $\delta g_{\mu\nu}$. Since
$T^{\mu\nu}_{em}=O(B^2)$, linearizing the equation
$T^{\mu\nu}_{~~;\nu}=0$ (and using the vanishing of the space
components of $u^\alpha$ when the magnetic field is absent), it
is easy to see that $\delta u^\alpha = \delta p= \delta \rho=O(B^2)$.
In a similar way, from the linearized Einstein equations it can be
shown that $\delta g_{\mu\nu}= O(B^2)$.  Thus, from eq.  (\ref{fphi})
we see that, since $\delta u^a=O(B^2)$, the $\phi$-component of the
Lorentz force is $f_\phi = O(B^4)$ and for this reason  hereafter we
shall set it equal to zero. This condition will be used to further
simplify the expression of the vector potential.  We stress that the
condition $f_\phi=0$ comes from the fact that $f_\phi = O(B^4)$, but
we do not assume that meridional current are zero.  If we compute
$f_\phi$ from Maxwell's equations and impose $f_\phi=0$ we find
\begin{equation}
f_\phi=\left(\Sigma_{,\theta\theta}+\cot\theta\Sigma_{,\theta}\right)
\Psi_{,r}-\Sigma_{,\theta r}\Psi_{,\theta}=0\,,
\end{equation}
therefore, if we define
\begin{equation}
\bar\Psi\equiv\sin\theta\Sigma_{,\theta}\,,
\end{equation}
we have $\bar\Psi_{,\theta}\Psi_{,r}-\bar\Psi_{,r}\Psi_{,\theta}=0$;
this equation implies $\bar\Psi=\bar\Psi(\Psi)$, and consequently, since
$\bar\Psi=O(B)$ and $\Psi=O(B)$,  we can write
$\bar\Psi=\zeta\Psi$, where $\zeta$ is a constant of
order $O(1)$. The equation
\begin{equation}
\zeta\Psi=\sin\theta\Sigma_{,\theta}
\end{equation}
is satisfied by
\begin{eqnarray}
\Sigma&=&\zeta a\nn\\
\Psi&=&\sin\theta a_{,\theta}\,,
\end{eqnarray}
with $a=a(r,\theta)$. Thus, the vector potential can be written as
\begin{equation}
A_\mu=(0,\zeta e^{(\lambda-\nu)/2}a,0,\sin\theta a_{,\theta})\,.\label{Amurth}
\end{equation}
As a consequence, the magnetic field takes the following form
\begin{eqnarray}
B_\mu=\frac{e^{-\lambda/2}}{\sin\theta}
(0, \frac{e^{\lambda}}{r^2}\left(\sin\theta a_{,\theta}\right)_{,\theta},
-\left(\sin\theta a_{,\theta}\right)_r,
-\zeta \sin^2\theta e^{(\lambda-\nu)/2}a_{,\theta})\,.
\label{BTOT}
\end{eqnarray}
From this expression we see that the coefficient $\zeta$ (or the
dimensionless quantity $\zeta R$, where $R$ is the radius of the star)
represents the ratio between the toroidal and the poloidal components
of the magnetic field. Since, as discussed in the introduction, a
magnetic field configuration with prevailing toroidal component is
expected to be unstable, we will not consider 
configurations with $\zeta R\gg 1$.

Assuming the form (\ref{Amurth}) of the vector potential, we find
(neglecting the metric perturbations, which contribute to higher
orders of $B$)
\begin{eqnarray}
f_a=(\sin\theta a_{,\theta})_{,a}\frac{\tilde J_\phi}{r^2\sin^2\theta}\,,
\label{def_fa}
\end{eqnarray}
where 
\begin{equation}
\tilde J_\phi\equiv J_\phi-\zeta^2\frac{e^{-\nu}}{4\pi}\sin\theta
a_{,\theta}\,.
\end{equation}

We shall now show that $f_a$ can be written as $(\rho+p)$ times the
gradient of a function of $(r,\theta)$.  Let us consider the
$a$-components of Euler's equation:
\begin{equation}
(\rho+p)a_a+p_{,a}+u_a u^b p_{,b}
-f_a=0\,.\label{Eulera}
\end{equation}
We remind that:
\[
u^i\equiv \delta u^i = O(B^2),\qquad g^{0i}\equiv \delta g^{0i}= O(B^2),\qquad
f_a= F_{a\mu}J^\mu=O(B^2)\quad i=1,2,3\,.
\]
Consequently, the term $u_a u^bp_{,b}$ is $O(B^4)$.  We shall now
compute $a_a$ and $p_{,a}$ up to terms of order $O(B^2)$.

The acceleration is 
\begin{eqnarray}
a_a&=&u^\mu u_{a;\mu} =u^b u_{a,b}-u^\mu
u^\nu\Gamma_{a\mu\nu}\simeq  \frac{1}{2}u^\mu u^\nu g_{\mu\nu,a}\nonumber\\
&=& \frac{1}{2}\left[ (u^0)^2 g_{00,a}+ 2 u^0 u^i g_{0i,a} + 
u^i u^j g_{ij,a}\right]
\simeq \frac{1}{2}(u^0)^2 g_{00,a}~.
\end{eqnarray}
We also find
\begin{equation}
g_{\mu\nu}u^\mu u^\nu =-1 = (u^0)^2 g_{00}+ 2 u^0 u^i g_{0i} + u^i u^j g_{ij}
\simeq (u^0)^2 g_{00}~,
\end{equation}
then
\begin{equation}
a_a=\frac{1}{2}\left(\ln (-g_{00})\right)_{,a}\,.
\label{def_aa}
\end{equation}
From the first principle of thermodynamics, written for a barotropic
equation of state $p = p(\rho)$, we find
\begin{equation}
p_{,a}=(\rho+p)
\left(\ln\frac{\rho+p}{n}\right)_{,a}\,.
\label{def_pa}
\end{equation}
If we introduce the function
\begin{equation}
\chi= \ln\left(\sqrt{-g_{00}} \frac{\rho+p}{n}\right)\,,
\label{defchi}
\end{equation}
using eqs. (\ref{def_fa}), (\ref{def_aa}) and (\ref{def_pa}),
eq. (\ref{Eulera}) becomes
\begin{equation}
(\rho+p)\chi_{,a}=(\sin\theta a_{,\theta})_{,a}
\frac{\tilde J_\phi}{r^2\sin^2\theta}\label{int3}\,.
\end{equation}
This equation is equivalent to eq. (12) of \cite{BBGN}.
From (\ref{int3}) we find
\begin{equation}
\chi_{,12}-\chi_{,21}=
(\sin\theta a_{,\theta})_{,1}
\left(\frac{\tilde J_\phi}{r^2\sin^2\theta(\rho+p)}\right)_{,2}
-(\sin\theta a_{,\theta})_{,2}
\left(\frac{\tilde J_\phi}{r^2\sin^2\theta(\rho+p)}\right)_{,1}
=0~,
\end{equation}
hence 
\begin{equation}
\frac{\tilde J_\phi}{r^2\sin^2\theta(\rho+p)} = F(\sin\theta a_{,\theta})\,.
\end{equation}
By expanding in powers of $B$ we find
\begin{equation}
\frac{\tilde J_\phi}{r^2\sin^2\theta
(\rho^{(0)}+p^{(0)})}=c_0+c_1\sin\theta a_{,\theta}+ O(B^2)\,,
\label{defc01}
\end{equation}
thus, the $\phi$ component of the electromagnetic current can be
written as follows
\begin{equation}
J_{\phi}=\zeta^2\frac{e^{-\nu}}{4\pi}\sin\theta
a_{,\theta}+\left[c_0+c_1\sin\theta a_{,\theta}\right]
(\rho^{(0)}+p^{(0)})r^2\sin^2\theta+ O(B^2)\,.\label{allJ}
\end{equation}
In the next section we will expand $a(r,\theta)$ in Legendre's
polynomials; if we assume $c_1\neq 0$, different harmonic components
of the field couple.  Following \cite{IS,KOK,KOKrot,Haskell},
hereafter we shall assume $c_1=0$.  With this simplification,
\begin{equation}
J_{\phi}=\zeta^2\frac{e^{-\nu}}{4\pi}\sin\theta
a_{,\theta}+c_0(\rho^{(0)}+p^{(0)})r^2\sin^2\theta\,,
\label{Jph}
\end{equation}
where $c_0$ is a constant of order $O(B)$.
The  $r,\theta$ components of the current are simply:
\begin{equation}
J_a=\frac{\zeta e^{-(\lambda+\nu)/2}}{4\pi\sin\theta}
\left(-\frac{e^\lambda}{r^2}(\sin\theta a_{,\theta})_{,\theta},(\sin\theta
a_{,\theta})_{,r}\right)\,.
\end{equation}
The electromagnetic current is the sum of two parts:
\begin{equation}
J_\mu=J_\mu^p+J_\mu^t
\end{equation}
with
\begin{eqnarray}
J_\mu^p&=&(0,0,0,c_0r^2\sin^2\theta(\rho^{(0)}+p^{(0)}))\nn\\
J_\mu^t&=&-\frac{\zeta e^{-\nu/2}}{4\pi}B_\mu\,.\label{Jpt}
\end{eqnarray}
$J_\mu^p$ is the source of the poloidal field (which does not depend
on $\zeta$); $J_\mu^t$ is the source of the toroidal field
(proportional to $\zeta$) and it is parallel to the magnetic
field. Note that
\begin{itemize}
\item{} since $J_\mu^t \propto B^\mu$, it follows that
$F_{\mu\nu}J^{t \nu}=0$;
\item{} when $J_\mu^p=0$ (i.e. when $c_0=0$),
then 
$f_\mu= F_{\mu\nu}J^{\nu} = 0$;
 therefore in this case the magnetic field is 
{\it force free}.
\end{itemize}

If outside the star we assume there is vacuum, currents must
vanish. As the poloidal current is proportional to
$\rho^{(0)}+p^{(0)}$, it automatically vanishes; conversely, the toroidal
current vanishes only if $\zeta=0$, i.e. if the toroidal field
vanishes. Therefore, in vacuum only  poloidal fields (with no current)
are allowed.

If outside the star there is a magnetosphere, the situation is
different because currents can be present, and consequently toroidal
fields can exist.  In any event, since the energy density in the
magnetosphere is negligible with respect to that prevailing in the
stellar interior, $J_\mu^p$ is negligible and the magnetic field is
force free.

%%%%%%%%%%%%%%%%%%%%%%%%%%%%%%%%%%%%%%%%%%%%%%%%%%%%%%%%%%%%
\subsection{The relativistic Grad-Shafranov equation}\label{secgrad}
%%%%%%%%%%%%%%%%%%%%%%%%%%%%%%%%%%%%%%%%%%%%%%%%%%%%%%%%%%%%
If we expand the function $a(r,\theta)$ in Legendre's polynomials:
\begin{equation}
a(r,\theta)=\sum_{l=1}^\infty a_l(r)P_l(\theta)\,,
\label{expandinga}
\end{equation}
the vector potential (\ref{Amurth}) and the magnetic field (\ref{BTOT}) become
\begin{eqnarray}
A_\mu&=&(0,\zeta e^{(\lambda-\nu)/2}\sum_l a_l
P_l,0,\sum_la_l\sin\theta P_{l,\theta})\\
\label{Amuall}
B_\mu&=&\sum_l\left(0,-\frac{e^{\lambda/2}}{r^2}l(l+1)a_lP_l.
-e^{-\lambda/2}a_{l,r}P_{l,\theta},-\zeta e^{-\nu/2}
a_l\sin\theta P_{l,\theta}\right)\,.\label{Bmuall}
\end{eqnarray}
From Maxwell's equations we find
$J_\phi=\frac{1}{4\pi}F_{\phi~~;\mu}^{~\mu}$, which gives
\begin{equation}
J_\phi=-\frac{1}{4\pi}\sin\theta\sum_l P_{l,\theta}\left(
e^{-\lambda}a_{l,rr}+\frac{\nu_{,r}-\lambda_{,r}}{2}e^{-\lambda}a_{l,r}
-\frac{l(l+1)}{r^2}a_l\right)\,.\label{Jp1}
\end{equation}
Using the expansion (\ref{expandinga}), eq. (\ref{Jph}) gives
\begin{equation}
J_{\phi}=\zeta^2\frac{e^{-\nu}}{4\pi}\sum_l a_l\sin\theta 
P_{l,\theta}+c_0(\rho+p)r^2\sin^2\theta
=\zeta^2\frac{e^{-\nu}}{4\pi}\sum_l\sin\theta a
P_{l,\theta}-c_0(\rho+p)r^2\sin\theta P_{1,\theta}\,.\label{Jp2}
\end{equation}
Notice that the poloidal current (i.e. the term in $c_0$) introduces
an $l=1$ dipole component.  The linearized relativistic Grad-Shafranov
equation \cite{IS} is found by equating eqs. (\ref{Jp1}) and
(\ref{Jp2}) (see also \cite{KOK}):
\begin{eqnarray}
e^{-\lambda}a_1''+\frac{\nu'-\lambda'}{2}e^{-\lambda}a_1'+\left(
\zeta^2e^{-\nu}-\frac{2}{r^2}\right)a_1&=&4\pi(\rho+p)r^2c_0\,,
\label{eqa1}\\
e^{-\lambda}a_l''+\frac{\nu'-\lambda'}{2}e^{-\lambda}a_l'+\left(
\zeta^2e^{-\nu}-\frac{l(l+1)}{r^2}\right)a_l&=&0~~~~~(l>1)\,.\label{eqal}
\end{eqnarray}
Hereafter we will consider only the solution of equation (\ref{eqa1})
corresponding to $l=1$, in which case the vector potential
(\ref{Amuall}) and the magnetic field (\ref{Bmuall}) become
\begin{eqnarray}
A_\mu&=&\left(0,\zeta e^{(\lambda-\nu)/2}a_1\cos\theta,
0,-a_1\sin^2\theta\right)\label{Amu}\\
B_\mu&=&\left(0,-2\frac{e^{\lambda/2}}{r^2}a_1\cos\theta,e^{-\lambda/2}
a_1'\sin\theta,\zeta e^{-\nu/2}a_1\sin^2\theta\right)\,.\label{Bmu}
\end{eqnarray}
It is convenient to express the magnetic field in terms of the
orthonormal tetrad components (i.e. those measured in a locally
inertial frame) in the background metric (\ref{g0}), i.e.
\begin{eqnarray}
B_{(r)}&=&-\frac{2a_1}{r^2}\cos\theta\label{Br}\\
B_{(\theta)}&=&\frac{e^{-\lambda/2}a_1'}{r}\sin\theta\label{Btheta}\\
B_{(\phi)}&=&\zeta\frac{e^{-\nu/2}a_1}{r}\sin\theta\,.\label{Bphi}
\end{eqnarray}
%%%%%%%%%%%%%%%%%%%%%%%%%%%%%%%%%%%%%%%%%%%%%%%%%%%%%%%%%%%%
\subsection{Boundary conditions and matching with the exterior}
\label{bcs}
%%%%%%%%%%%%%%%%%%%%%%%%%%%%%%%%%%%%%%%%%%%%%%%%%%%%%%%%%%%%
Different choices of the boundary conditions and of the matching
conditions of the interior and exterior solutions of the
Grad-Shafranov equation, correspond to different physical
configurations. We shall consider the following cases.
\begin{itemize} 
\item Magnetic field extending throughout the star.  This
configuration has been studied in the literature in several papers
(for instance, in \cite{KOK,IS}); however, it conflicts with the
common belief that the neutron star core is superconductor. Actually,
if the superconductor is of type II, the magnetic field extends
throughout the star, but it has a very complicated structure (it is
``quantized'' in flux tubes).  Thus, the smooth magnetic field we
consider in this paper is a rough representation of such
configuration.

If we impose a regular behavior at the origin (which implies
$a_1(r\simeq 0)=\alpha_0r^2+O(r^4)$), for each pair of assigned
constants $\alpha_0,c_0$ the solution $a_1(r)$ is unique.

\item Crustal fields.  If matter in the  core is
a type I superconductor, the magnetic field is confined in the crust,
i.e. within
\begin{equation}
r_c\le r\le R\,,
\end{equation}
where $r_c$ is the inner boundary of the crust and $R$ is the stellar
radius. We choose $r_c=0.9\,R$. By imposing a regular behavior
near $r_c$, i.e.  $a_1(r\gtrsim r_c)=\alpha_0(r-r_c)+O((r-r_c)^2)$,
for each pair of assigned constants $\alpha_0,c_0$ the solution
$a_1(r)$ is unique.
\end{itemize}
We shall assume that outside the star there is vacuum,
currents vanish and $\zeta=0$ (see eq. (\ref{Jph})). Equation
(\ref{eqa1}) then reduces to
\begin{equation}
\left(1-\frac{2M}{r}\right)a_1''+\frac{2M}{r^2}a_1'-\frac{2}{r^2}a_1=0\,;
\label{eqa1out}
\end{equation}
its general solution (decaying at infinity) is a pure dipole
\begin{equation}
a_1(r)=-\frac{3\mu}{8M^3}r^2\left[\ln\left(1-\frac{2M}{r}\right)+\frac{2M}{r}
+\frac{2M^2}{r^2}\right]\,,
\label{dipole}
\end{equation}
where the constant $\mu$ is the magnetic dipole moment in geometrical
units. The corresponding magnetic field has the form
\begin{equation}
B_\mu=\left(0,-2\frac{e^{\lambda/2}}{r^2}a_1\cos\theta,e^{-\lambda/2}
a_1'\sin\theta,0\right)\,.\label{Bmuout}
\end{equation}
On the surface of the star, the function $a_1(r)$ solution of
eq. (\ref{eqa1}) has to be matched with the  exterior
solution (\ref{dipole}), imposing the continuity of  $a_1$ and $a_1'$.
The ratio $\alpha_0/c_0$ is fixed by matching the quantity
$a_1'/a_1$ (which does not depend on $\mu$). Once this ratio has been
determined, the constants $\alpha_0,c_0$ are rescaled by a common
factor, which changes the constant $\mu$ (and then the global
normalization of the field) by the same amount. We fix this constant
by assuming that the magnetic field  at the pole is $B_{pole}=10^{15}$ G.
%which is the order of magnitude of the external magnetic field of
%magnetars \cite{WT}.  
In this way, for each assigned value of $\zeta$
we determine $\alpha_0$ and $c_0$.

In previous papers on magnetized stars \cite{IS,Haskell}, boundary
conditions have been imposed in such a way that not only the toroidal,
but also the poloidal component of the magnetic field vanishes outside
the star; as a consequence, the parameter $\zeta$ can take only a
discrete set of values, a fact for which we do not see a reasonable,
physical explanation.

The matching conditions we impose at the boundaries are different, and
should be considered as an attempt to better approximate realistic
boundary conditions.  Let us see why. We remind that outside the star
we assume there is vacuum and $\zeta=0$.  By comparing (\ref{Bmu}),
(\ref{Bmuout}) we see that if we choose $a_1,a_1'$ to be continuous
across the stellar surface then $B_r$, $B_\theta$ are continuous.
However, if $\zeta\neq0$ inside the star, $B_\phi$ is discontinuous
because, having set $\zeta=0$ outside, it vanishes there.
Such discontinuity corresponds to a surface current
\begin{equation}
J_\mu^{surf}=\left(0,0,-\zeta\frac{e^{-(\lambda+\nu)/2}}{4\pi}
a_1\sin\theta\delta(r-R),0\right)\,.
\end{equation}
A true neutron star is surrounded by a magnetosphere, where fluid
energy density and pressure are small, but currents do not
vanish. There, the magnetic field has both poloidal and toroidal
components, and both match continuously across the stellar surface with
their interior correspondent.  The values of $a_1$, $a_1'$ which would
ensure the continuity of both components, would be different from
those we choose by imposing $B_\phi=0$ outside the star; however, with
our choice at least we allow the poloidal field, which extends all over the
space and decays as $r^{-l-2}$, to be continuous, whereas outside
the star we switch off the toroidal component which
extends only in the magnetosphere and tends to zero smoothly at its
edges.  A more precise characterization of the field behavior at the
stellar surface would require the modelling of the magnetosphere,
which is beyond the scope of this paper.

The function $a_1(r)$, solution of eq. (\ref{eqa1}), which describes
the vector potential inside the star, can have zeros at some points
$r=\bar r_i$; conversely, the exterior, vacuum solution (\ref{dipole})
never vanishes.  This happens also in the Newtonian limit; for
example, in \cite{PMP} it has been shown that the solution of the
equation corresponding to eq. (\ref{eqa1}) with $c_0=0$ is
a linear combination of the spherical Bessel functions
\begin{eqnarray}
j_1(x)&=&\frac{\sin x}{x^2}-\frac{\cos x}{x}\nn\\
n_1(x)&=&-\frac{\cos x}{x^2}-\frac{\sin x}{x}\,,\nn\\
\end{eqnarray}
where $x=\zeta r$.  Such combination vanishes at given values of $x$.
In this case $\zeta$, which has the dimensions of an inverse
length, can be interpreted as a sort of wavenumber of the solution.

In our case also, though we set $c_0\neq 0$, the location of the
points where $a_1(r)$ vanishes depend on $\zeta$. If $r=\bar r<R$ is a
zero of $a_1$, then $B_r(\bar r)=0$ (see eq. (\ref{Bmu})), and the
magnetic flux is confined within the spherical surface $r=\bar
r$. This means that the field lines inside the star are defined in
disjoint domains.  Although we do not have a physical interpretation
for this configuration, in our study we will not exclude this
possibility.
%%%%%%%%%%%%%%%%%%%%%%%%%%%%%%%%%%%%%%%%%%%%%%%%%%%%%%%%%%%%
\subsection{The ellipticity of the star}
\label{ellip}
%%%%%%%%%%%%%%%%%%%%%%%%%%%%%%%%%%%%%%%%%%%%%%%%%%%%%%%%%%%%
The stellar deformation, which we determine by solving the perturbed
Einstein equations given in Appendix \ref{deform}, can be expressed in
terms of the stellar ellipticity.
In the current literature there are two different definitions of
ellipticity, which correspond to two conceptually different
quantities.  The {\it surface ellipticity}, $e_{surf}$, is
\cite{CM,KOK,KOKrot}
\begin{equation}
e_{surf}=\frac{\hbox{(equatorial radius)
-(polar radius)}}{\hbox{( polar radius)}}\,.\label{esurf}
\end{equation}
It describes the geometrical shape of the star.  It should be
mentioned that a slightly different definition has been employed in
\cite{Hartle1,HartleThorne,BFGM}, i.e. $\tilde e_{surf}=
\sqrt{(e_{surf})^2+2e_{surf}}$.  The surface ellipticity describes the
external appearance of the star.

A different quantity is the {\it quadrupole ellipticity}, $e_Q$,
which is a measure of the mass quadrupole of
the star \cite{BG,Cutler,Haskell}:
\begin{equation}
e_Q=-\frac{Q}{I}\label{eQ}
\end{equation}
where $I$ is the mean value of the moment of inertia of the star
$I_{ij}$, and $Q$ is its mass-energy quadrupole moment.
For a stationary, axisymmetric compact object, $Q$ can be extracted by
the far field limit of the metric \cite{Thorne,HartleThorne}. Indeed,
it is the coefficient of the $1/r^3 P_2(\cos\theta)$ term in the
expansion of $g_{00}$ in powers of $1/r$ and in Legendre polynomials
$P_l(\theta)$:
\begin{equation}
g_{00}\rightarrow\dots-2Q\frac{1}{r^3}P_2(\cos\theta)\,.
\end{equation}
As discussed in \cite{Thorne,LP}, in the weak field limit the
mass-energy quadrupole moment reduces to
\begin{equation}
Q=\int_V\rho(r,\theta)r^2P_2(\cos\theta)dV\,,
\end{equation}
where $V$ is the star volume.  In this limit, the quadrupole tensor of
the axially symmetric star can be expressed in terms of $Q$: $Q_{ij}=
{\rm diag}(-Q/3,-Q/3,2/3\,Q)$, and the quadrupole ellipticity can also
be written in terms of the inertia tensor
\begin{equation}
e_Q=\frac{I_{zz}-I_{yy}}{I_{zz}}\,.
\end{equation}
In the general case, the quadrupole ellipticity is a measure of the
entire stellar bulk deformation.

Since $e_Q$ and $e_{surf}$ are quantities with different physical
meaning, they are in general different. They coincide only in the case
of a constant density star, in the Newtonian limit, as shown in
Chapter 16 of \cite{ST}.

It is worth stressing that the quadrupole ellipticity is the quantity
that should be used to evaluate the gravitational emission of a
rotating star; moreover, it has been used to study the spin-flip
mechanism proposed by Jones and Cutler \cite{Jones,Cutler}.

%%%%%%%%%%%%%%%%%%%%%%%%%%%%%%%%%%%%%%%%%%%%%%%%%%%%%%%%%%%%%%%%%%%%%%%%%%% 
\section{Results}\label{results}
%%%%%%%%%%%%%%%%%%%%%%%%%%%%%%%%%%%%%%%%%%%%%%%%%%%%%%%%%%%%%%%%%%%%%%%%%%% 

In this section we present the results of the numerical integration of
eqs. (\ref{eqa1}), (\ref{y2primo}) and (\ref{h2primo}).

As a test, we have first ran our codes for the polytropic star used
in \cite{IS}, endowed with mixed (poloidal and toroidal) magnetic
field, which vanishes outside the star.  Thus, we impose $a_1=0$ on
the stellar surface $r=R$, and solve the eigenvalue problem to find
the set of values $\zeta_i$ for which this condition is satisfied.  We
have reproduced the values of $\zeta_{i}$ given in Table I of
\cite{IS} for different values of the stellar compactness, with an
accuracy better than $1\%$. The corresponding magnetic field profiles
and stellar deformations (surface ellipticity and mass-energy
quadrupole) are also in full agreement with \cite{IS}.

Furthermore, we have integrated the equations for the models
considered in Ref.~\cite{BG}; there, non rotating, magnetized stars
with only poloidal fields have been modeled by solving numerically the
full set of non-linear Einstein's equations; magnetic fields are
either defined throughout the star, or confined in the crust.
Following \cite{BG}, we introduce
the magnetic distortion factor $\beta$, given by 
\begin{equation}\label{beta}
e_Q=\beta \frac{{\cal M}^2}{{\cal M}_0^2}~,
\end{equation}
where ${\cal M}$ is the magnetic dipole moment,  related to
magnetic field at the pole, $B_{pol}$, by
\begin{equation}\label{mommag}
{\cal M}\equiv B_{pole}R^3 \frac{4\pi}{2\mu_0}~.
\end{equation}
Here, $\mu_0$ is the magnetic permeability. The normalization
factor ${\cal M}_0$ is given by
\begin{equation}\label{mommag0}
{\cal M}_0\equiv\frac{4\pi}{\mu_0}\frac{GI^2}{R^2} \,.
\end{equation}
With this normalization, the coefficient $\beta$ is dimensionless.
Moreover, as $e_Q=O(B^2)$ and ${\cal M} = O(B)$, $\beta$ is nearly
independent of $B$, and indicates to what extent a star can be
deformed by the magnetic field.  It is worth mentioning that the
magnetic dipole ${\cal M}$ defined in (\ref{mommag}) differs from the
quantity $\mu$ defined in eq. (\ref{dipole}), since eq. (\ref{mommag})
has been derived in the context of Newtonian theory.  As in \cite{BG},
we use the equation of state (EOS) of Wiringa Fix and Fabrocini
\cite{WFF} (WFF), and consider an $M=1.4\,M_\odot$ star.

When the magnetic field extends throughout the star, we find
$\beta=0.505$, while the authors of \cite{BG} find $\beta=1.01$; when
the field is confined to the crust, we find $\beta\sim 5$, while the
authors of \cite{BG} find a very large value: $\beta\sim 10^2$.
However, if we compute $\beta$ from the same equation, but using
$e_{surf}$ instead of $e_Q$, we find $\beta=1.01$ when the magnetic
field extends throughout the star, and $\beta\sim 10^2$ in the case of
crustal fields, in agreement with \cite{BG}.

The reason why, when crustal fields are present, the factor $\beta$
computed using $e_{surf}$ is much larger than that computed using
$e_Q$, is the following.  The crust contains a very small fraction of
stellar matter therefore, although its deformation is large (because
the field lines are squeezed in a small region), it does not induce a
big change in the distribution of matter in the stellar bulk. As a
consequence, $e_{surf} \gg e_Q$.

%%%%%%%%%%%%%%%%%%%%%%%%%%%%%%%%%%%%%%%%%%%%%%%%%%%%%%%%%%%%%%%%%%%%%%%%%%% 
\subsection{Deformations induced by different magnetic field configurations}
\label{diffconf}
%%%%%%%%%%%%%%%%%%%%%%%%%%%%%%%%%%%%%%%%%%%%%%%%%%%%%%%%%%%%%%%%%%%%%%%%%%% 
We shall now study how  the stellar deformations induced by a mixed
(poloidal and toroidal) magnetic field depend on the field
configuration.  To describe matter in the stellar core we use the
equation of state of Akmal, Pandharipande and Ravenhall \cite{APR}
(denoted as APR2); we choose a star with mass $M=1.4\,M_\odot$ and a
radius $R=11.58$ km.  The magnetic field is normalized assuming that
its value at the pole is $B_{pole}=10^{15}$ G.  For the different
configurations discussed in Section \ref{bcs}, we find the magnetic
field structure, the surface and quadrupole ellipticities $e_{surf}$,
$e_Q$, and the maximal values of the internal poloidal and toroidal
fields, $B_p^{max}$ and $B_t^{max}$. The equations for the stellar
deformation and the procedure to compute $e_{surf}$, $e_Q$ are
described in Appendix \ref{deform}.

We stress that it is important to determine if the magnetic star has a
an oblate or prolate shape, i.e. to determine the sign of $e_Q$;
indeed, as suggested by Jones and Cutler \cite{Jones,Cutler}, if
$e_Q<0$ the star could change its rotation axis due to viscous forces
(``spin flip'') becoming an orthogonal rotator (with magnetic axis
orthogonal to the rotation axis), and the process could be associated
to a large gravitational wave emission.  In this respect, it is also
important to determine the absolute values of the allowed quadrupole
ellipticities, because if the star rotates around an axis different from the
magnetic field symmetry axis, it emits gravitational waves with
amplitude \cite{BG}
\begin{equation}
h_0\sim\frac{4G}{rc^4}\Omega^2I |e_Q|\,,
\end{equation}
and frequency $\nu_{GW}=\Omega/(2\pi)$, where $\Omega$ is the angular
velocity.

It is worth stressing that the current upper bound on neutron star
ellipticity, i.e.  $|e_Q|\lesssim 10^{-6}$, is obtained by evaluating
the maximal strain that the crust of an old and cold neutron star can
sustain \cite{UCB,HJS}.  However, a large deformation may be induced
by the effect of strong magnetic fields in the very early phases of
the stellar life, when the crust has not formed yet.  These
deformation may persist as the star cools down, leading to final
configurations having an ellipticity larger than the above limit.  Let
us now discuss the two field configurations described in section
\ref{bcs}.
%%%%%%%%%%%%%%%%%%%%%%%%%%%%%%%%%%%%%%%%%%%%%%%%%%%%%%%%%%%%%%%
%\begin{center}
\begin{figure}[htbp]
\includegraphics[width=6cm,angle=270]{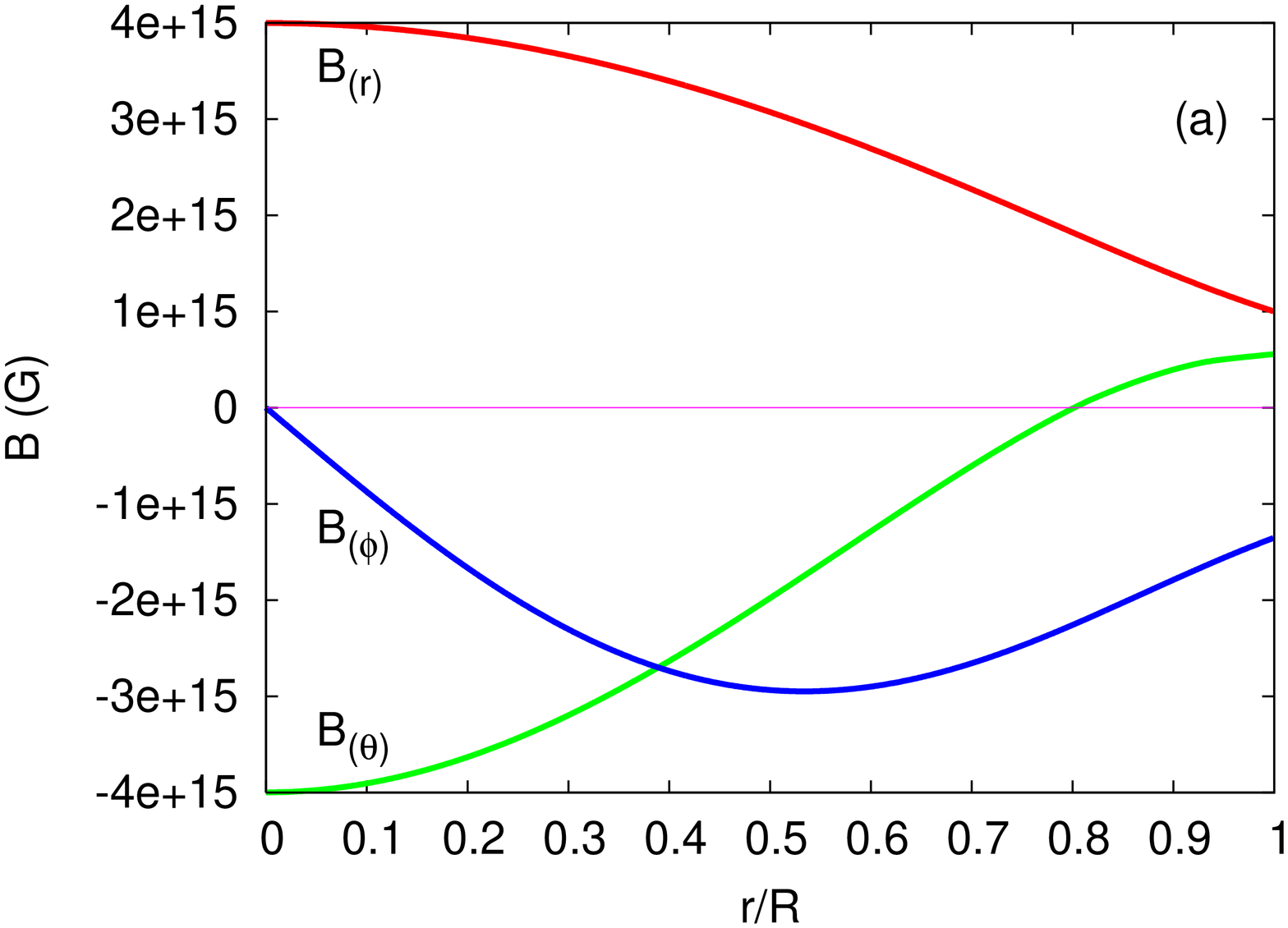}
\includegraphics[width=6cm,angle=270]{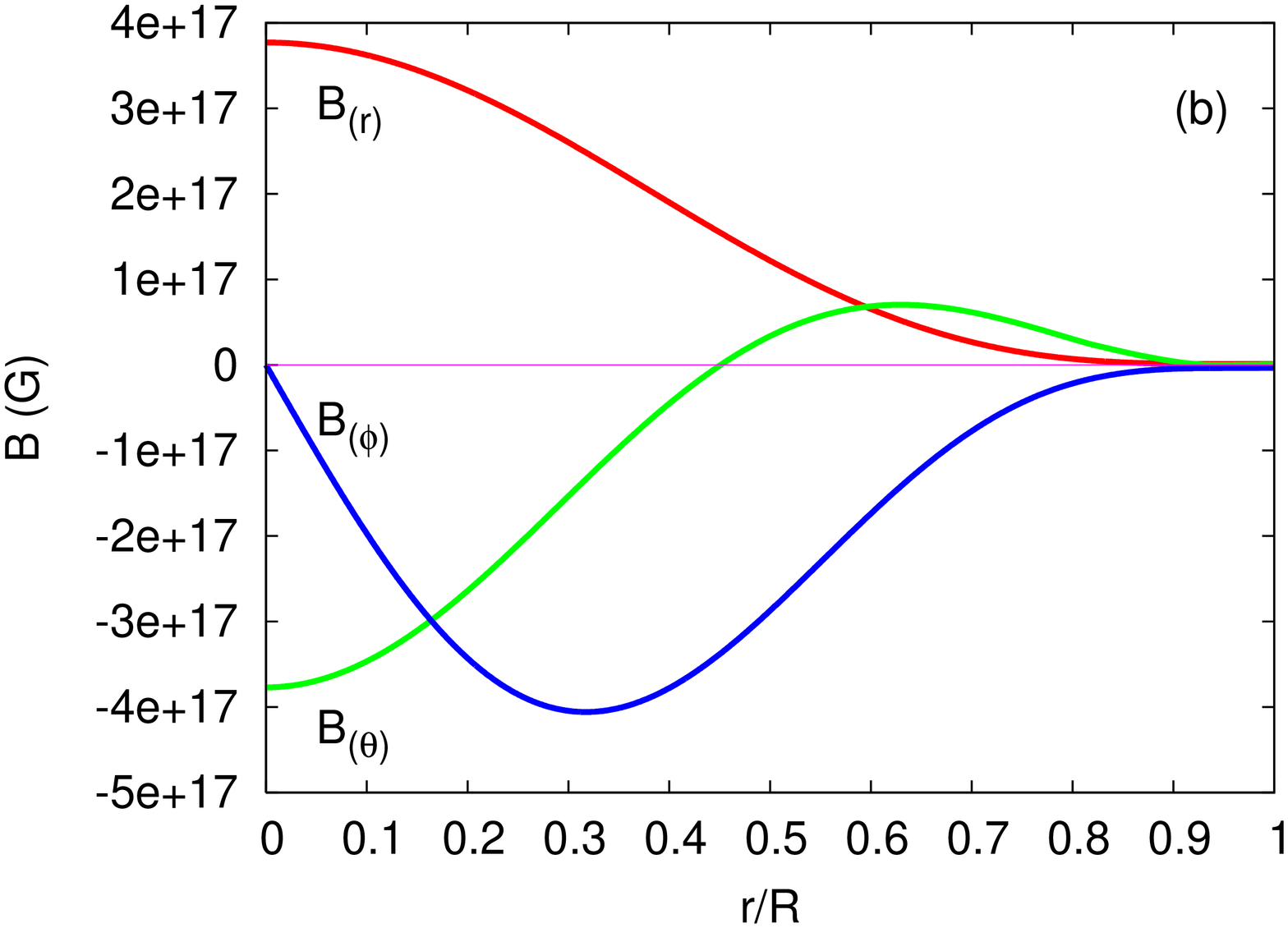}
\includegraphics[width=6cm,angle=270]{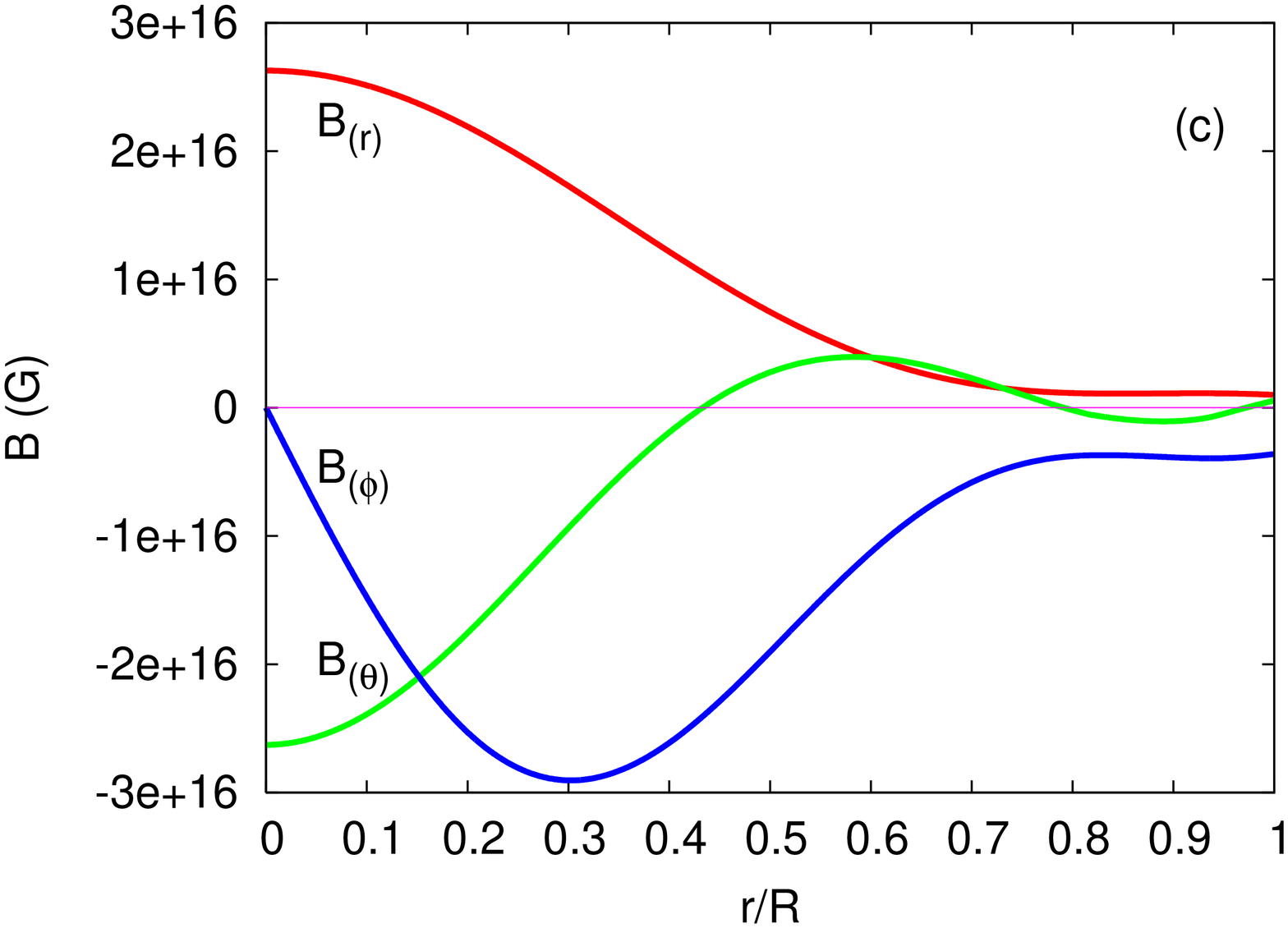}
\includegraphics[width=6cm,angle=270]{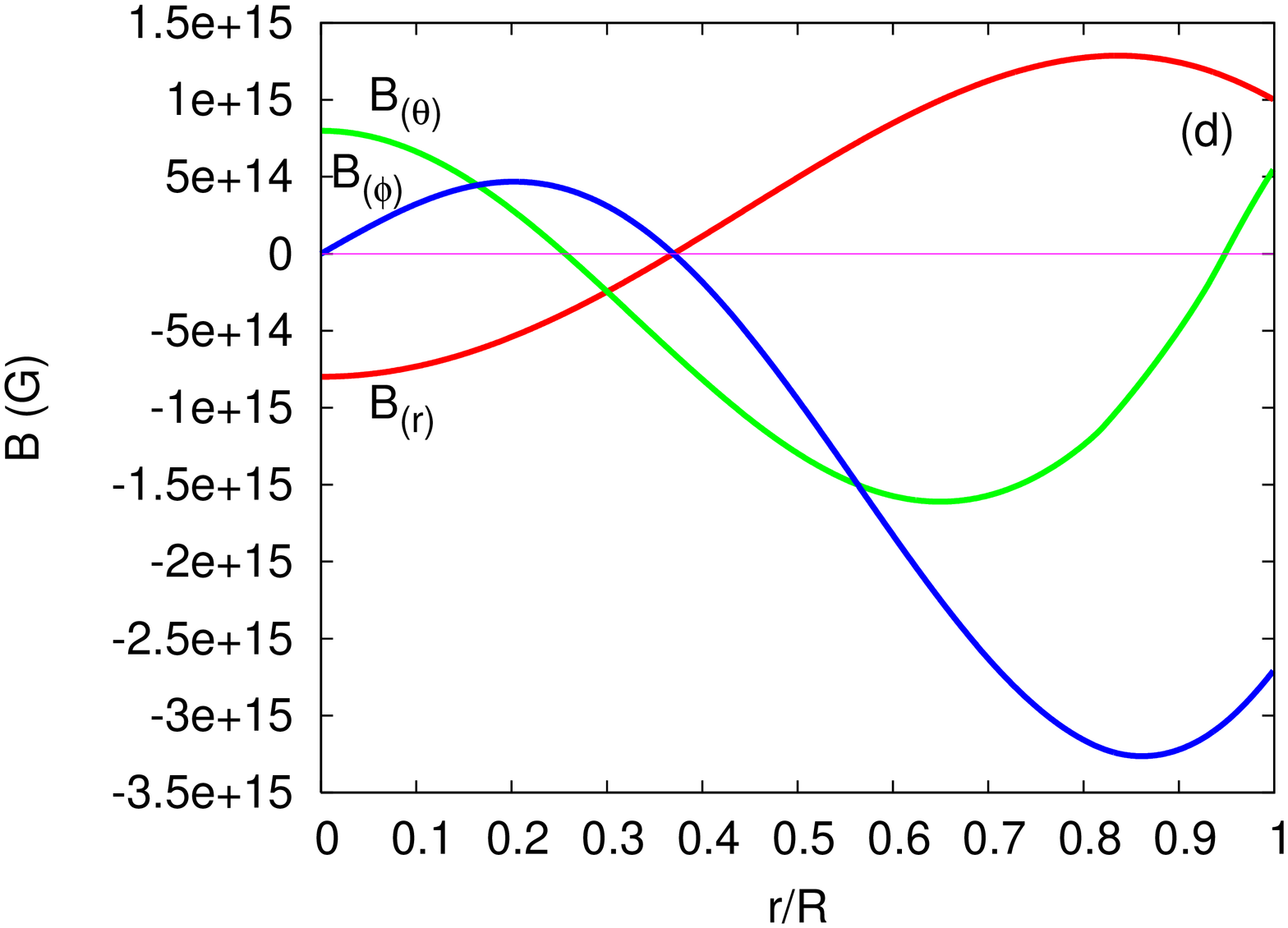}
\caption{The profiles of $B_{(r)}$ evaluated at $\theta=0$, and of
$B_{(\theta)}$ and $B_{(\phi)}$, evaluated at $\theta=\pi/2$, are
plotted as functions of the normalized radius inside the star.  The
magnetic field is defined through the whole star.  Each panel corresponds
to a value of $\zeta$: $\zeta=0.15$ km$^{-1}$ in panel (a), $\zeta=0.37$
km$^{-1}$ in (b), $\zeta=0.40$ km$^{-1}$ in (c) and $\zeta=0.30$
km$^{-1}$ in (d). Panel (d) refers to a value of $\zeta$ exterior to
the ranges (\ref{ranges}); thus in this case the magnetic field lines
are defined in disjoint domains (see text).
\label{profile1}}
\end{figure}
%\end{center}
%%%%%%%%%%%%%%%%%%%%%%%%%%%%%%%%%%%%%%%%%%%%%%%%%%%%%%%%%%%%%%%
%\begin{center}
\begin{figure}[htbp]
\includegraphics[width=7cm]{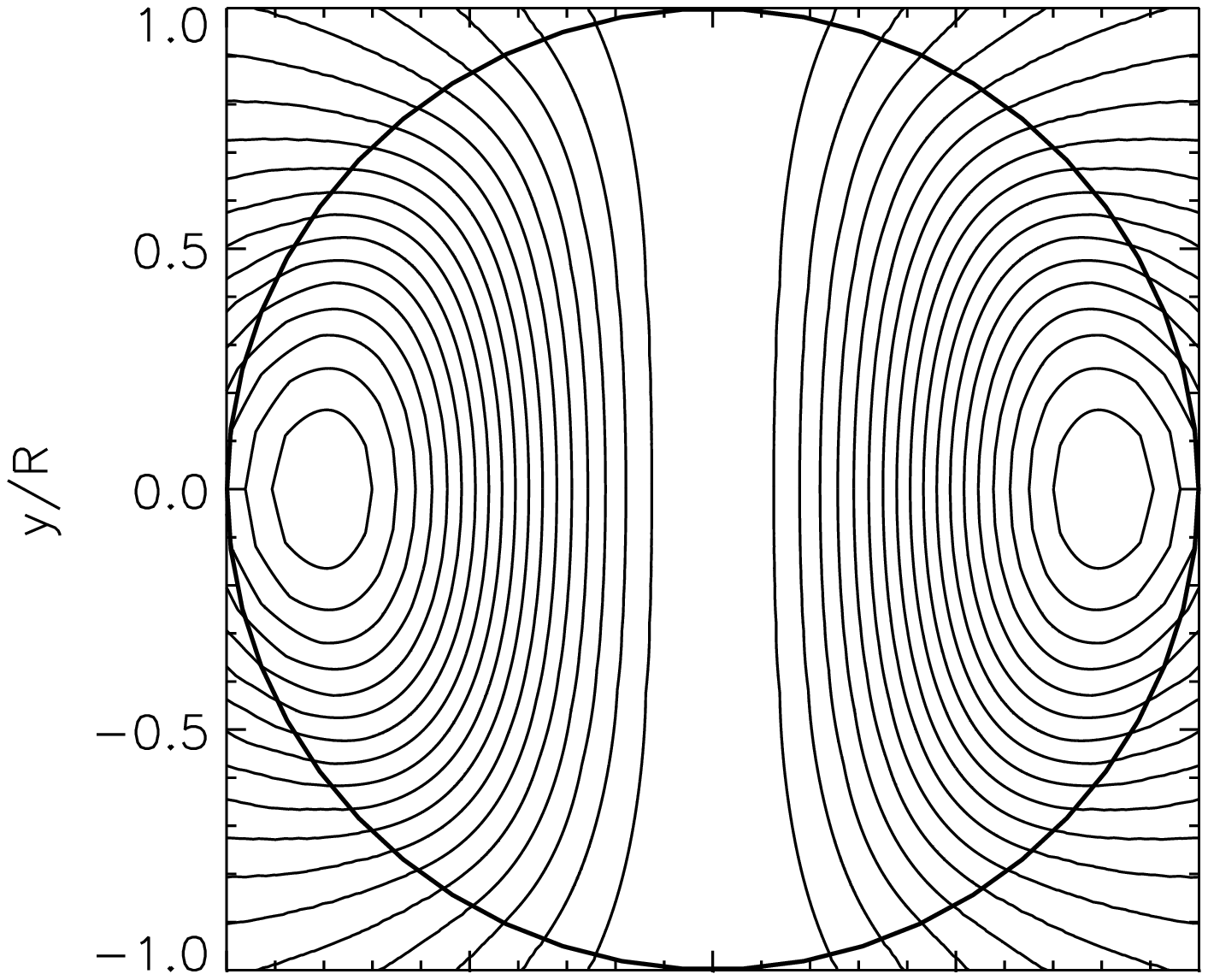}
\includegraphics[width=7cm]{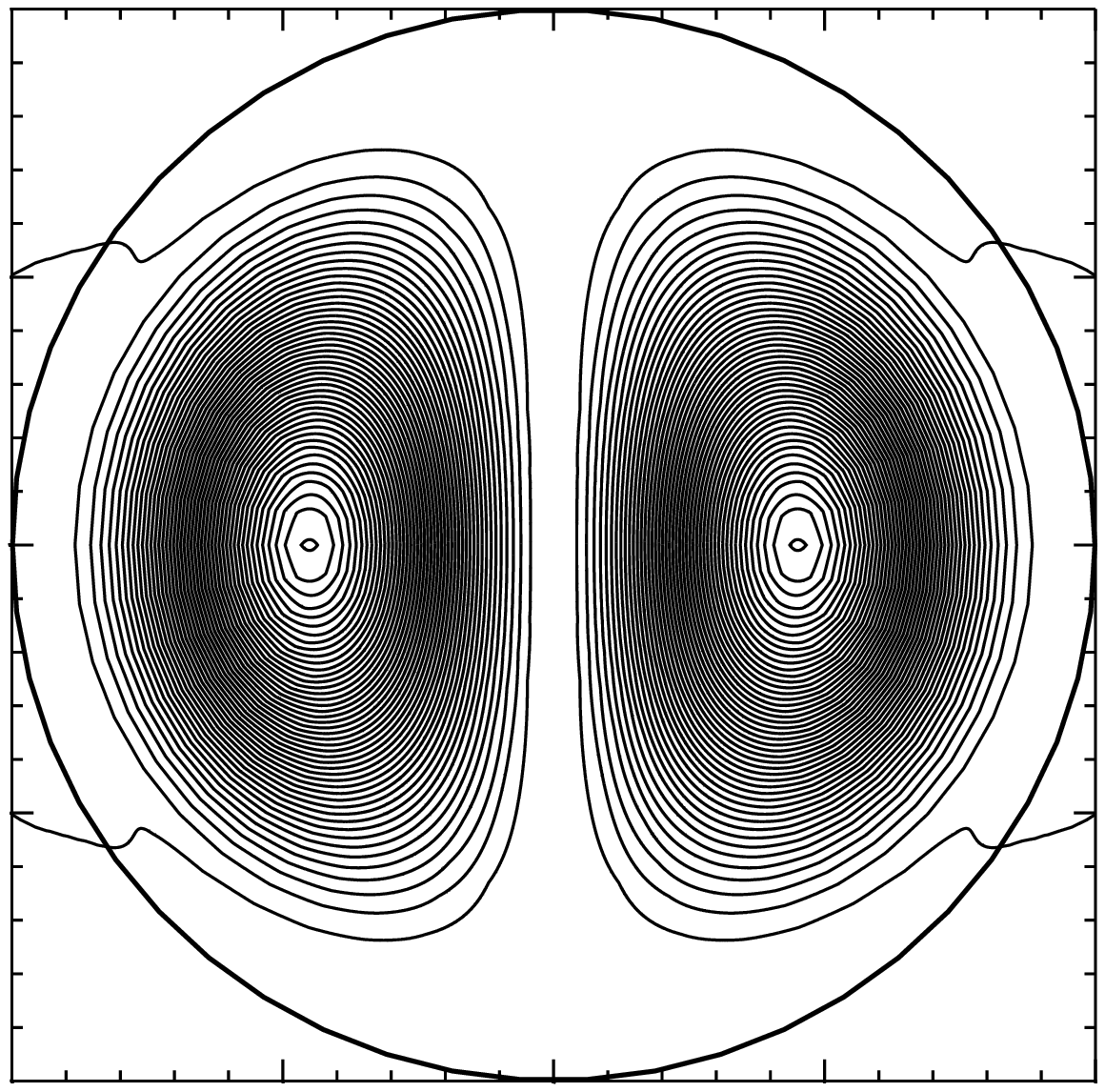}
\includegraphics[width=7cm]{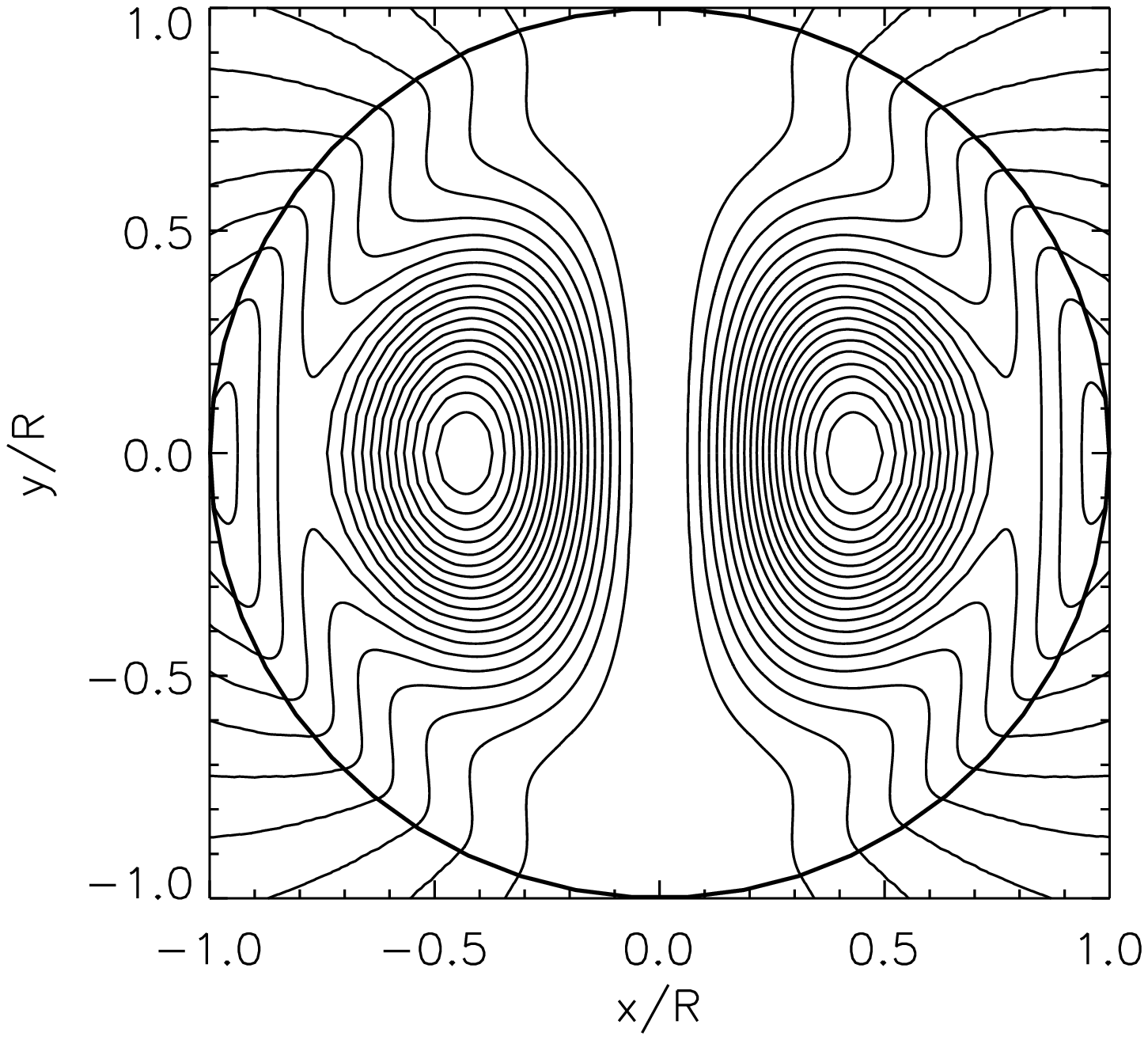}
\includegraphics[width=7cm]{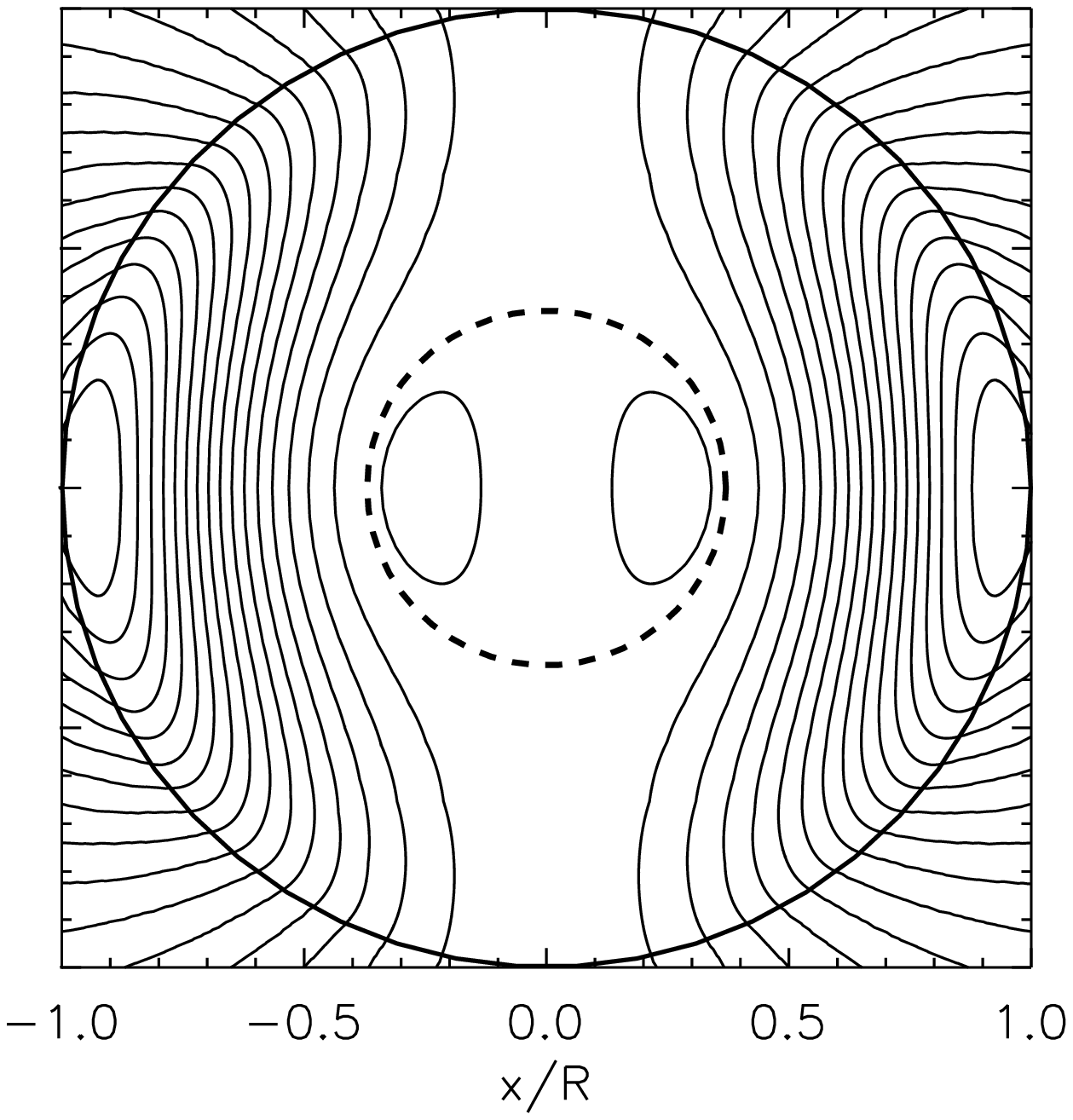}
\caption{
The projection of the field lines in the meridional plane is
shown for $\zeta=0.15$ km$^{-1}$ (upper panel, left), $\zeta=0.37$ km$^{-1}$
(upper panel, right), $\zeta=0.40$ km$^{-1}$ (lower panel, left) and
$\zeta=0.30$ km$^{-1}$ (lower panel, right). The dashed circle in the
lower panel on the right, separates two disjoint domains.
The magnetic field extends throughout the star.
\label{contour1}}
\end{figure}
%\end{center}
%%%%%%%%%%%%%%%%%%%%%%%%%%%%%%%%%%%%%%%%%%%%%%%%%%%%%%%%%%%%%%%
%%%%%%%%%%%%%%%%%%%%%%%%%%%%%%%%%%%%%%%%%%%%%%%%%%%%%%%%%%%%
\subsubsection{Magnetic field defined throughout the star}
\label{corefields2}
%%%%%%%%%%%%%%%%%%%%%%%%%%%%%%%%%%%%%%%%%%%%%%%%%%%%%%%%%%%%
If the magnetic field is non vanishing through the whole  star (see
Section \ref{bcs}), we find that $a_1(r)$ has no nodes for $r < R$ in
two ranges
\begin{eqnarray}
\hbox{range~ 1}&:&
(0\le\zeta\le0.2915) \\
\hbox{range~ 2}&:&
 (0.369\le\zeta\le0.46)\,.
\nonumber
\label{ranges}
\end{eqnarray}
If $\zeta$ lies outside these ranges, the field lines inside the star
are defined in disjoint domains, as discussed in Section \ref{bcs}.

In Figure \ref{profile1} we plot the magnetic field components versus
the radial distance, for $r \leq R$.  $B_{(r)}$ is evaluated at
$\theta=0$, $B_{(\theta)}$ and $B_{(\phi)}$ are evaluated at
$\theta=\pi/2$.  The plots are shown for four values of $\zeta$ (in
km$^{-1}$): $\zeta= 0.15$ in Fig.~\ref{profile1}a) (this value is in
the range 1), $\zeta= 0.37$ in Fig.~\ref{profile1}b), $\zeta=0.4$ in
Fig.~\ref{profile1}c), (both values are in the range 2) and
$\zeta=0.3$ in Fig.~\ref{profile1}d), which is outside the ranges 1
and 2.  Different values of $\zeta$ give qualitatively similar
behaviours.  We see that when $\zeta$ approaches the lower bound of
range 2, as in Fig.~\ref{profile1}b), the field components become much
larger than in the other cases.  Of course they cannot be arbitrarily
large, since they must not exceed the virial theorem limit
$B\lesssim10^{18}$ G \cite{BBGN}.  This is a peculiar behaviour, which
is not observed if one approaches the other bounds of range 1 and 2,
either from inside or from outside.  The reason for such behaviour is
that the configuration with $\zeta=0.369$ km$^{-1}$ is a singular
limit. It corresponds to a configuration in which $a_1(R)=0$, i.e. the
magnetic field is confined inside the star and vanishes outside. This
is inconsistent with the boundary condition we impose, i.e.
$B_{pole}=10^{15}$ G.  Thus, this singular value is unacceptable.
However, values of $\zeta$ approaching this limit can be accepted,
provided the virial limit is not violated. We mention that, as long as
$B$ is smaller than the virial limit, the stress-energy tensor of the
electromagnetic field is smaller than that of the fluid, and the
perturbative approach we use is appropriate.

The field profiles shown in Fig.~\ref{profile1}d) refers to a case in
which inside the star the field lines are defined in disjoint domains:
indeed, they cannot cross the sphere $r=0.37~R$ since
$B_r(r=0.37~R)=0$. 

The projection of the field lines in the meridional plane is shown in
Figure \ref{contour1}; the four panels refer to the same values of
$\zeta$ considered in Figure \ref{profile1}.  Fig.~\ref{contour1}d)
corresponds to $\zeta=0.3$, i.e. to the case of disjoint domains:
field lines do not cross the dashed circle in the picture.

\begin{center}
\begin{figure}[htbp]
\includegraphics[width=6cm,angle=270]{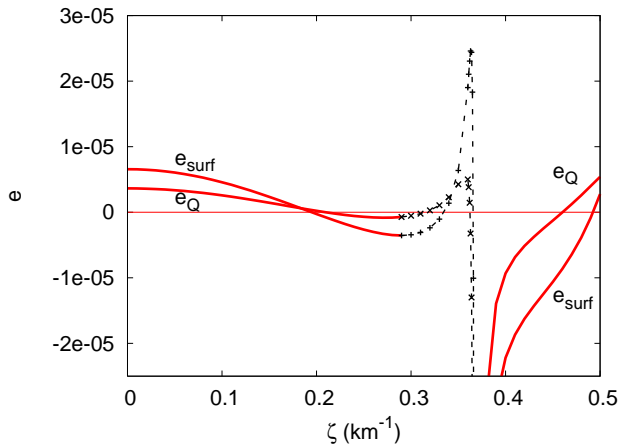}
\caption{Surface and quadrupole ellipticities as functions of $\zeta$
for a star with mass $M=1.4\,M_\odot$, and  equation of state
APR2. The magnetic field extends throughout the star.  The dashed
(solid) lines correspond to models for which $a_1(r)$ has nodes (has
no nodes) inside the star.
\label{ell}
}
\end{figure}
\end{center}

The ellipticities $e_{surf}$ and $e_{Q}$ are plotted in Figure
\ref{ell} as functions of $\zeta\in[0,0.5]$.  Continuous lines
correspond to values of $\zeta$ inside the ranges (\ref{ranges}) (no
nodes inside the star), while the dashed lines correspond to values of
$\zeta$ for which there is a node inside the star.

For small values of $\zeta$ (i.e. if the poloidal field prevails) the
star is oblate ($e_{surf,Q}>0$). As $\zeta$ increases, the toroidal
part becomes more important and the star becomes prolate
($e_{surf,Q}<0)$.  In other words, the toroidal field tends to make
the star prolate, while the poloidal field tends to make it oblate;
this behavior has already been discussed in the literature, see for
instance \cite{IS}.  For larger values of $\zeta$ the behavior is
different.  If we exclude values close to the singular point
$\zeta\simeq0.369$ km$^{-1}$, we find that the ellipticity is
\begin{equation}
|e_{surf,Q}|\simeq10^{-6}\,-\,10^{-5}\,.
\end{equation}
If we approach the value $\zeta = 0.369$ from either sides, then
$e_{surf,Q}<0$, and the deformation can be much larger; the virial
theorem constraint, $B\lesssim10^{18}$ G, corresponds to
\begin{equation}
|e_{surf}|\lesssim 2\times 10^{-3}~~\,,~~~
|e_Q|\lesssim10^{-3}\,.
\end{equation}
Thus, for a large range of values of $\zeta$, the magnetic field
induces a shape, either prolate or oblate, with $|e_Q| \sim
10^{-6}\,-\,10^{-5}$; however, for very particular values of $\zeta$,
the star can have a strongly prolate shape ($e_Q<0$), with $|e_Q|$ as
large as $10^{-3}$.

In Table \ref{Tabell1} we give, for selected values of $\zeta$ in the
range $[0,0.5]$ km$^{-1}$, the surface and quadrupole ellipticities,
and the maximal values of the internal poloidal and toroidal fields.
It is interesting to note that for values of $\zeta \lesssim 0.1$,
$e_{surf}\simeq2e_Q$. We find a similar behaviour when the ellipticity
is induced by rotation and no magnetic field is present. Indeed, by
integrating the equations of stellar deformation to second order in
the angular velocity as in \cite{BFGM}, we find that, for a large
variety of neutron stars EOS, $e_{surf}\simeq2e_Q$ for $\Omega
\lesssim 0.1~\Omega_{ms}$, where $\Omega_{ms}=\sqrt{M/R^3}$.
%%%%%%%%%%%%%%%%%%%%%%%%%%%%%%%%%%%%%%%%%%%%%%%%%%%%%%%%%%%%
%%%%%%%%%%%%%%%%%%%%%%%%%%%%%%%%%%%%%%%%%%%%%%%%%%%%%%%%%%%%%%%
%%%%%%%%%%%%%%%%%%%%%%%%%%%%%%%%%%%%%%%%%%%%%%%%%%%%%%%%%%%%%%%
\begin{table}[hbtp]
\begin{center}
\begin{tabular}{|c|c|c|c|c|}
\hline
$\zeta$ (km$^{-1}$)&$e_{surf}$&$e_Q$&$B^p_{max}/(10^{15}G)$
&$B^t_{max}/(10^{15}G)$\\
\hline 
$0$& $6.572\times 10^{-6}$&$3.642\times 10^{-6}$&$4.579$&$0$\\
$0.05$&$6.057\times 10^{-6}$&$3.364\times 10^{-6}$&$4.522$&$1.112$\\
$0.1$ &$4.580\times 10^{-6}$&$2.582\times 10^{-6}$&$4.341$&$2.132$\\
$0.15$ &$2.349\times 10^{-6}$&$1.447\times 10^{-6}$&$3.998$&$2.948$\\
$0.2$ &$-2.661\times 10^{-7}$&$2.199\times 10^{-7}$&$3.391$&$3.404$\\
$0.25$& $-2.643\times 10^{-6}$&$-6.945\times 10^{-7}$&$2.219$&$3.303$\\
$0.30$&$-3.433\times 10^{-6}$&$-5.343\times 10^{-7}$&$1.610$&$3.264$\\
$0.37$&$-1.106\times 10^{-3}$&$-2.250\times 10^{-3}$&$518.0$&$557.0$\\
$0.35$&$6.375\times 10^{-6}$&$4.273\times 10^{-6}$&$21.52$&$22.15$\\
$0.40$&$-2.220\times 10^{-5}$&$-9.313\times 10^{-6}$&$26.28$&$29.05$\\
$0.45$&$-1.062\times 10^{-5}$&$-1.263\times 10^{-6}$&$18.89$&$20.51$\\
$0.50$&$2.773\times 10^{-6}$&$5.410\times 10^{-6}$&$26.68$&$27.50$\\
\hline
\end{tabular}
\caption{Surface and quadrupole ellipticities, and maximal values of
the internal poloidal and toroidal magnetic fields, are tabulated for
different values of $\zeta$.
\label{Tabell1}}
\end{center}
\end{table}
%%%%%%%%%%%%%%%%%%%%%%%%%%%%%%%%%%%%%%%%%%%%%%%%%%%%%%%%%%%%%%%

%%%%%%%%%%%%%%%%%%%%%%%%%%%%%%%%%%%%%%%%%%%%%%%%%%%%%%%%%%%%
\subsubsection{Crustal fields}
%%%%%%%%%%%%%%%%%%%%%%%%%%%%%%%%%%%%%%%%%%%%%%%%%%%%%%%%%%%%
When the  magnetic field is confined to the crust, we
find that $a_1(r)$ has no nodes inside the star for
\begin{equation}
0\le\zeta\le1.085\,.\label{rangecrustal}
\end{equation}
$a_1(R)\neq 0$ for all values of $\zeta$, therefore 
crustal field do not exhibit the singular behavior discussed 
in Section \ref{corefields2}.

%%%%%%%%%%%%%%%%%%%%%%%%%%%%%%%%%%%%%%%%%%%%%%%%%%%%%%%%%%%%%%%
\begin{figure}[ht]
\includegraphics[width=8.5cm]{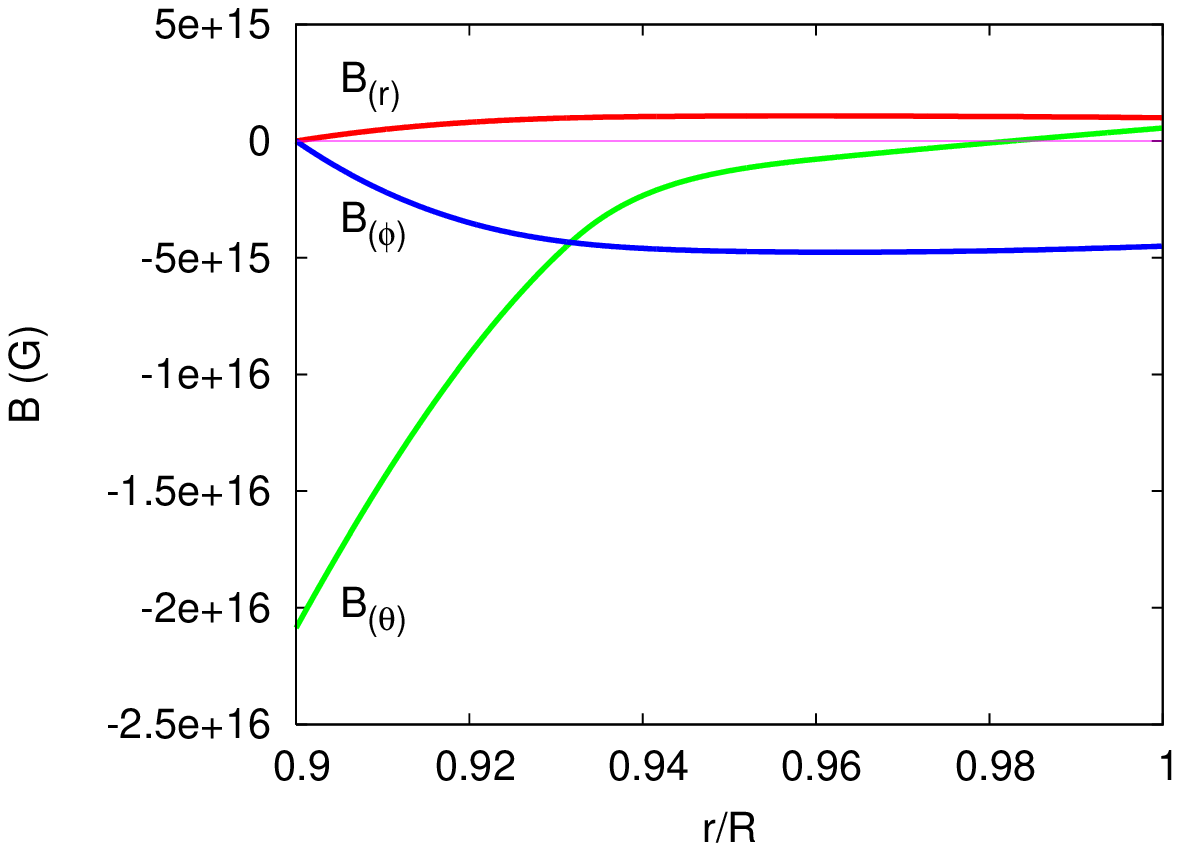}~~~~~
\includegraphics[width=7cm]{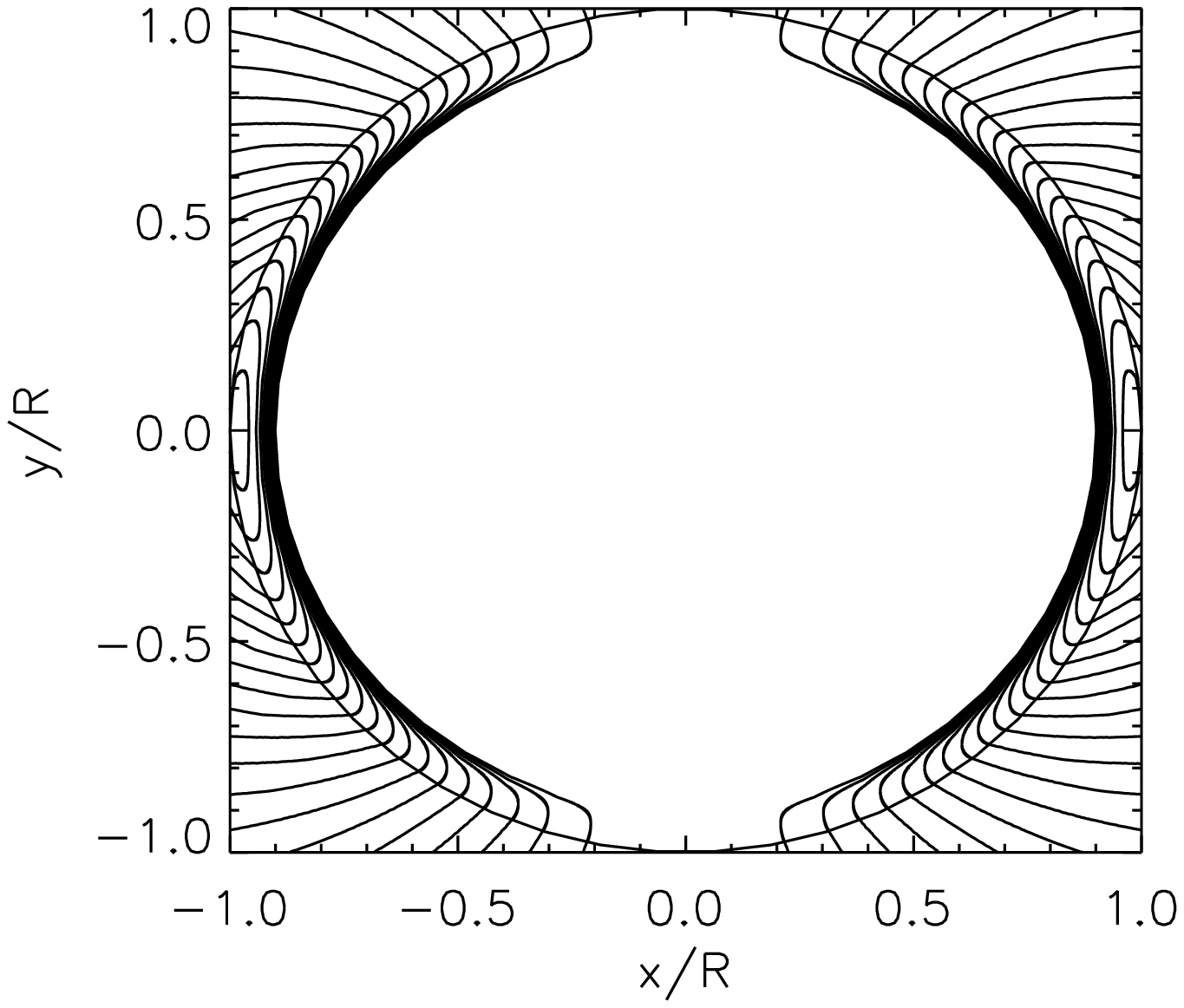}
\caption{The profiles of $B_{(r)}$, evaluated at $\theta=0$, and of
$B_{(\theta)}$ and $B_{(\phi)}$ evaluated at $\theta=\pi/2$, are
plotted for $\zeta=0.5$ km$^{-1}$ in the crust (left panel).
The projection of the  field lines in the meridional plane 
is shown in the right panel, for the same value of $\zeta$.}
\label{profile2}
\end{figure}
%%%%%%%%%%%%%%%%%%%%%%%%%%%%%%%%%%%%%%%%%%%%%%%%%%%%%%%%%%%%%%%

In the left panel of Figure \ref{profile2} we show, for $r \leq R$,
the profiles of $B_{(r)}$ evaluated at $\theta=0$, and of
$B_{(\theta)}$ and $B_{(\phi)}$ evaluated at $\theta=\pi/2$), for
$\zeta=0.5$ km$^{-1}$.  Different values of $\zeta$ correspond to
qualitatively similar behaviours. We see that the interior field is
one order of magnitude larger than the surface field; this behavior,
peculiar of crustal fields, is common to all values of $\zeta$.  
The projection of the  field lines in the meridional plane 
is shown in the right panel.
%%%%%%%%%%%%%%%%%%%%%%%%%%%%%%%%%%%%%%%%%%%%%%%%%%%%%%%%%%%%%%%
%%%%%%%%%%%%%%%%%%%%%%%%%%%%%%%%%%%%%%%%%%%%%%%%%%%%%%%%%%%%%%%
\begin{figure}[ht]
\centering
\includegraphics[width=6cm,angle=270]{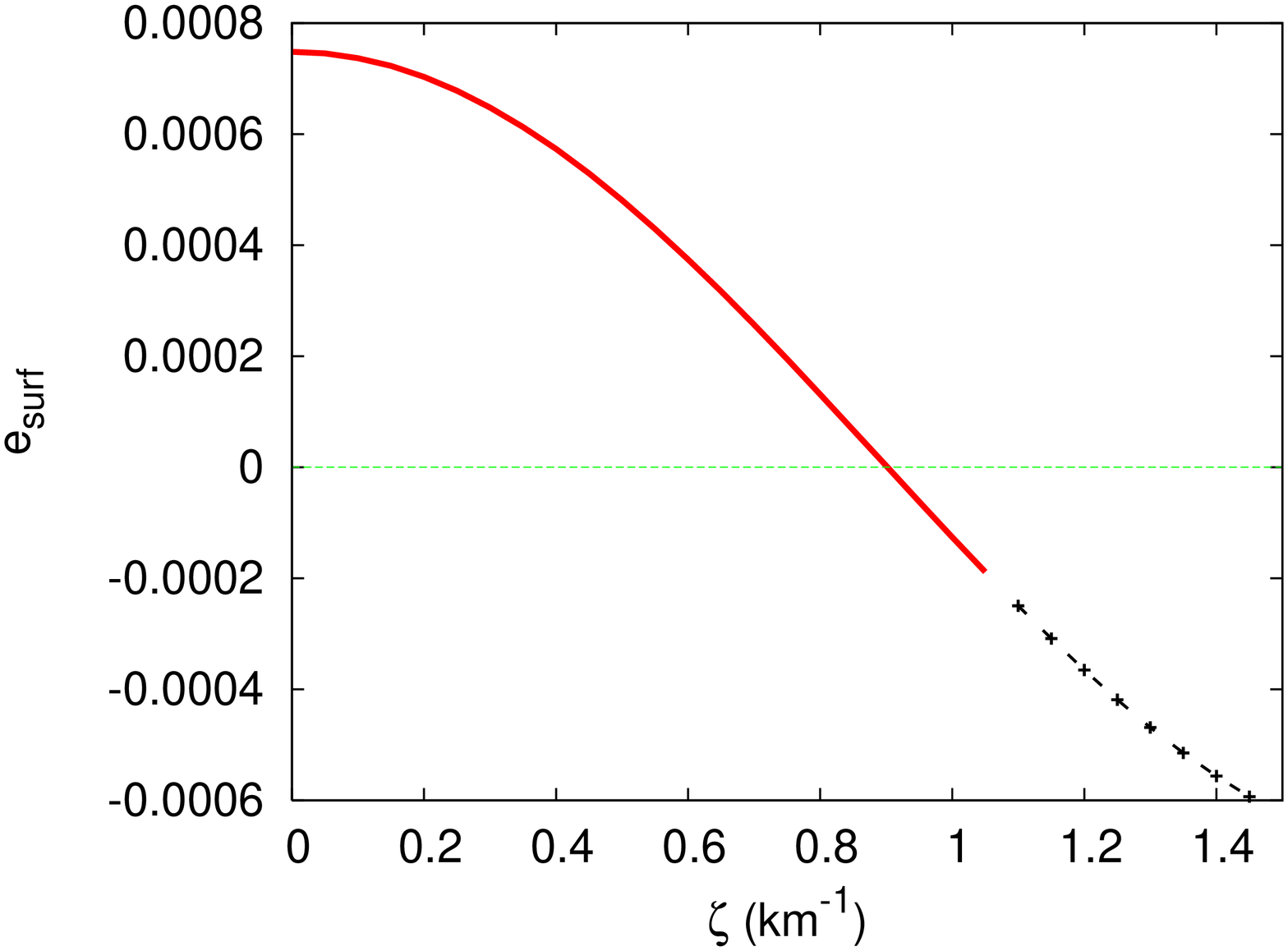}
\includegraphics[width=6cm,angle=270]{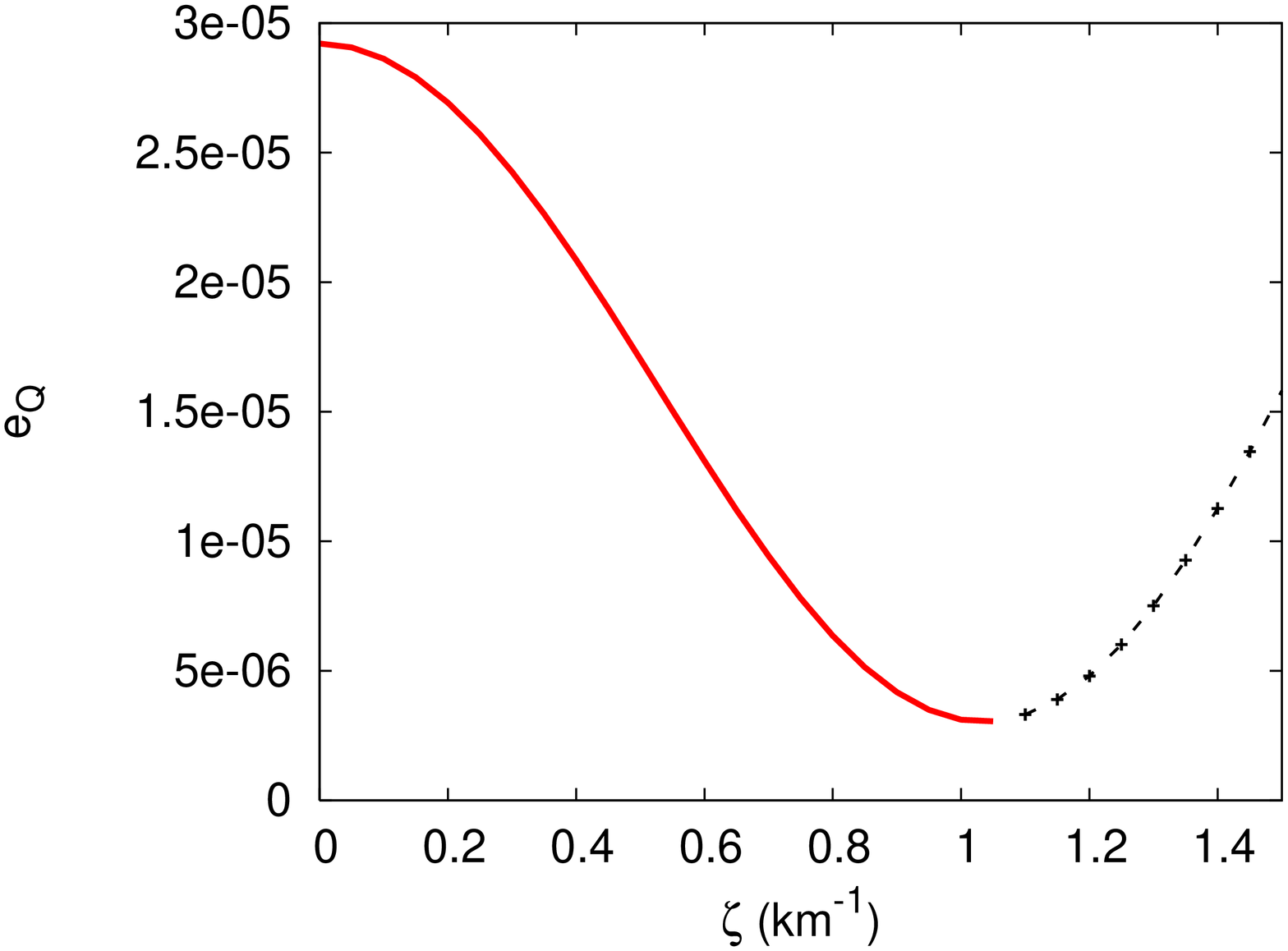}
\caption{The surface (left panel) and quadrupole (right panel)
ellipticities are plotted as functions of $\zeta$ for a star with mass
$M=1.4\,M_\odot$ and equation of state APR2, when the magnetic field
is confined to the crust.  The dashed (solid) lines correspond to
models for which $a_1(r)$ has nodes (has no nodes) inside the star.}
\label{ell2}
\end{figure}
%%%%%%%%%%%%%%%%%%%%%%%%%%%%%%%%%%%%%%%%%%%%%%%%%%%%%%%%%%%%%%%
\begin{table}[htbp]
\begin{center}
\begin{tabular}{|c|c|c|c|c|}
\hline
$\zeta$ (km$^{-1}$)&$e_{surf}$&$e_Q$&$B^p_{max}/(10^{15}G)$
&$B^t_{max}/(10^{15}G)$\\
\hline 
$0$& $7.483\times 10^{-4}$&$2.921\times 10^{-5}$&$27.80$&$0$\\
$0.2$ &$7.031\times 10^{-4}$&$2.693\times 10^{-5}$&$26.65$&$2.081$\\
$0.4$ &$5.731\times 10^{-4}$&$2.086\times 10^{-5}$&$23.29$&$3.924$\\
$0.6$ &$3.744\times 10^{-4}$&$1.310\times 10^{-5}$&$18.00$&$5.616$\\
$0.8$ &$1.313\times 10^{-4}$&$6.352\times 10^{-6}$&$11.19$&$7.361$\\
$1.0$ &$-1.259\times 10^{-4}$&$3.113\times 10^{-6}$&$6.580$&$9.130$\\
$1.2$&$-3.653\times 10^{-4}$&$4.798\times 10^{-6}$&$8.216$&$10.91$\\
$1.4$&$-5.564\times 10^{-4}$&$1.127\times 10^{-5}$&$12.30$&$12.70$\\
$1.5$&$-6.257\times 10^{-4}$&$1.581\times 10^{-5}$&$15.79$&$13.59$\\
\hline
\end{tabular}
\caption{Surface and quadrupole ellipticities and maximal values of
the internal poloidal and toroidal magnetic fields, are given for different
values of $\zeta$, in the case of crustal fields.
\label{Tabell2}}
\end{center}
\end{table}
%%%%%%%%%%%%%%%%%%%%%%%%%%%%%%%%%%%%%%%%%%%%%%%%%%%%%%%%%%%%%%%

In Figure \ref{ell2} we show the ellipticities as functions of
$\zeta$; continuous lines correspond to values of $\zeta$ inside the
range (\ref{rangecrustal}), dashed lines to values outside that range.
We see that, as  discussed in the previous section, the
geometrical shape of the star is oblate for small values of $\zeta$
(for which $e_{surf} > 0$) and prolate for larger values: the surface
ellipticity is a monotonically decreasing function of $\zeta$.
Conversely, the quadrupole ellipticity is always positive and, in
modulus, much smaller than $e_{surf}$, even for values of $\zeta$
larger than those considered in Figure \ref{ell2}.  We note that these
results rule out the Jones-Cutler mechanism in the case of crustal
fields, since it can only occur when $e_Q<0$.  As explained in
Section \ref{results}, the reason why $e_{surf}\gg e_Q$ is that,
though the crust deformation is large since the field lines are
squeezed in a small region, it does not induce a big change in the
distribution of matter in the stellar bulk.

In Table \ref{Tabell2} we give, for selected values of $\zeta$ in the
range $[0,1.5]$ km$^{-1}$ , the surface and quadrupole ellipticities,
and the maximal values of the internal poloidal and toroidal fields.
Comparing Tables \ref{Tabell1} and \ref{Tabell2} we see that, for
crustal fields, typical values of $e_{surf}$ are two orders of
magnitude larger than for fields extending through the whole star.  The
quadrupole ellipticity is, typically,  one order of magnitude
larger:
\begin{equation}
|e_{surf}|\sim10^{-4}\,-\,10^{-3}\,,~~~e_Q\sim10^{-5}\,-\,10^{-4}\,,
\end{equation}
with the exception of the models with $\zeta$ close to $0.369$, for which 
the deformation is larger in the case of fields extending throughout
the star.

%%%%%%%%%%%%%%%%%%%%%%%%%%%%%%%%%%%%%%%%%%%%%%%%%%%%%%%%%%%%%%%%%%%%%%%%%%% 
\subsection{Comparison between magnetic and rotational deformations}
%%%%%%%%%%%%%%%%%%%%%%%%%%%%%%%%%%%%%%%%%%%%%%%%%%%%%%%%%%%%%%%%%%%%%%%%%%% 
Both rotation and magnetic field contribute to the ellipticity of the star,
i.e. $e_{surf,Q}= e_{surf,Q}^\Omega+e_{surf,Q}^B$.  It is interesting
to compare the two contributions, evaluated in the range of parameters
typical of observed magnetars (SGR and AXP), i.e. \cite{WT}
\begin{eqnarray}
0.6\times 10^{14}\,G\lesssim& B&\lesssim 7.8\times 10^{14}\,G\label{rangeB}\\
5.2\,s\lesssim &T&\lesssim 11.8\,s\,,\label{rangeT}
\end{eqnarray}
where $T$ is the rotational period.  It should be mentioned that, as
explained in \cite{Cutler}, only $e_Q^B$ contributes to the spin-flip
process, which occurs when $e_Q^B<0$.

We have computed $e_{surf,Q}^\Omega$ for an $M=1.4\,M_\odot$ star with
equation of state APR2, using the codes, developed by some of us
\cite{BFGM}, which describe the structure of a non magnetized,
rotating star, up to $O(\Omega^3)$; $e_{surf,Q}^B$ have been computed
using the approach described in this paper.

\begin{center}
\begin{figure}[htbp]
\includegraphics[width=6cm,angle=270]{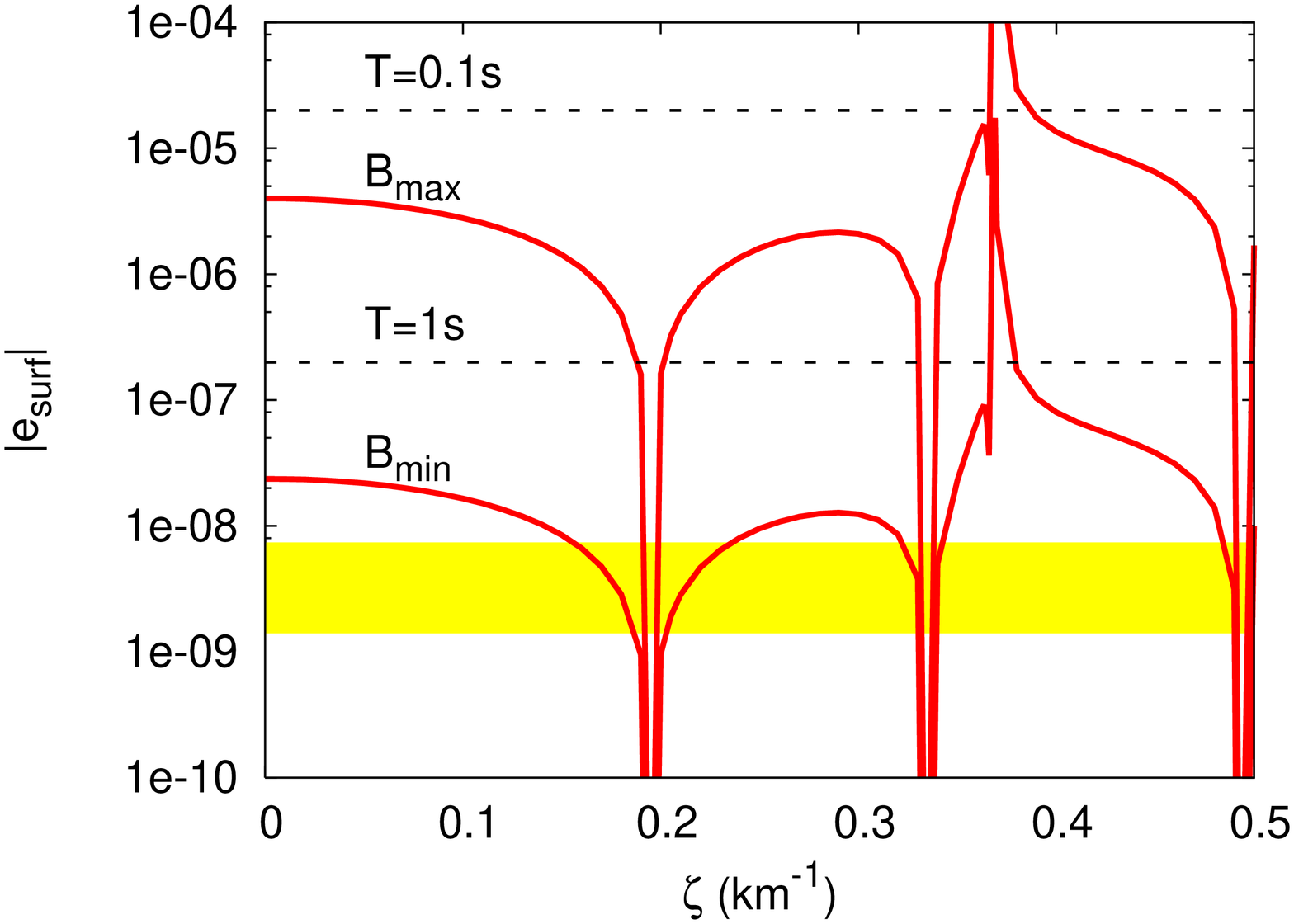}
\includegraphics[width=6cm,angle=270]{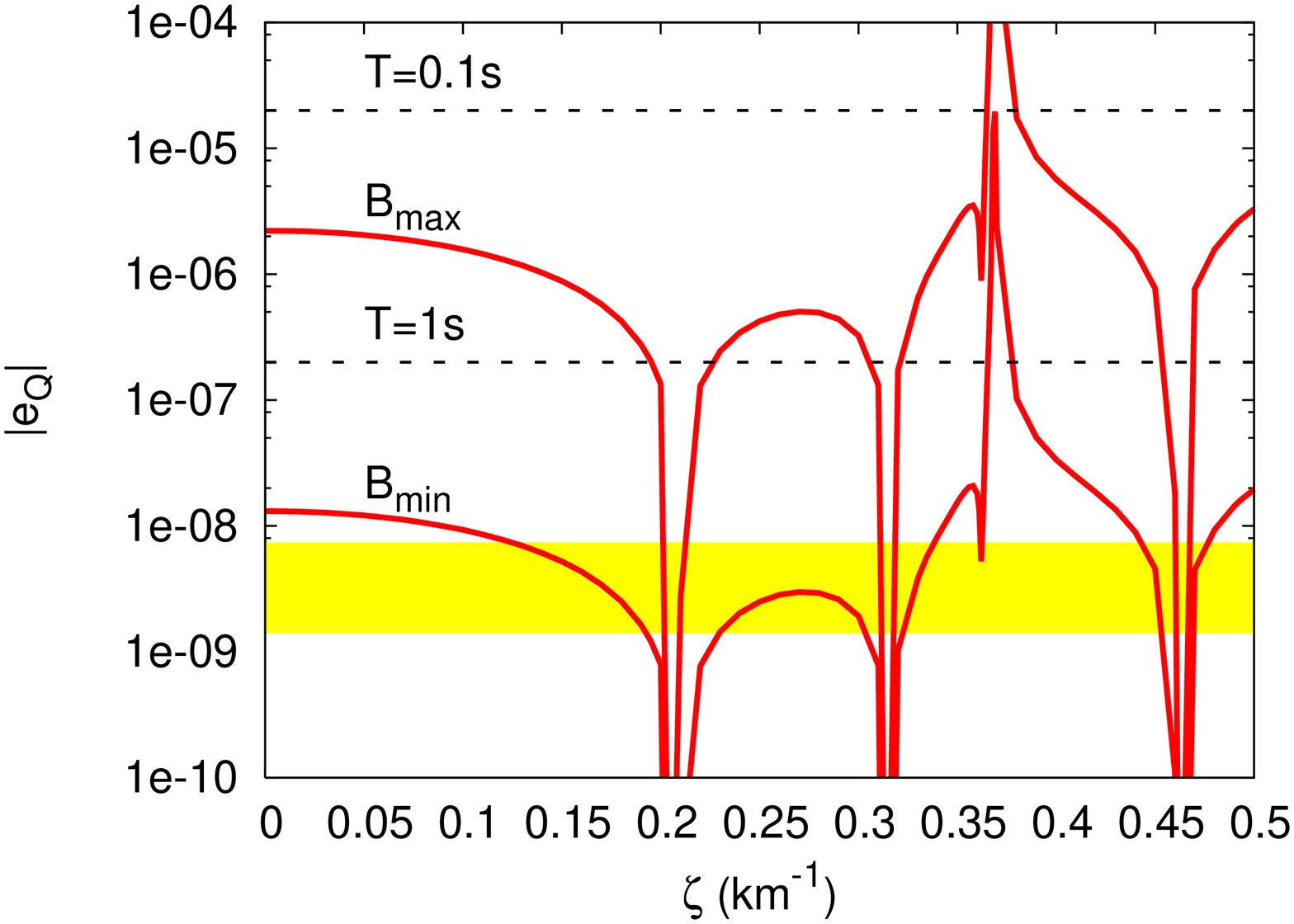}
\includegraphics[width=6cm,angle=270]{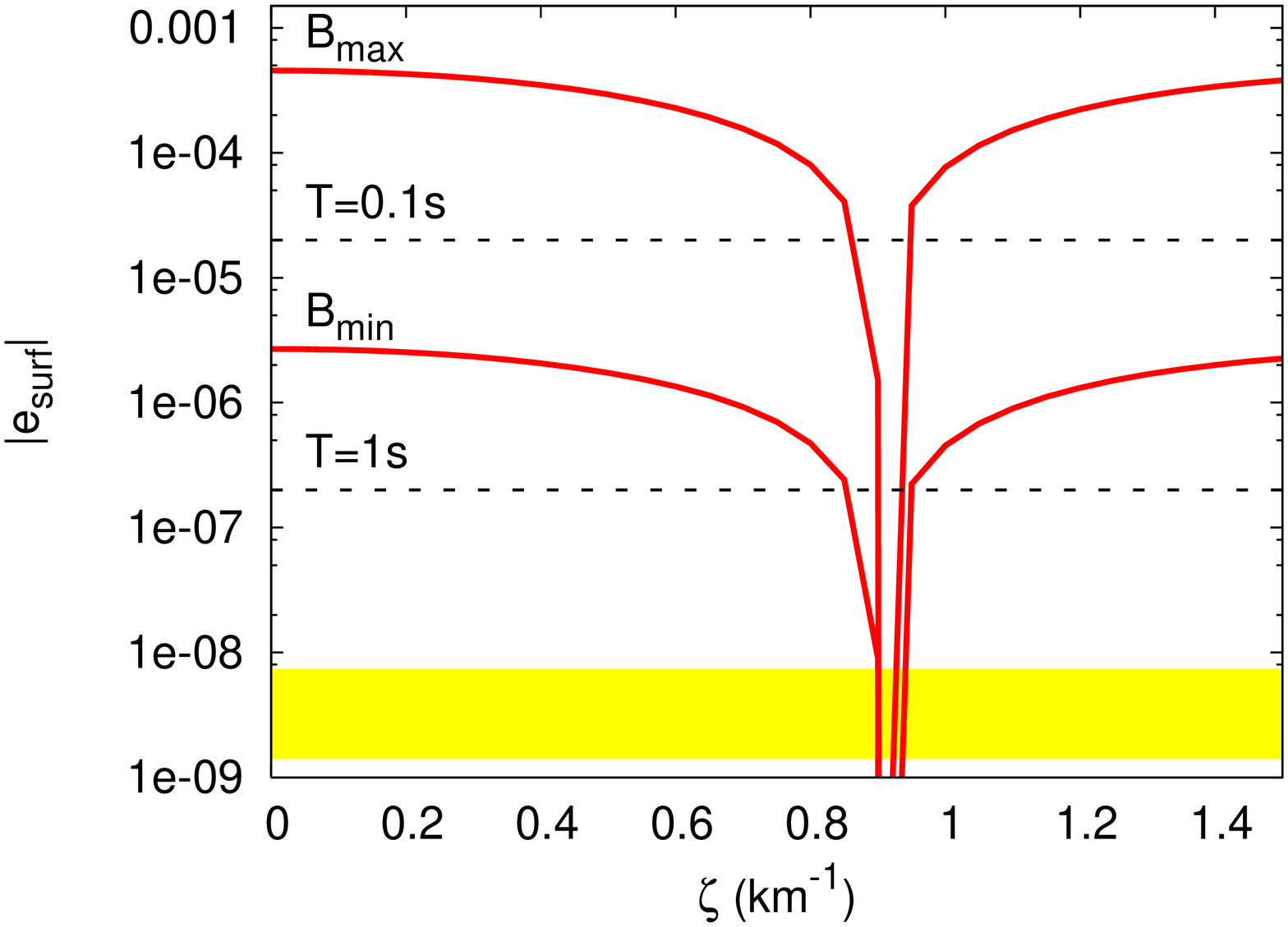}
\includegraphics[width=6cm,angle=270]{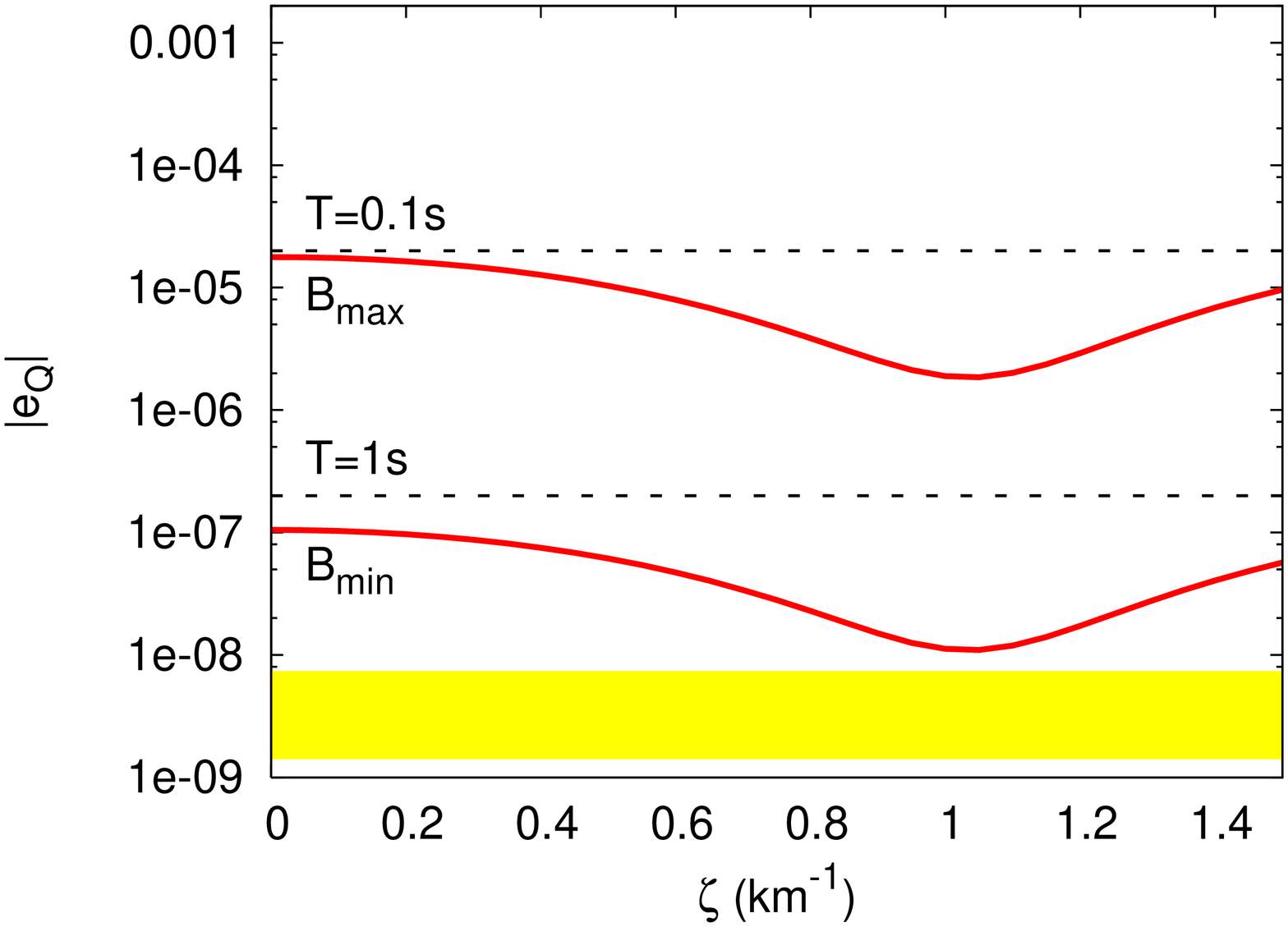}
\caption{$|e_{surf}|$ and $|e_{Q}|$ are plotted as functions of
$\zeta$, for two magnetic field configurations: field throughout the
star (upper panels), and crustal fields (lower panels).  The two solid
lines refer to $|e_{surf,Q}^B|$ computed for $B_{pole}= B_{min,max}$
corresponding to the extrema of the range (\ref{rangeB}).  The
shadowed region corresponds to $e_{surf,Q}^\Omega$ evaluated for
rotation periods in the range (\ref{rangeT}).  The dashed lines
correspond to smaller rotation periods.
\label{ellrot}}
\end{figure}
\end{center}

In Figure \ref{ellrot} we show $|e_{surf}|$ and $|e_{Q}|$ as functions
of $\zeta$, for the two magnetic field configurations described in
Section \ref{bcs}: field throughout the star (upper panels), and
crustal fields (lower panels).  The two solid lines correspond the
$|e_{surf,Q}^B|$ computed for $B_{pole}$ equal to the minimum and
maximum values of the range (\ref{rangeB}).  The shadowed
region corresponds to the rotation contribution, $e_{surf,Q}^\Omega$,
for rotation periods in the range (\ref{rangeT}).  The dashed lines
correspond to $T=1$ s and $T=0.1$ s, outside that range and smaller
than the observed periods of SGR's and AXP's: we show these values
since they may possibly occur in young magnetars.  From Figure
\ref{ellrot} we see that for the observed magnetars $|e_{Q,surf}^B|$
is typically larger than $|e_{Q,surf}^\Omega|$.  This behaviour is
magnified when crustal field are present (lower panels in Figure
\ref{ellrot}).  The rotational contribution may significantly exceed
that of magnetic field only for stars rotating faster (dashed
lines). The solid line minima in the pictures correspond to the points
where $e_{Q,surf}^B=0$; there the effects of the poloidal and toroidal
fields balance and the ellipticity changes sign.

%%%%%%%%%%%%%%%%%%%%%%%%%%%%%%%%%%%%%%%%%%%%%%%%%%%%%%%%%%%%%%%
%%%%%%%%%%%%%%%%%%%%%%%%%%%%%%%%%%%%%%%%%%%%%%%%%%%%%%%%%%%%%%%

%%%%%%%%%%%%%%%%%%%%%%%%%%%%%%%%%%%%%%%%%%%%%%%%%%%%%%%%%%%%%%%%%%%%%%%%%%% 
\subsection{Deformation of magnetized stars with different masses and EOS}
%%%%%%%%%%%%%%%%%%%%%%%%%%%%%%%%%%%%%%%%%%%%%%%%%%%%%%%%%%%%%%%%%%%%%%%%%%% 
The results discussed in previous sections where obtained for a star
with mass $M=1.4\,M_\odot$ and equation of state (EOS) APR2.  We shall
now see how the results depend on the EOS and on the stellar mass.  To
this purpose, as an example we shall consider three different EOS:
\begin{itemize}
\item APR2 \cite{APR}, derived within the non-relativistic nuclear
many-body theory, assuming that the star is made of ordinary nuclear
matter; the maximum mass is $M_{max}=2.202\,M_\odot$.
\item G240 \cite{Glen}, derived within the relativistic mean-field theory
and allowing for the presence of hyperons in coexistence with ordinary
nuclear matter; $M_{max}=1.553\,M_\odot$.
\item QS, based on the MIT bag model \cite{Bag} (with $B=95$
MeV/fm$^3$, $\alpha_s=0.4$, $m_s=100$ MeV), assuming that the star is
a bare quark star, i.e. composed entirely of deconfined quark matter;
$M_{max}=1.445\,M_\odot$.
\end{itemize}
G240 with hyperons is a very soft EOS, QS is very stiff (for a comparative
discussion of these EOS see refs.  \cite{BFG,BFGM}).  Furthermore we
shall consider the two magnetic field configurations discussed in
Section \ref{bcs}, and two values of mass, $M=1.20\,M_\odot$ and
$M=1.40\,M_\odot$.

\begin{center}
\begin{figure}[htbp]
\includegraphics[width=6cm,angle=270]{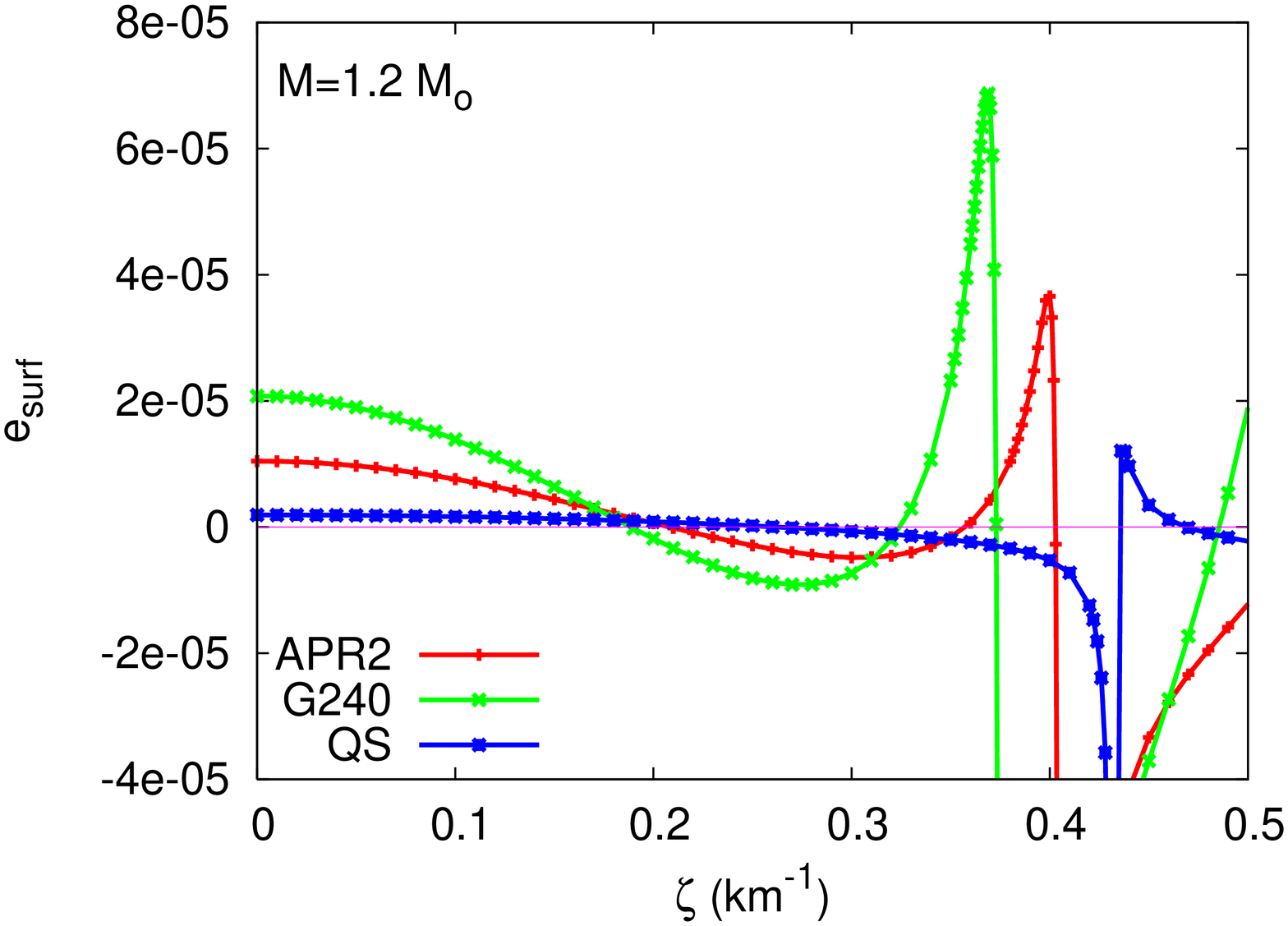}
\includegraphics[width=6cm,angle=270]{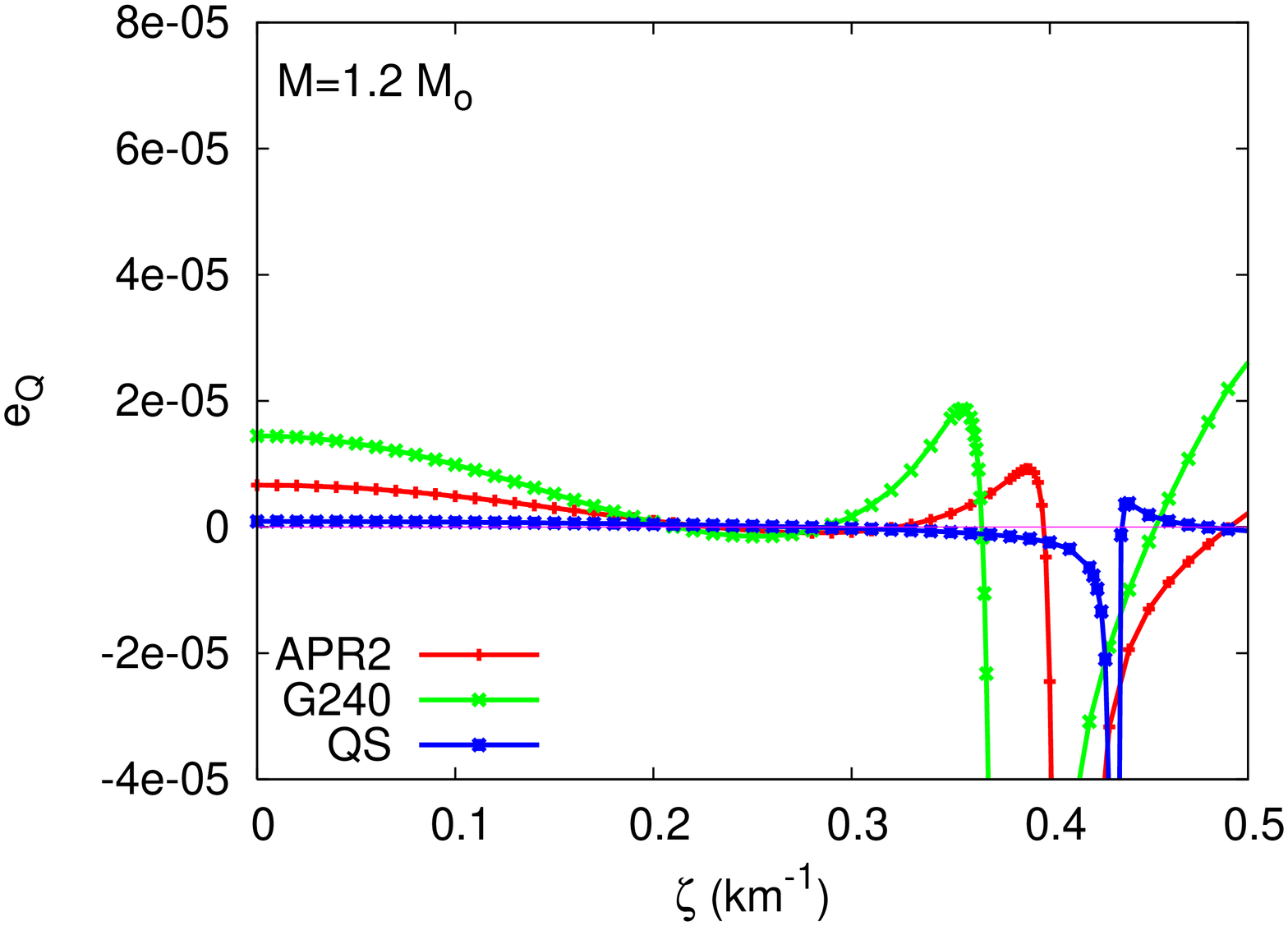}
\includegraphics[width=6cm,angle=270]{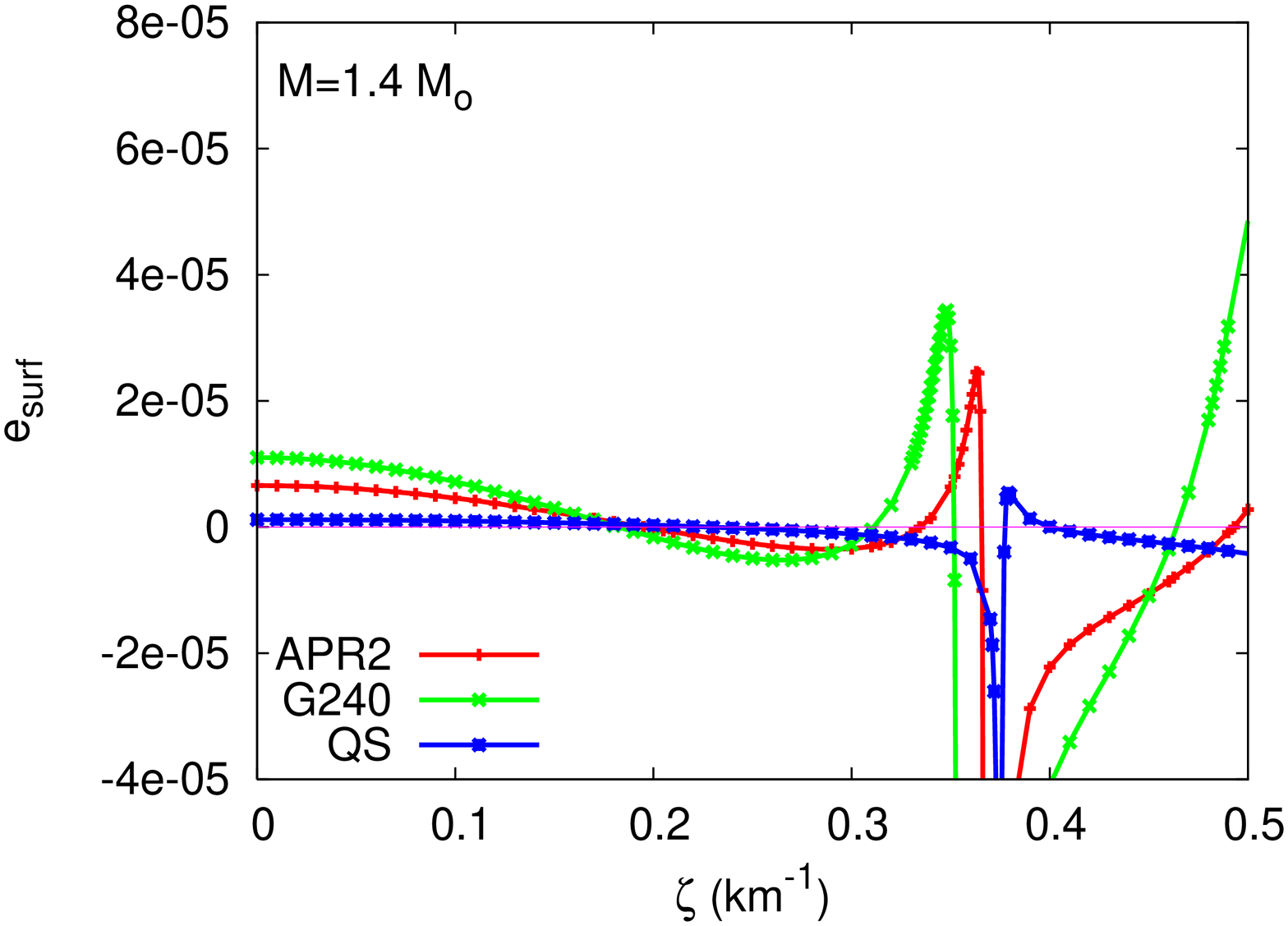}
\includegraphics[width=6cm,angle=270]{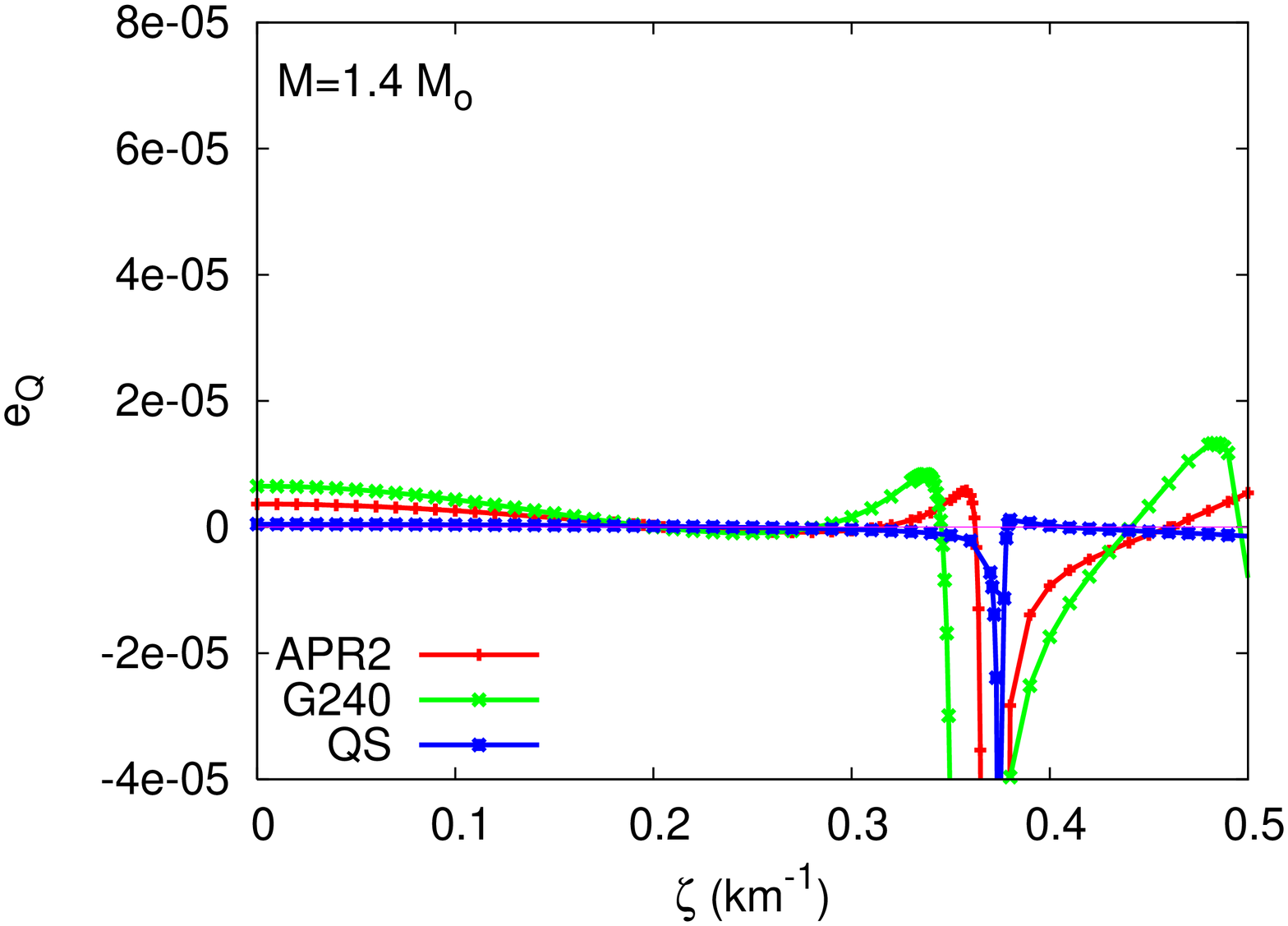}
\caption{Surface and quadrupole ellipticities as functions of $\zeta$,
for different equations of state, and magnetic fields extending
throughout the star. The stellar mass is $M=1.20\,M_\odot$ (upper
panels) and $M=1.40\,M_\odot$ (lower panels).
\label{elleos}
}
\end{figure}
\end{center}

In Figure \ref{elleos} we show $e_{surf}$ and $e_Q$ as functions of
$\zeta$, for $M=1.2\,M_\odot$ (upper panels) and for $M=1.4\,M_\odot$
(lower panels), for the selected EOS, when the magnetic field extends
throughout the star.  We see that, as expected, softer EOS and smaller
mass correspond to larger deformations.  For all masses and EOS, we
find the same qualitative behaviour shown in Figure\,\ref{ell}.

\begin{center}
\begin{figure}[htbp]
\includegraphics[width=6cm,angle=270]{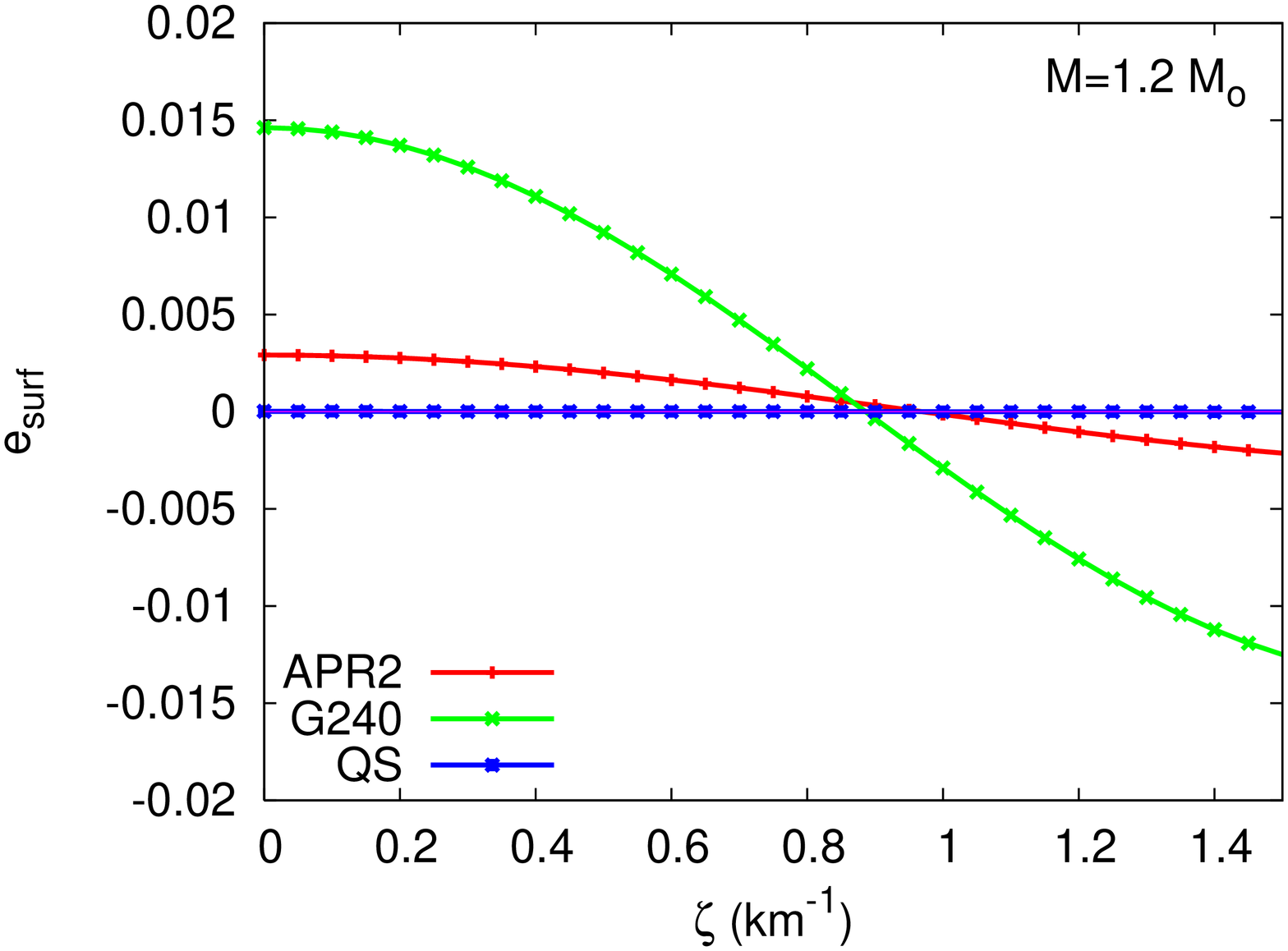}
\includegraphics[width=6cm,angle=270]{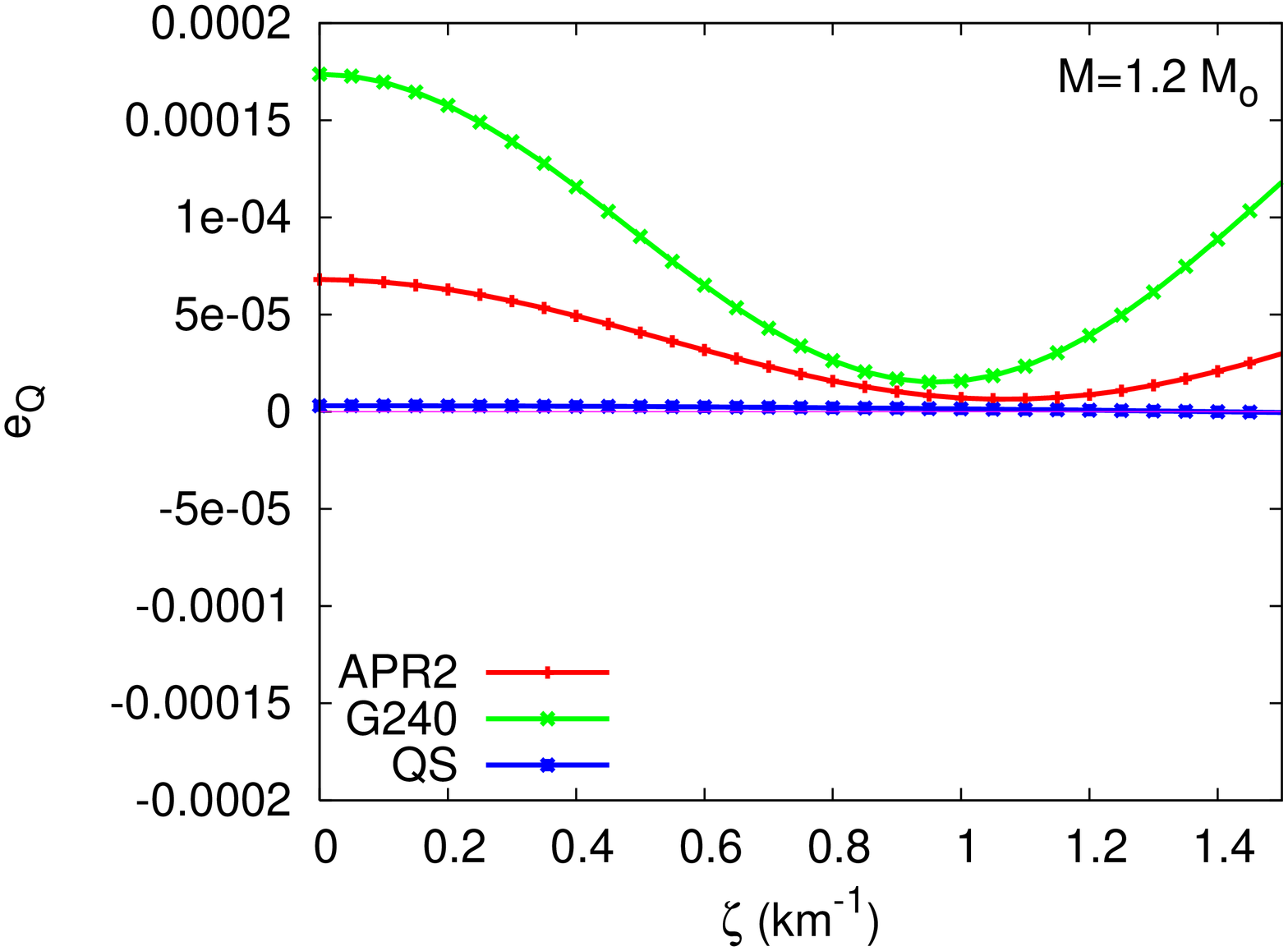}
\includegraphics[width=6cm,angle=270]{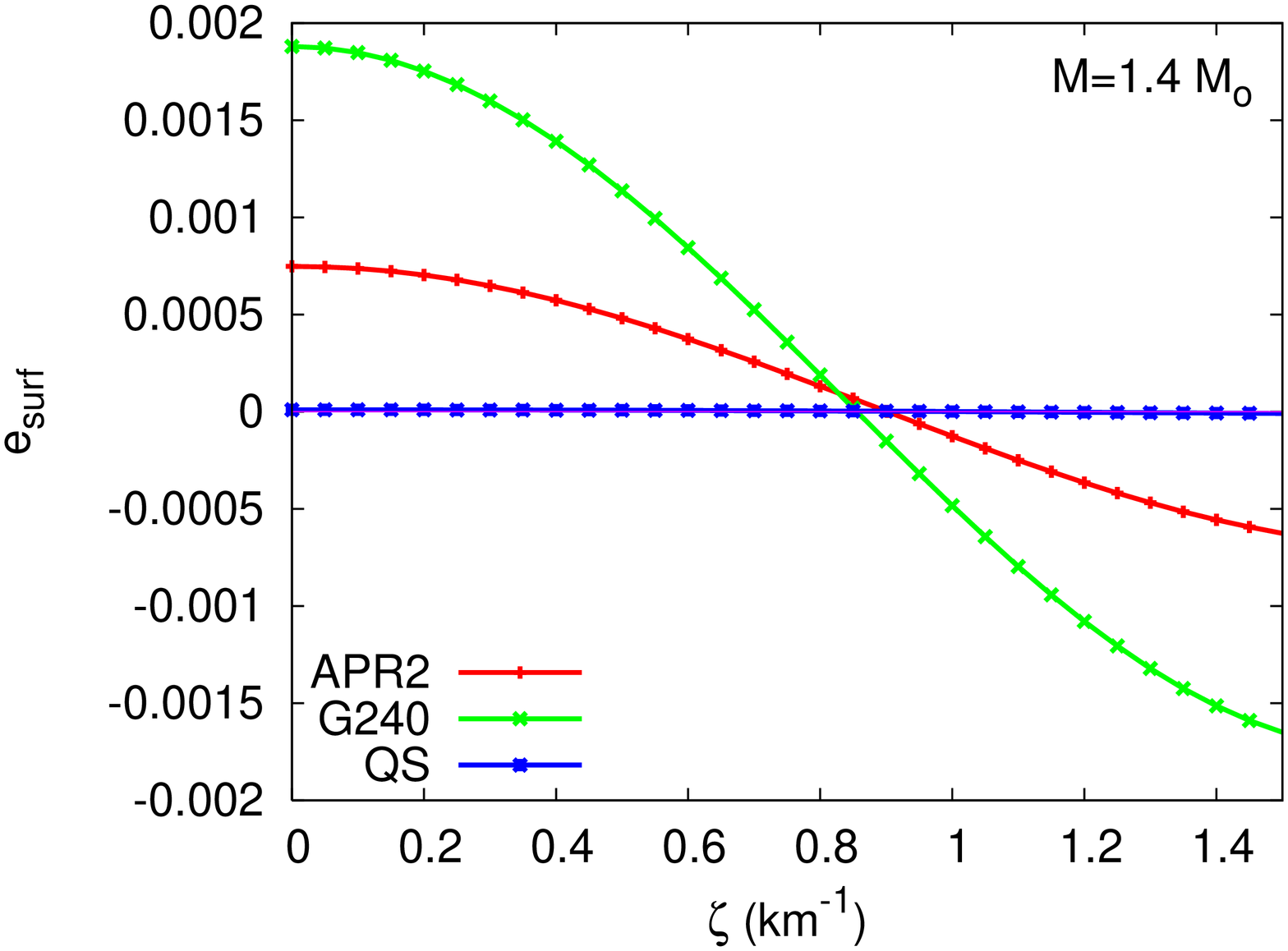}
\includegraphics[width=6cm,angle=270]{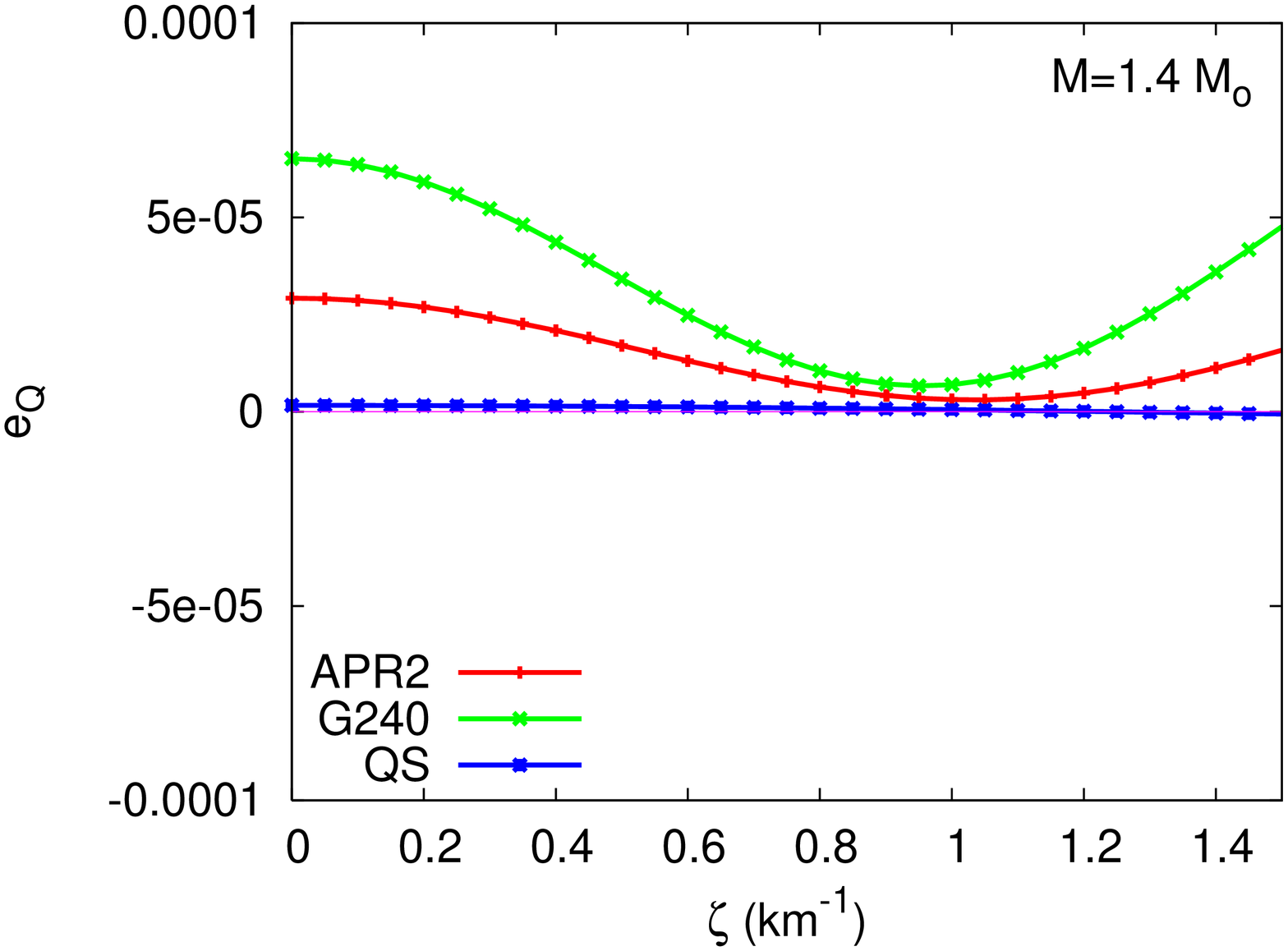}
\caption{Surface and quadrupole ellipticities are plotted as in
Figure~\ref{elleos} in the case of crustal fields.
\label{elleoscrust}
}
\end{figure}
\end{center}

In Figure \ref{elleoscrust}, $e_{surf}$ and $e_Q$ are shown in the case
of crustal fields.  We find that $e_{surf}$ depends strongly on the
mass and on the EOS.  For an assigned EOS, changing the mass from
$1.4\,M_\odot$ to $1.2\,M_\odot$, $e_{surf}$ increases by a factor
$\sim 10$, and $e_Q$ by factor $\sim2$. If we fix the mass and change
the EOS we find
\[
e_{surf,Q}(G240)/e_{surf,Q}(APR2)\sim [2-4]~,  
\]
whereas
\[
e_{surf,Q}(APR2)/e_{surf,Q}(QS)\sim [10-100]~.
\]
Finally, we find that for the stiffest EOS we consider (QS), for some
values of $\zeta$ the quadrupole ellipticity can become negative,
whereas this never occurs for APR2 and G240.

It should be stressed that when magnetic fields extend throughout the
star the dependence of the ellipticities on the EOS and on the mass is
considerably weaker (Figure \ref{elleos}).
%%%%%%%%%%%%%%%%%%%%%%%%%%%%%%%%%%%%%%%%%%%%%%%%%%%%%%%%%%%%%%%
%%%%%%%%%%%%%%%%%%%%%%%%%%%%%%%%%%%%%%%%%%%%%%%%%%%%%%%%%%%%%%%
%%%%%%%%%%%%%%%%%%%%%%%%%%%%%%%%%%%%%%%%%%%%%%%%%%%%%%%%%%%%%%%
%%%%%%%%%%%%%%%%%%%%%%%%%%%%%%%%%%%%%%%%%%%%%%%%%%%%%%%%%%%%%%%
%\begin{center}
\begin{figure}[htbp]
\includegraphics[width=7cm]{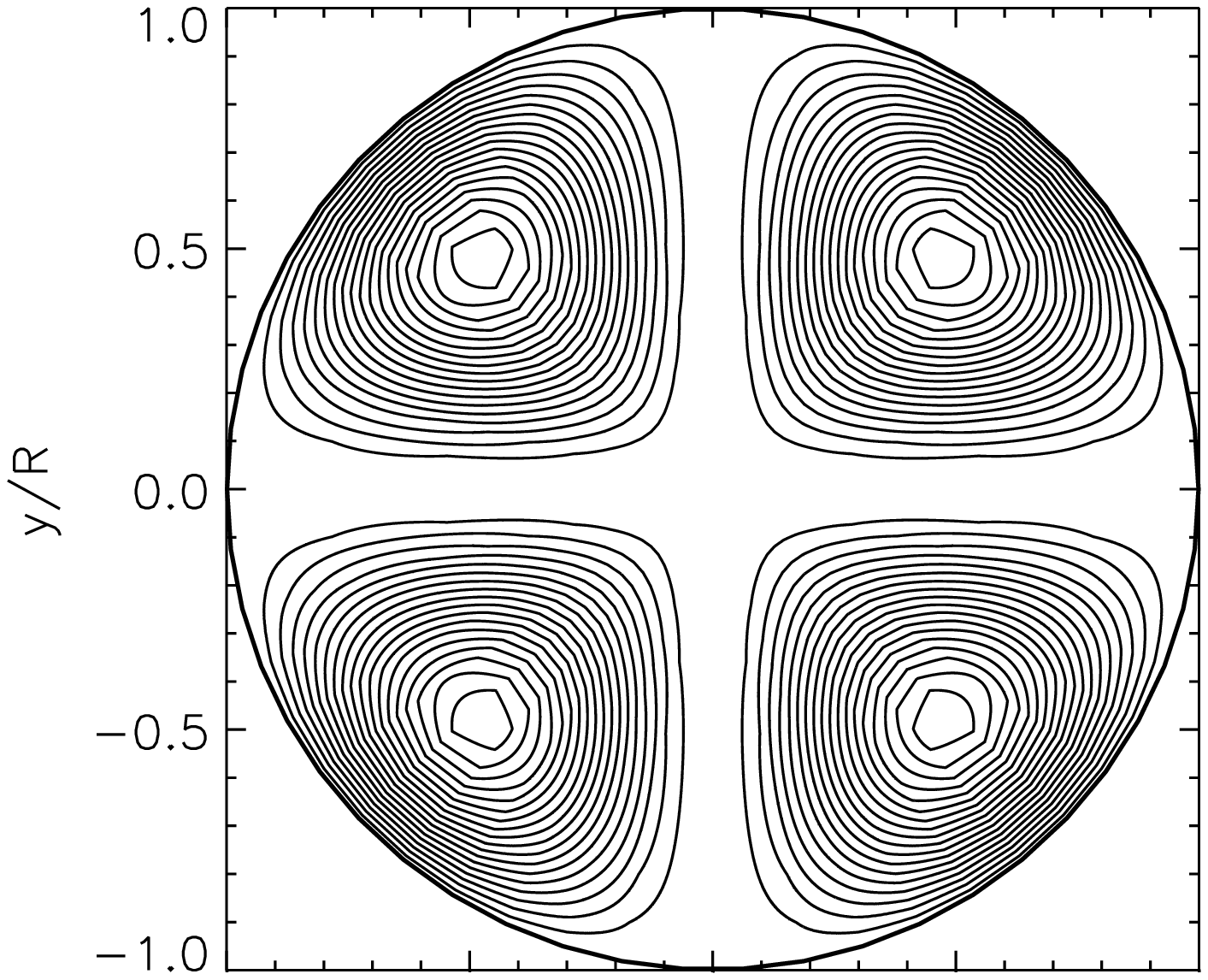}
\includegraphics[width=7cm]{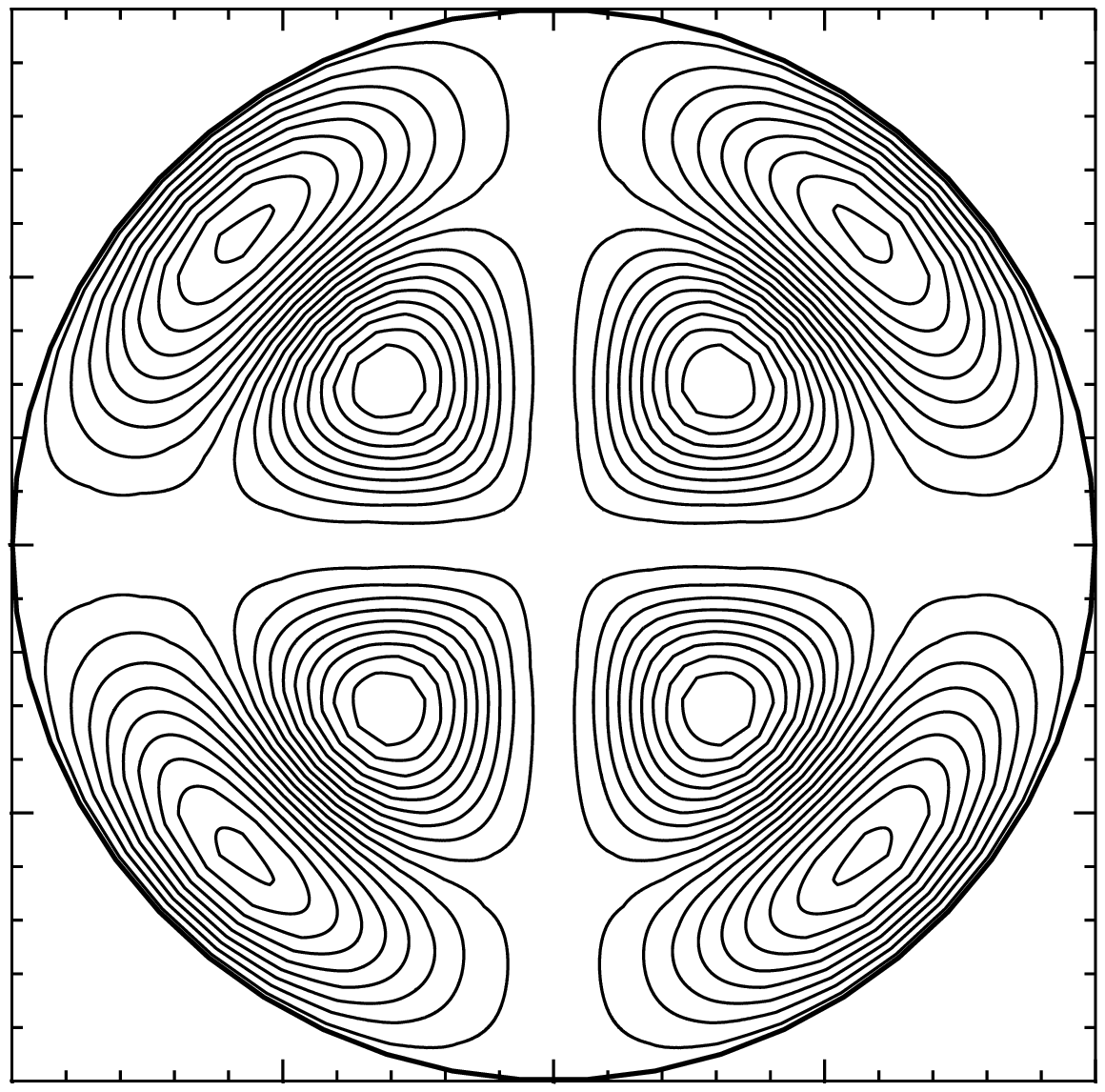}
\vskip 12pt\noindent
\includegraphics[width=7cm]{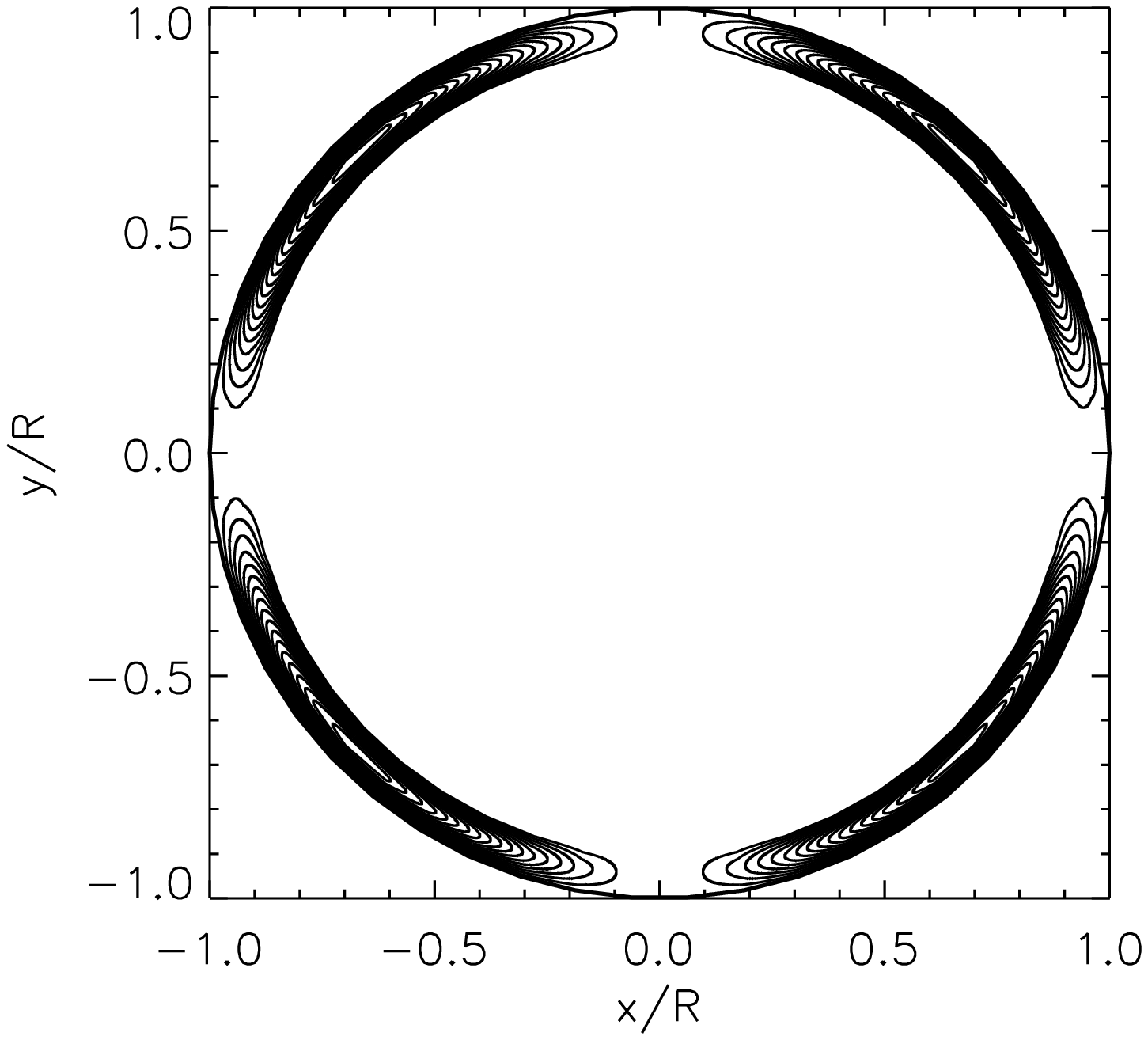}
\includegraphics[width=7cm]{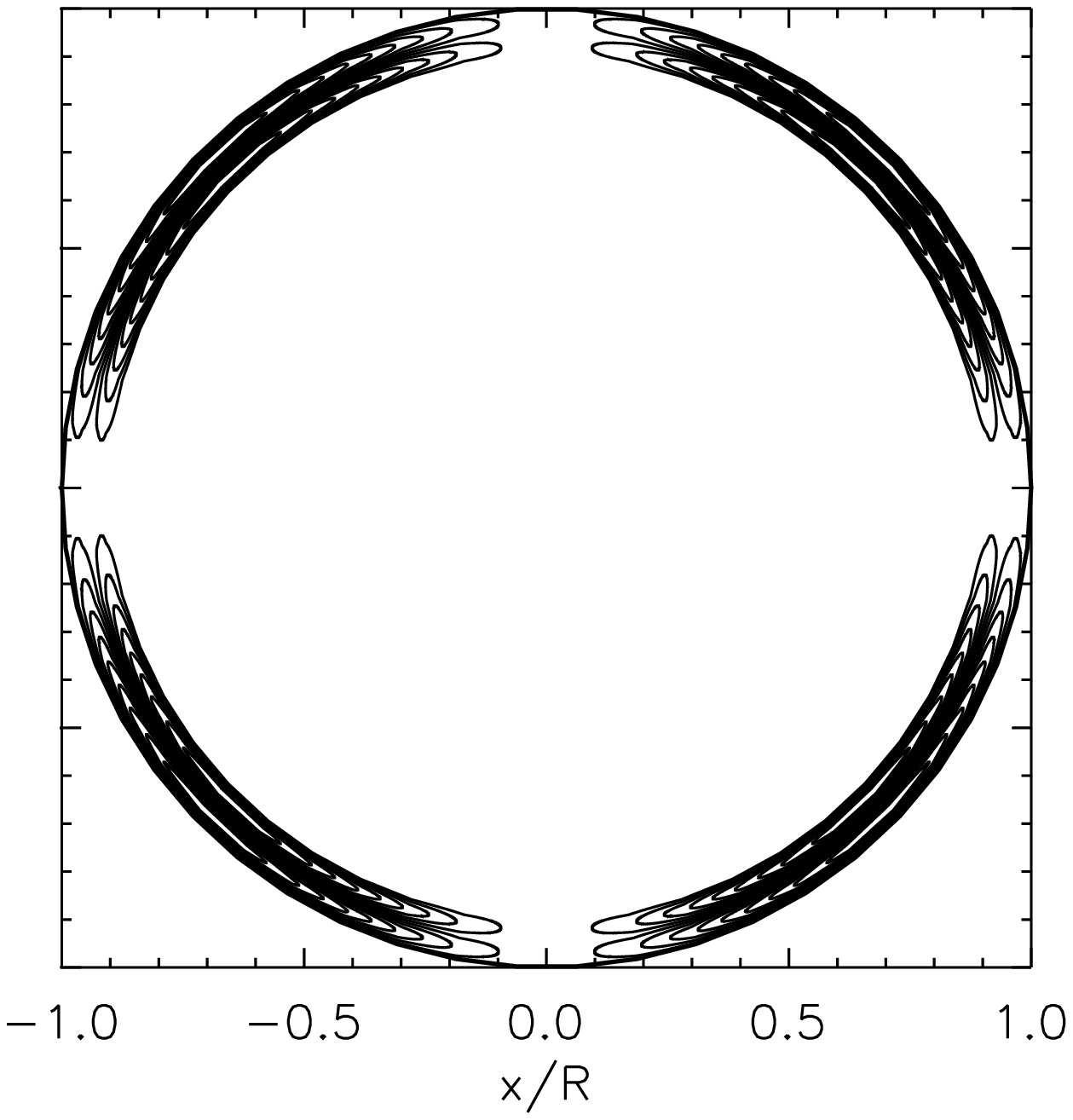}
\caption{
The projection of the $l=2$ field lines in the meridional plane 
is shown for $\zeta=\zeta_1$ (left panels), and
$\zeta=\zeta_2$ (right panels). The upper panels refer to fields
extending throughout the star, the lower panels to crustal fields.
\label{contourl2}}
\end{figure}
%\end{center}
%%%%%%%%%%%%%%%%%%%%%%%%%%%%%%%%%%%%%%%%%%%%%%%%%%%%%%%%%%%%
%%%%%%%%%%%%%%%%%%%%%%%%%%%%%%%%%%%%%%%%%%%%%%%%%%%%%%%%%%%%%%%%%%%%%%%%%%% 
\subsection{Higher order multipoles}
%%%%%%%%%%%%%%%%%%%%%%%%%%%%%%%%%%%%%%%%%%%%%%%%%%%%%%%%%%%%%%%%%%%%%%%%%%% 
In this paper we have focused on the study of dipole ($l=1$) magnetic
fields, which decay as $r^{-l-2}$ and therefore dominate far away from
the star.
In this Section we solve the Grad-Shafranov equation (\ref{eqal}) for  $l=2$,
including both poloidal and toroidal components:
\begin{equation}
e^{-\lambda}a_2''+\frac{\nu'-\lambda'}{2}e^{-\lambda}a_2'+\left(
\zeta^2e^{-\nu}-\frac{6}{r^2}\right)a_2=0\label{eqa22}\,.
\end{equation}
For fields extending throughout the star, we impose a regular
behaviour near the origin: $a_2(r\simeq0)=\alpha_0r^3 +O(r^5)$. In the
case of crustal fields the regularity condition is imposed near the
crust-core interface, i.e. $a_2(r\gtrsim r_c)=\alpha_2(r-r_c)
+O((r-r_c)^2)$.

We assume that the field vanishes outside the star, i.e.
$a_2(r>R)=0$.  Continuity of $a_2$ (and then of $B_r$) on the stellar
surface, implies $a_2(R)=0$; thus we have to solve an eigenvalue
problem (like in \cite{IS,Haskell}), to select the discrete set of
values $\zeta=\zeta_i$ for which the boundary conditions are
satisfied.  The eigenfunction $a_2(r)$, which corresponds to $\zeta_i$, has
$i$-nodes, one of which is located at the stellar surface.  We note
that, since as mentioned in Section \ref{secgrad}, only the current
$J^t_\mu$ contributes to $a_l$ when $l>1$, we do not have as much
freedom as in the $l=1$ case, when we used the constant $c_0$ to
impose $a_2'(R)=0$; consequently,  $a_2'$ is discontinuous (and so is
$B_\theta$) on the stellar surface.

For a star with $M=1.4\,M_\odot$, described by the EOS APR2, we have
determined the field configurations corresponding to the first five
eigenvalues $\zeta_i$.  Since the field vanishes on the stellar
surface, we normalize $B$ by choosing $\alpha_0$ such that the maximum
value of the magnetic field inside the star is $B_{max}=10^{16}$ G,
i.e.  of the same order of magnitude of the field considered in
Section \ref{diffconf} for $l=1$ (see
Tables\,\ref{Tabell1},\ref{Tabell2}).  Then we have solved the equations of stellar
deformation given in Appendix (\ref{deforml2}), finding the
surface and quadrupole ellipticities.

The projection of the field lines in the meridional plane is shown in
figure \ref{contourl2} for the first two eigenvalues.  The upper panels
refer to fields defined throughout the star, the lower panels to
crustal fields.
%%%%%%%%%%%%%%%%%%%%%%%%%%%%%%%%%%%%%%%%%%%%%%%%%%%%%%%%%%%%%%%
\begin{table}[htbp]
%\begin{center}
\begin{tabular}{|c|c|c|}
\hline
$\zeta$ (km$^{-1}$)&$e_{surf}$&$e_Q$\\
\hline 
$0.325$&$-5.07\times 10^{-6}$&$-6.24\times 10^{-6}$\\
$0.495$&$-1.78\times 10^{-6}$&$-2.19\times 10^{-6}$\\
$0.662$&$-9.37\times 10^{-7}$&$-1.15\times 10^{-6}$\\
$0.829$&$-5.80\times 10^{-7}$&$-7.13\times 10^{-7}$\\
$0.996$&$-3.95\times 10^{-7}$&$-4.85\times 10^{-7}$\\
\hline
\end{tabular}
\caption{First eigenvalues $\zeta_i$, and the corresponding surface
and quadrupole ellipticities, for $l=2$ magnetic fields extending
throughout the star.\label{Tabell3}}
\begin{tabular}{|c|c|c|}
\hline
$\zeta$ (km$^{-1}$)&$e_{surf}$&$e_Q$\\
\hline 
$1.705$&$1.75\times 10^{-5}$&$2.16\times 10^{-5}$\\
$3.397$&$1.64\times 10^{-5}$&$2.01\times 10^{-5}$\\
$5.091$&$1.60\times 10^{-5}$&$1.96\times 10^{-5}$\\
$6.787$&$1.58\times 10^{-5}$&$1.94\times 10^{-5}$\\
$8.482$&$1.56\times 10^{-5}$&$1.92\times 10^{-5}$\\
\hline
\end{tabular}
\caption{First eigenvalues $\zeta_i$, and the corresponding surface
and quadrupole ellipticities, for $l=2$ crustal
fields. \label{Tabell4}}
%\end{center}
\end{table}
%%%%%%%%%%%%%%%%%%%%%%%%%%%%%%%%%%%%%%%%%%%%%%%%%%%%%%%%%%%%%%%

In Tables \ref{Tabell3} and \ref{Tabell4} we give the first five
eigenvalues $\zeta_i$ and the corresponding ellipticities, for fields
extending throughout the star and for crustal fields, respectively.
In the first case the ellipticities are always negative and of the
order
\begin{equation}
|e_{surf,Q}|\sim10^{-7}\,-\,10^{-6}
\end{equation}
i.e. smaller than for the $l=1$ fields.
For crustal fields, ellipticities are always positive and 
of the same order of magnitude as for $l=1$, i.e.
\begin{equation}
 e_{surf,Q} \sim10^{-5}\,.
\end{equation}

%%%%%%%%%%%%%%%%%%%%%%%%%%%%%%%%%%%%%%%%%%%%%%%%%%%%%%%%%%%%%%%%%%%%%%%%%%% 
\section{Concluding Remarks}\label{conclusions}
%%%%%%%%%%%%%%%%%%%%%%%%%%%%%%%%%%%%%%%%%%%%%%%%%%%%%%%%%%%%%%%%%%%%%%%%%%% 
In this paper we 
solve Einstein-Maxwell's equations, using  a perturbative approach, to
study the structure of the magnetic field of magnetars,
and to find the deformation it induces on the star.
We extend previous works on the
subject \cite{IS,KOK,KOKrot,Haskell} in several respects: we include
toroidal fields inside the star, thus removing the assumption,
used in \cite{BBGN,BG,CPL}, of circular spacetime; we determine both
the surface ellipticity and the quadrupole ellipticity; we explore
various field configurations, corresponding to different boundary
conditions; we compare the effects produced by a magnetic field and by
rotation on the stellar structure; we study how different equations of state and masses affect
the magnetic field structure and the quadrupole ellipticity it
induces; we solve the equations for higher order ($l=2$) multipoles.

In summary, the main results of our study are the following.
\begin{itemize}
\item Crustal fields induce surface deformations much larger than
fields extending throughout the star, but the quadrupole deformations
are comparable in the two cases. Typically, crustal fields produce
oblate, rather than prolate shapes.
\item For particular values of the parameter $\zeta$, representing the
ratio between toroidal and poloidal components, the magnetic field inside the
star and the deformation can be extremely large; such configurations
correspond to prolate shapes.
\item  For the typical rotation rates of observed magnetars,
the deformation induced by rotation is much smaller than that 
induced by the magnetic field. 
\item Neutron stars with the same magnetic field, but with
softer EOS or smaller mass, exhibit larger deformations.
\item If the magnetic field extends throughout the star, the
deformations induced by higher order ($l=2$) multipoles are one order
of magnitude smaller than the dipolar contributions; for crustal
fields, they are comparable.
\end{itemize}
As a future extension of this work, we plan to study the effect of
couplings between different multipoles, which we have neglected in the
present paper, and to determine their relative weights.  Furthermore,
the equilibrium configurations we have found will be used as
background models to study the oscillations of highly magnetized
neutron stars.

%%%%%%%%%%%%%%%%%%%%%%%%%%%%%%%%%%%%%%%%%%%%%%%%%%%%%%%%%%%%
\begin{acknowledgments} 
We thank Nils Andersson, Juan Antonio Miralles, Luciano Rezzolla, Lars
Samuelsson, Kostas Glampedakis and Riccardo Ciolfi for useful
suggestions and discussions.
\end{acknowledgments}
%%%%%%%%%%%%%%%%%%%%%%%%%%%%%%%%%%%%%%%%%%%%%%%%%%%%%%%%%%%%
\appendix
%%%%%%%%%%%%%%%%%%%%%%%%%%%%%%%%%%%%%%%%%%%%%%%%%%%%%%%%%%%%
\section{The deformations of the star}\label{deform}
%%%%%%%%%%%%%%%%%%%%%%%%%%%%%%%%%%%%%%%%%%%%%%%%%%%%%%%%%%%%
The metric of a non rotating star deformed by a magnetic field can be
written, up to $O(B^2)$, as \cite{IS}
\begin{eqnarray}
ds^2&=&-e^{\nu}\left(1+2[h_0+h_2P_2(\cos\theta)]\right)dt^2+
e^{\lambda}\left(1+\frac{2e^{\lambda}}{r}[m_0+m_2P_2(\cos\theta)]
\right)dr^2\nn\\
&&+ r^2[1+2 k_2P_2(\cos\theta)]\left(d\theta^2+sin^2 
\theta d\phi^2\right)\nn\\
&&+2\left[i_1 P_1(\cos\theta)+i_2 P_2(\cos \theta)+
i_3 P_3 (\cos \theta)\right]]dtdr\nn\\
&&+2 \sin \theta \left(v_1  \frac{\partial}{\partial \theta}
P_1(\cos \theta)+v_2  \frac{\partial}{\partial \theta}P_2(\cos \theta)
+v_3 \frac{\partial}{\partial\theta}P_3(\cos \theta)\right)dtd\phi\nn\\
&&+2\sin \theta\left(w_2\frac{\partial}{\partial \theta}P_2(\cos \theta)
+w_3\frac{\partial}{\partial \theta}P_3(\cos \theta)\right)drd\phi\,.
\label{metricall}
\end{eqnarray}
The  perturbed quantities ($h_i(r)$, $m_i(r)$,
$m_i(r)$, $k_i(r)$) (i=0,2) and ($i_i(r)$, $v_i(r)$ $i_i(r)$)
(i=1,2,3) are found by solving  the linearized Einstein equations
\begin{equation}\label{einsteineq}
\delta G_{\mu \nu}=8\pi \delta T_{\mu \nu}\,.
\end{equation}
The pressure $p$ and the energy density $\rho$ can be expanded 
as $p=p^{(0)}+\delta p$, $\rho=\rho^{(0)}+\delta\rho$, with
\begin{eqnarray}
\delta p(r,\theta)&=&(\delta p_{0}+\delta p_2P_2(\cos \theta))
\label{Pmulti}\\
\delta \rho(r,\theta)&=&\frac{\rho^{(0)\prime}}{P^{(0)\prime}}
(\delta p_{0}+\delta p_{2}P_2(\cos \theta))\,.\label{denmulti}
\end{eqnarray}
%%%%%%%%%%%%%%%%%%%%%%%%%%%%%%%%%%%%%%%%%%%%%%%%%%%%%%%%%%%%
\subsection{Deformation induced by a  dipole ($l=1$) magnetic field}
\label{deforml1}
%%%%%%%%%%%%%%%%%%%%%%%%%%%%%%%%%%%%%%%%%%%%%%%%%%%%%%%%%%%%
As discussed in section \ref{secgrad},
$a(r,\theta)=a_1(r)P_1(\cos\theta)$; by expanding the components
$(rr)$, $(r\theta)$, $(\theta\theta)-\sin^{-2}\theta(\phi\phi)$ of the
perturbed Einstein equations (\ref{einsteineq}) in spherical, tensor
harmonics and by considering the $l=2$ equations, which give the
stellar deformation, we have \cite{IS}
\begin{eqnarray}
&&h'_2+\frac{4e^{\lambda}}{\nu'r^2}y_2
+\left[\nu'-\frac{8\pi e^{\lambda}}{\nu'}
(p^{(0)}+\rho^{(0)})+\frac{2}{r^2\nu'}(e^{\lambda}-1)
\right]h_2\nn\\
&&=\frac{\nu'}{3}e^{-\lambda}a_1^{\prime 2}+\frac{4}{3r^2}a_1a_1'+\frac{1}{3}
\left(-\nu'+\frac{2}{\nu'r^2}e^{\lambda}\right)\zeta^2 e^{-\nu}(a_1)^2
-\frac{16\pi}{3\nu'r^2}e^{\lambda}j_1a_1\label{h2primo}\\
&&y'_2+\nu'h_2=\frac{\nu'}{2}e^{-\lambda}a_1^{\prime 2}
+\frac{1}{3}\left[\frac{e^{-\lambda}}{r}\left(\nu'+\lambda'+\frac{2}{r}\right)
+e^{-\nu}\zeta^2-\frac{2}{r^2}\right]a_1a_1'\nn\\
&&-\frac{\nu'}{3}e^{-\nu}\zeta^2a_1^2-\frac{4\pi}{3}j_1
\left(a_1'+\frac{2}{r}a_1\right)\label{y2primo}~,
\end{eqnarray}
where $j_1=c_0(\rho+p)r^2$ and 
\begin{equation}
y_2\equiv h_2+k_2-\frac{e^{-\lambda}}{6}a_1^{\prime 2}
-\frac{2e^{-\lambda}}{3r}a_1a_1'-\frac{2}{3r^2}a_1^2\,.\label{definy2}
\end{equation}
Assuming regularity of $h_2$ and $y_2$ as $r\rightarrow 0$ implies
that near the origin
\begin{equation}\label{y2h2in}
h_{2}\simeq Ar^2\,,~~~~~y_2 \simeq Br^4~,
\end{equation}
where 
\begin{equation}\label{By2}
B=\left(-2\pi A+\frac{16}{3}\pi \alpha_0^2 \right)
\left(p_c^{(0)}+\frac{\rho_c^{(0)}}{3}\right)
-\frac{4\pi}{3}\alpha_0c_0(\rho^{(0)}_c+p^{(0)}_c)
+\frac{\alpha_0^2\zeta^2}{6e^{\nu_c}}\,.
\end{equation}
It is worth mentioning that the terms in $a_1,a_1'$ which appear in the
definition of $y_2$ (eq. \ref{definy2}) are important: if they are not
included (i.e. if we define $y_2\equiv h_2+k_2$), the asymptotic
behaviour (\ref{y2h2in}) is not satisfied.

The quantities $h_2$, $y_2$ inside the star can be decomposed as
follows:
\begin{eqnarray}
h_2&=&c_1 h_2^h+h_2^p\nn\\
y_2&=&c_1 y_2^h+y_2^p\label{hy2somma}\,.
\end{eqnarray}
For magnetic fields extending throughout the star, $h_2^p$ and $y_2^p$
can be found by integrating (\ref{h2primo}), (\ref{y2primo}) from
$r=0$ with, for instance, $A=1$ and $B$ given by (\ref{By2}); $h_2^h$
and $y_2^h$ are the solutions of the associated homogeneous equations
(i.e. with $a_1=a'_1=0$).

When the magnetic field is confined to the crust, in the core (defined
conventionally by $0\le r\le r_c$) $a_1\equiv0$,
and eqs. (\ref{h2primo}), (\ref{y2primo}) are homogeneous;
thus in this region $h_2=h_2^h$, $y_2=y_2^h$.
On the crust-core interface $r=r_c$, we impose $a_1=0$, $a_1'=const$.  
We integrate the non-homogeneous equations starting at $r=r_c$ with the initial
conditions
\begin{equation}
h_2^p(r_c)=0\,,~~~~~y_2^p(r_c)=-\frac{e^{-\lambda(r_c)}}{6}(a_1'(r_c))^2\,.
\end{equation}
The non-vanishing value for $y_2^p(r_c)$ follows from the requirement
of continuity of $h_2+k_2$ at the crust-core interface (see
eq. (\ref{definy2})).

The constant $c_1$ in (\ref{hy2somma}) can be determined by matching
the solution inside the star with the analytical solution in vacuum
\cite{KOK}:
\begin{eqnarray}
h_2&=&KQ^2_2(z)+\hat h_2(z)\nn\\
y_2&=&-\frac{2K}{\sqrt{z^2-1}} Q^1_2(z)+\hat y_2(z)
-\frac{e^{-\lambda}}{6}(a_1')^2-\frac{2}{3r}e^{-\lambda}(a_1'a_1)
-\frac{2}{3r^2}(a_1)^2\,.\label{solexth2y2dip}
\end{eqnarray}
Here $K$ is an integration constant, $a_1(r)$ is given by
eq. (\ref{dipole})
\begin{equation}
a_1=-\frac{3\mu}{8M^3}r^2\left[\ln\left(1-\frac{2M}{r}\right)+\frac{2M}{r}
+\frac{2M^2}{r^2}\right]\,,
\end{equation}
$Q^n_m$ are the associated Legendre functions of the second kind
\begin{eqnarray}
Q^2_2(z)&\equiv&\frac{z(5-3z^2)}{z^2-1}+\frac{3}{2}(z^2-1)\ln
\left(\frac{z+1}{z-1}\right)\label{q2}\\
Q^1_2(z)&\equiv&\frac{2-3z^2}{\sqrt{z^2-1}}
+\frac{3}{2}z(\sqrt{z^2-1})\ln
\left(\frac{z+1}{z-1}\right)~,
\label{q1}
\end{eqnarray}
with $z\equiv\frac{r}{M}-1$, and
\begin{eqnarray}
\hat y_2&\equiv&\frac{3\mu^2}{8M^4}\frac{7z^2-4}{z^2-1}
+\frac{3\mu^2}{16M^4}\frac{z(11z^2-7)}{z^2-1}\ln\left(\frac{z-1}{z+1}
\right)
+\frac{3\mu^2}{16M^4}(2z^2+1)\left(\ln\frac{z-1}{z+1}\right)^2
\label{y2partsol}\\
\hat h_2&\equiv&-\frac{3\mu^2}{16M^4}
\left[3z-\frac{4z^2+2z}{z^2-1}\right]
-\frac{3\mu^2}{32M^4}\left[3z^2-8z-3-\frac{8}{z^2-1}\right]
\ln\left(\frac{z-1}{z+1}\right)\nn\\
&&+\frac{3\mu^2}{16M^4}(z^2-1)\left(\ln\frac{z-1}{z+1}\right)^2\,.
\label{h2partsol}
\end{eqnarray}
We have checked, both analytically and numerically, that
(\ref{solexth2y2dip}) is actually solution of (\ref{h2primo}),
(\ref{y2primo}) in vacuum. Matching $h_2$ and $y_2$ at $r=R$ allows to
fix the constants $c_1$ and $K$.

The integration constant $K$ is related to the mass-energy quadrupole moment
of the star (see section \ref{ellip}) by the relation
\begin{equation}
K=\frac{5Q}{8M^3}+\frac{3\mu^2}{4M^4}\,.
\end{equation}
Indeed, the asymptotic limit of
$h_2(r)$ for $r\rightarrow\infty$ is
\begin{equation}
h_2\rightarrow\frac{Q}{r^3}\,.
\end{equation}
Finally, we can compute the surface ellipticity of the star
(\ref{esurf}) following the definitions of \cite{CM,KOK}:
\begin{equation}\label{eccentricita}
e_{surf}=-\frac{3}{2}\left(\frac{\delta r_2}{r}-k_2\right)
=-\frac{3}{2}\left(\frac{\delta p_2}{rp^{(0)\prime}}-k_2\right)=
\left(-\frac{2c_0a_1}{r\nu'}+\frac{3h_2}{r\nu'}-\frac{3k_2}{2}\right)_{r=R}
\end{equation}
where $\delta p=\sum_l\delta p_lP_l$, $\delta r=\sum_l\delta r_lP_l$ and 
\begin{equation}
\delta p_2=-(\rho^{(0)}+p^{(0)})h_2+\frac{2}{3r^2}a_1j_1.\label{dph2aj}
\end{equation}
The relation (\ref{dph2aj}) is a consequence of Euler's
equation. Indeed, from equations (\ref{defchi}), (\ref{int3}),
(\ref{defc01}), it follows that
\begin{equation}
\ln\left(\sqrt{-g_{00}}\frac{\rho+p}{n}\right)=c_0\sin\theta a_{,\theta}+const.
\label{chigc0}
\end{equation}
If we perturb (\ref{chigc0}), using the following relation which holds
for a barotropic EOS
\[
\delta p=n\delta\left(\frac{\rho+p}{n}\right)\,,
\]
we find (\ref{dph2aj}).

%%%%%%%%%%%%%%%%%%%%%%%%%%%%%%%%%%%%%%%%%%%%%%%%%%%%%%%%%%%%
\subsection{Deformations induced by a quadrupole ($l=2$) magnetic field}
\label{deforml2}
%%%%%%%%%%%%%%%%%%%%%%%%%%%%%%%%%%%%%%%%%%%%%%%%%%%%%%%%%%%%
We assume $a(r,\theta)=a_2(r)P_2(\cos\theta)$, and expand in
spherical, tensor harmonics the $(rr)$-, $(r\theta)$-,
$(\theta\theta)-\sin^{-2}\theta(\phi\phi)$-components of the perturbed
Einstein equations (\ref{einsteineq}). We find that the $l=2$
equations are:
\begin{eqnarray}
&&h'_2+\frac{4e^{\lambda}}{\nu'r^2}y_2
+\left[\nu'-\frac{8\pi e^{\lambda}}{\nu'}
(p^{(0)}+\rho^{(0)})+\frac{2}{r^2\nu'}(e^{\lambda}-1)
\right]h_2\nn\\
&&=\frac{3}{7}\nu'e^{-\lambda}a_2^{\prime 2}+\frac{12}{7r^2}a_2a_2'-\frac{3}{7}
\left(\nu'+\frac{2}{\nu'r^2}e^{\lambda}\right)\zeta^2 e^{-\nu}(a_2)^2
\label{h2primol2}\\
&&y'_2+\nu'h_2=\frac{3}{14}\nu'e^{-\lambda}a_2^{\prime 2}
+\frac{3}{7}\left[\frac{e^{-\lambda}}{r}\left(\nu'+\lambda'+\frac{2}{r}\right)
-\frac{3}{7}e^{-\nu}\zeta^2-\frac{2}{r^2}\right]a_2a_2'\nn\\
&&-\frac{3}{7}\nu'e^{-\nu}\zeta^2a_1^2~,
\label{y2primol2}
\end{eqnarray}
where we have defined 
\begin{equation}
y_2\equiv h_2+k_2+\frac{3}{14}e^{-\lambda}a_2^{\prime 2}
-\frac{6e^{-\lambda}}{7r}a_2a_2'-\frac{18}{7r^2}a_2^2\,.\label{definy2l2}
\end{equation}
Assuming regularity of $h_2$ and $y_2$ as $r\rightarrow 0$ implies
that near the origin
\begin{equation}\label{y2h2inl2}
h_{2}\simeq Ar^2\,,~~~~~y_2 \simeq Br^4\,,
\end{equation}
where 
\begin{equation}\label{By2l2}
B=-2\pi A\left(p_c^{(0)}+\frac{\rho_c^{(0)}}{3}\right)\,.
\end{equation}
The integration of eqs. (\ref{h2primol2}), (\ref{y2primol2}) and the
determination of $Q$ and $e_{surf}$ can be performed as in the
previous section.
%%%%%%%%%%%%%%%%%%%%%%%%%%%%%%%%%%%%%%%%%%%%%%%%%%%%%%%%%%%%%%%%%%%%%%%%%%% 
 

\begin{thebibliography}{99} 
\bibitem{SGRAXP} R.C. Duncan, C. Thompson, Astrophys. J. 392, L9
(1992); C. Thompson, R.C. Duncan, Astrophys. J. 408, 194 (1993);
S. Mereghetti, L. Stella, Astrophys. J. 442, L17 (1995);
C. Kouvelioutou et al., Nature, 393, 235 (1998).
\bibitem{WT} P.M. Woods, C. Thompson, in {\it Compact stellar X-ray
sources}, Cambridge Astrophysics Series, No. 39, p.547-586, Cambridge
University Press (2006).
\bibitem{GRB} V.V. Usov, Nature, 357, 472 (1992);
W. Kluzniak, M. Ruderman, Astrophys. J. 505, L113 (1998);
J.C. Wheeler at al., Astrophys. J. 537, 810 (2000).
\bibitem{KOUV} C. Kouveliotou, S. Dieter, T. Strohmayer, J. van
Paradijs, G.J. Fishman, C.A. Meegan, K. Hurley, Nature 393, 235
(1998).
\bibitem{FLARES} G.L. Israel et al., Astrophys. J. 628, L53 (2005);
T.E. Strohmayer, A.L. Watts, Astrophys. J. 632, L111 (2005);
L. Samuelsson, N. Andersson, Mon. Not. Roy. Astron. Soc. 374, 256
(2005); H. Sotani, K.D. Kokkotas, N. Stergioulas,
Mon. Not. Roy. Astron. Soc.  375, 261 (2007).
\bibitem{Jones} P.B. Jones, Astrophys. Space Sci. 33, 215 (1975). 
\bibitem{Cutler} C. Cutler, Phys. Rev. D 66, 084025 (2002).
\bibitem{Oron02} A. Oron, Phys. Rev. D 66, 023006 (2002).
\bibitem{BR} A. Bonanno, L. Rezzolla, Astron. \& Astrophys.  410, L33
(2003).
\bibitem{FR} E. Flowers, M. Ruderman, Astrophys. J. 215, 302 (1977).
\bibitem{BS} J. Braithwaite, H.C. Spruit, Astron. \& Astrophys. 450,
1097 (2006).
\bibitem{PG07} J.A. Pons, U. Geppert, Astron. \& Astrophys.
470, 303 (2007).
\bibitem{BBGN} M. Boquet, S. Bonazzola, E. Gourgoulhon, J. Novak,
Astron. \& Astrophys. 301, 757 (1995).
\bibitem{BG} S. Bonazzola, E. Gourgoulhon, Astron. \& Astrophys. 
312, 675 (1996).
\bibitem{CPL} C.Y. Cardall, M. Prakash, L.M. Lattimer, 
Astrophys. J. 554, 322 (2001).
\bibitem{IS} K. Ioka, M. Sasaki, Astrophys. J. 600, 296 (2004).
\bibitem{KOK} K. Konno, T. Obata, Y. Kojima, Astron. \& Astrophys. 
352, 211 (1999).
\bibitem{KOKrot} K. Konno, T. Obata, Y. Kojima, Astron. \& Astrophys. 
356, 234 (2000).
\bibitem{Haskell} B. Haskell, L. Samuelsson, K. Glampedakis, N. Andersson,
arXiv:0705.1780 [astro-ph].
\bibitem{Carter} B. Carter, in: C. DeWitt, B. S. DeWitt, {\it Black
 holes} - Les Houches 1972. Gordon \& Breack, New York (1973).
\bibitem{Oron} J. D. Bekenstein, E. Oron, Phys. Rev. D18, 1809 (1978);
J. D. Bekenstein, E. Oron, Phys. Rev. D19, 2827 (1979).
\bibitem{BGSM} S. Bonazzola, E. Gourgoulhon, M. Salgado, J. A. Marck, 
Astron. \& Astrophys., 278, 421 (1993).
\bibitem{PMP} J. F. Perez-Azorin, J. A. Miralles, J. A. Pons, 
Astron. \& Astrophys., 451, 1009 (2006). 
\bibitem{CM} S. Chandrasekhar, J. C. Miller,
Mon. Not. Roy. Astron. Soc. 167, 63 (1974).
\bibitem{Hartle1} J. B. Hartle, Astrophys. J. 150, 1005 (1967).
\bibitem{HartleThorne} J. B. Hartle, K. S. Thorne, Astrophys. J. 153,
807 (1968).
\bibitem{BFGM} O. Benhar, V. Ferrari, L. Gualtieri, S. Marassi, 
Phys. Rev. D72, 044028 (2005).
\bibitem{Thorne} K.S. Thorne, Rev. Mod. Phys. 52, 299 (1980).
\bibitem{LP} W.G. Laarakkers, E. Poisson, Astrophys. J. 512, 282 (1999).
\bibitem{ST} S. L. Shapiro, S. A. Teukolsky, {\it Black Holes, White
dwarfs and Neutron Stars}, John Wiley \& Sons (1983).
\bibitem{WFF} R.B. Wiringa, V. Fiks, A. Fabrocini, Phys. Rev. C 38, 
1010 (1988).
\bibitem{APR} A. Akmal, V.R. Pandharipande and D.G. Ravenhall, 
Phys. Rev. C58, 1804 (1998).
\bibitem{UCB} G. Ushomirsky, C. Cutler, L. Bildsten,
Mon. Not. Roy. Astron. Soc. 319, 902 (2000).
\bibitem{HJS} B. Haskell, D.I. Jones, N. Andersson,
Mon. Not. Roy. Astron. Soc. 373, 1423 (2006).  
\bibitem{Glen} N.K. Glendenning, {\it Compact Stars} (Springer, 
New York, 2000). 
\bibitem{Bag} A. Chodos, R.L. Jaffe, K. Johnson, C.B. Thorne and
V.F. Weiskopf, Phys. Rev. D 9, (1974) 3471.
\bibitem{BFG} O. Benhar, V. Ferrari, L. Gualtieri, 
Phys. Rev. D70, 124015 (2004).
\end{thebibliography}
\end{document}